\documentclass[11pt]{article}
\usepackage{jheppub}

\usepackage{graphicx}
\usepackage{dcolumn}
\usepackage{bm}
\usepackage{amssymb}
\usepackage{amsmath}
\usepackage{amsthm}
\usepackage{mathtools}
\usepackage[usenames,dvipsnames]{xcolor}
\usepackage{tikz}
\usepackage{tikz-cd}
\usetikzlibrary{arrows,shapes,positioning}
\usetikzlibrary{shapes.geometric, arrows}
\usepackage{appendix}
\usepackage[dvipsnames]{xcolor} 
\usepackage[utf8]{inputenc}
\usepackage{dsfont}
\usepackage{cancel}
\usepackage{comment}
\usepackage{graphicx}
\usepackage{graphbox}
\usepackage{mathrsfs}
\usepackage{fontawesome5}
\usepackage[normalem]{ulem}
\usepackage[skip=.5em, indent=1em]{parskip}
\usepackage{lscape}
\usepackage{ stmaryrd }

\usepackage{enumitem}
\setlist{  
  listparindent=\parindent,
  parsep=0pt,
}

\usepackage{hyperref}
\hypersetup{
	colorlinks = True,
   	citecolor = BrickRed,
   	linkcolor = MidnightBlue,
        urlcolor = MidnightBlue
}

\usepackage[multiple]{footmisc}

\usepackage[most]{tcolorbox}
\tcbset{
    enhanced jigsaw, 
    colback=gray!5!white,
    colframe=gray!20!white,
    coltitle=black,
    fonttitle=\bfseries
}

\makeatletter
\newcommand{\github}{%
   {\href{https://github.com/ASphericalCow/PhysicalBasisFromCuts.git}{\faGithub}}%
}
   
\makeatother

\newcommand{\nn}{\nonumber}
\renewcommand{\d}{\text{d}}
\newcommand{\D}{\text{D}}
\newcommand{\res}{\text{Res}}
\newcommand{\vep}{\varepsilon}
\newcommand{\vphi}{\varphi}

\newcommand{\la}{\langle}
\newcommand{\ra}{\rangle}
\newcommand{\mbf}[1]{\mathbf{#1}}
\newcommand{\bs}[1]{\boldsymbol{#1}} 
\newcommand{\mat}[1]{\underline{\boldsymbol{#1}}}
\newcommand{\del}[1]{\nabla}
\renewcommand{\c}[1]{\check{#1}}
\newcommand{\cdel}[1]{{\check{\nabla}}}

\newcommand{\cvphi}{\check{\vphi}}
\newcommand{\cphi}{\check{\phi}}
\renewcommand{\vec}[1]{\mbf{#1}}
\newcommand{\kin}{\mathrm{kin}}
\newcommand{\dtot}{\mathrm{D}}
\newcommand{\phys}{\mathrm{phys}}
\newcommand{\B}{\mathcal{B}}

\newcommand{\dlog}{\mathrm{dlog}}
\newcommand{\G}{\mathcal{G}}
\newcommand{\p}{\mathcal{P}}
\newcommand{\elll}{{(\ell)}}
\newcommand{\C}{\mathcal{C}}
\newcommand\fnsep{\textsuperscript{,}}

\newcommand{\tikzxmark}{%
\tikz[scale=0.23] {
    \draw[line width=0.7,line cap=round] (0,0) to [bend left=6] (1,1);
    \draw[line width=0.7,line cap=round] (0.2,0.95) to [bend right=3] (0.8,0.05);
}}

\title{A physical basis for cosmological correlators from cuts}

\author{Shounak De}
\emailAdd{shounak\_de@brown.edu}
\author{and Andrzej Pokraka}
\emailAdd{andrzej\_pokraka@brown.edu}

\affiliation{Department of Physics, Brown University, \\
182 Hope Street, Providence, RI 02912, U.S.A.}

\abstract{%
Significant progress has been made in our understanding of the analytic structure of FRW wavefunction coefficients, facilitated by the development of efficient algorithms to derive the differential equations they satisfy. 
Moreover, recent findings indicate that the twisted cohomology of the associated hyperplane arrangement defining FRW integrals overestimates the number of integrals required to define differential equations for the wavefunction coefficient. 
We demonstrate that the associated dual cohomology is automatically organized in a way that is ideal for understanding and exploiting the cut/residue structure of FRW integrals.
Utilizing this understanding, we develop a systematic approach to organize compatible sequential residues, which dictates the physical subspace of FRW integrals for \textit{any} $n$-site, $\ell$-loop graph. 
In particular, the physical subspace of tree-level FRW wavefunction coefficients is populated by differential forms associated to cuts/residues that factorize the integrand of the wavefunction coefficient into \emph{only} flat space amplitudes. 
After demonstrating the validity of our construction using intersection theory, we develop simple graphical rules for cut tubings that enumerate the space of physical cuts and, consequently, differential forms without any calculation. 
}

\begin{document}
\maketitle

\section{Introduction \label{sec:intro}}

Cosmological correlators, or primitively, the wavefunction of the universe, are important quantum mechanical observables in cosmology.
Recently, there has been tremendous progress in our perturbative understanding of the analytic structure of these observables in certain toy models of quantum cosmology, specifically for a theory of conformally coupled scalars with general polynomial interactions evolving in power-law Friedmann–Robertson–Walker (FRW) cosmologies. 

Underlying many of these advances are the cosmological bootstrap \cite{Arkani-Hamed:2018kmz, Baumann:2022jpr, Baumann:2019oyu, Baumann:2020dch, Pajer:2020wnj, Pajer:2020wxk} and polytope \cite{ArkaniHamed2017, Benincasa:2024leu, Benincasa:2019vqr} programs. 
The cosmological polytope provides a geometric and combinatorial way to compute wavefunction coefficients independent of the notions of locality, causality and unitarity baked into the formalism of Feynman diagrams.
Techniques from the Feynman integral community\footnote{Additional connections include  symbology \cite{Hillman:2019wgh} and Landau analysis \cite{Salcedo:2022aal, Lee:2023kno}.} (related to algebraic geometry) have been used to compute canonical differential equations for both massless and massive wavefunction coefficients 
\cite{De:2023xue, He:2024olr, gasparotto2024}. 
In particular, the kinematic flow algorithm has identified a powerful yet mysterious structure in these canonical differential equations 
\cite{Arkani-Hamed:2023kig, Arkani-Hamed:2023bsv, Baumann:2024mvm, Hang:2024xas}. 
After accounting for the loop-momentum integration, elliptic sectors in these differential equations were identified in 
\cite{Benincasa:2024ptf}; 
this is not surprising since the loop integrals defining wavefunction coefficients include all Feynman integrals, which are known to contain elliptic sectors. 
Ideas from tropical geometry and sector decomposition have been used to identify the origin of infrared divergences in the integration space and develop subtraction schemes \cite{Benincasa:2024lxe}. 
Hypergeometric solutions, relations to GKZ-systems and $D$-modules have also been explored  
\cite{Chen:2024glu, Grimm:2024tbg, Fevola:2024nzj,Fan:2024iek,Xianyu:2023ytd}. 
Cosmological cutting rules and implications from unitarity have been formalized 
\cite{Ema:2024hkj, Ghosh:2024aqd, AguiSalcedo:2023nds, Tong:2021wai, Baumann:2021fxj, Goodhew:2021oqg, Melville:2021lst, Goodhew:2020hob, Arkani-Hamed:2018bjr, Werth:2024mjg, Goodhew:2024eup} and will play an important role in this paper.

In this toy model of conformal scalars in power-law FRW cosmologies, the wavefunction coefficient corresponding to a given Feynman graph $\G$ is represented as a certain \textit{twisted} integral
\begin{align} \label{eq:wvfn}
    \psi^{(\ell)}_{n}
    = \int_0^\infty u \; \Psi_{n}^{(\ell)}
    \quad
    \Psi_{n}^{(\ell)} 
    := 
    \hat{\Omega}^{(\ell)}_{n} 
    (\mbf{x}+\mbf{X},\mbf{Y})\;
    \d^n\mbf{x}~,
\end{align}
where $n$ and $\ell$ are the number of vertices and loops in $\G$; for loop graphs, $\psi^{(\ell)}_{n}$ is the \textit{loop integrand}.
$\hat{\Omega}^{(\ell)}_{n}(\mbf{X},\mbf{Y})$ is the canonical function of the cosmological polytope $\p_{\G}$ associated to the graph $\G$ and is essentially the wavefunction coefficient for flat spacetimes. 
It depends on the kinematic parameters $X_i = \sum_{m=n_{i-1}+1}^{n_i} |\mbf{k}_m|$, the sum of external energies flowing from the vertex $i$ to the boundary, and $Y_i = |\sum_{m=n_{i-1}+1}^{n_i} \mbf{k}_m|$, the energy exchanged through the $i^\text{th}$ edge of a given graph $\G$:
\begin{center}
\includegraphics[align=c, width=.55\textwidth]{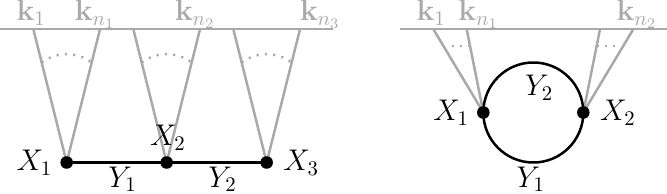}
\end{center}
The \emph{twist} $u$ is a multi-valued function associated to a given graph. 
It is a product of polynomials raised to specific powers that depend on the underlying power-law cosmology (see below \eqref{eq:psiphys}). 

The mathematical structure of these integrals is largely governed by the theory of twisted (co)homology \cite{aomoto2011theory, yoshida2013hypergeometric, Matsubara-Heo:2023ylc, De:2023xue, Caron-Huot:2021iev, Mastrolia:2018uzb, Caron-Huot:2021xqj, Frellesvig:2019uqt, Mizera:2019vvs}. 
Moreover, the twisted cohomology is a finite $|\chi|$-dimensional vector space, where $\chi$ is the Euler characteristic of the underlying hyperplane arrangement (equivalently, the number of bounded chambers). 
Due to the finite dimensionality of this vector space, the twisted integrals must satisfy coupled first-order linear differential equations (DEQs)
\begin{align}
    \d_\kin \int u\; \vphi_{i} = B_{ij} \int u\; \vphi_j
    \,.
\end{align}
Here, the $\vphi_{j=1, \dots, |\chi|}$ form a basis of differential forms for the twisted cohomology and $\d_\kin = \sum_{i=1}^n \d X_i\; \partial_{X_i} + \sum_{i=1}^{n+\ell-1} \d Y_i\; \partial_{Y_i}$ is the exterior derivative with respect to the kinematic parameters.

When the connection matrix $\mat{B}$ is proportional to the cosmological parameter $\vep$, $\mat{B} = \vep \mat{A}$, the associated basis is said to be \emph{canonical} \cite{Henn:2013pwa}. 
Such differential equations (DEQs) efficiently separate the kinematic dependence from the dependence on the cosmological parameter and are easy to integrate order-by-order in $\vep$. 
They play a central role in the evaluation of dimensionally regulated Feynman integrals in flat-space and now  cosmology.

Recently, efficient methods for computing the connection matrix $\mat{A}$ have been developed \cite{De:2023xue, Arkani-Hamed:2023bsv, Arkani-Hamed:2023kig, Benincasa:2024lxe, He:2024olr}. 
In particular, the kinematic flow algorithm \cite{Arkani-Hamed:2023bsv, Arkani-Hamed:2023kig, Hang:2024xas, Baumann:2024mvm} provides a diagrammatic way to choose a basis and evaluate the connection matrix $\mat{A}$. 
Surprisingly, these universal graphical rules pick out a distinguished subset of integrals that include the physical wavefunction coefficient \eqref{eq:wvfn} and whose DEQs close. 
This physical subspace of integrals is vastly smaller than the total size of the twisted cohomology, which is easy to see at tree-level where the kinematic flow basis has size $4^{n-1}$: 
\begin{center}
\begin{tabular}{c|cccc}
     {}
     &  \includegraphics[align=c, scale=.25]{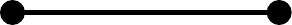}
     &  \includegraphics[align=c, scale=.25]{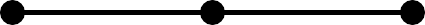}
     &  \includegraphics[align=c, scale=.25]{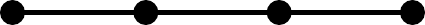}
     &  \includegraphics[align=c, scale=.25]{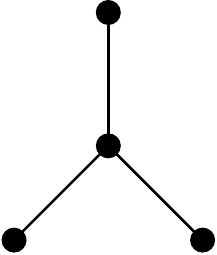}
     \\
     \hline
     $\chi$
     & 4
     & 25
     & 213
     & 312
    \\
    $4^{n-1}$
    & 4
    & 16
    & 64
    & 64
\end{tabular}
\end{center}
So far the kinematic flow algorithm is an empirical observation; there is no satisfactory characterization of the physical subspace. Using intuition from generalized unitarity and algebro-geometric tools, we provide physical and mathematical origin for the physical subset of the twisted cohomology.

Our paper is organized as follows. In section \ref{sec:review}, we review how to construct the relevant flat-space and FRW  wavefunction coefficients for any Feynman graph. 
We also introduce the minimal set of ideas from (relative) twisted cohomology and intersection theory---mathematics underpinning the structure of cosmological wavefunction coefficients---needed to identify the physical subspace.
In section \ref{sec:deq}, we demonstrate how the structure of the relative twisted cohomology---organized by cuts---clearly separates the space of all differential forms into three categories: physical, unphysical and degenerate.
As a pedagogical example, we illustrate how to construct a basis for the physical subspace of the 3-site chain graph by further separating the space of degenerate differential forms into physical and unphysical forms in section \ref{sec:3SitePed}. 
Section \ref{sec:linrel} systematically classifies the degeneracy of cosmological hyperplane arrangements by providing a simple graphical rule that generates all linear relations among hyperplane polynomials. 
Furthermore, we show how linear relations among hyperplane polynomials result in linear relations among residue operators and differential forms. 
In fact, the set of linear relations in conjunction with a cohomological condition is enough to classify all physical differential forms. 
Utilizing this, we develop graphical rules for constructing a basis of differential forms for the physical subspace by enumerating all physical cuts in section \ref{sec:goodTubes} for all tree and loop graphs.

\section{Correlators from twisted integrals \label{sec:review}}

As mentioned, our playground will be a theory of conformally-coupled scalars in a power-law FRW cosmology with (non-conformal) polynomial interactions. The action for such a theory in a $(d+1)$-dimensional spacetime is 
\begin{align} 
    \mathcal{S} = \int \d^d x \, \d\eta \sqrt{-g} \left[-\frac{1}{2} g^{\mu \nu} \partial_{\mu} \phi \partial_{\nu} \phi - \frac{d-1}{8d} R \phi^2 - \sum_{p \geq 3} \frac{\lambda_p}{p!} \phi^p \right]~,
    \label{eq:actioninFRW}
\end{align}
where the FRW metric $\d s^2 = a^2(\eta) \left[-\d \eta^2 + \d x_i \d x^i\right]$ is written in comoving coordinates with conformal time $\eta \in (-\infty, 0]$ and the index $i=1,\dots,d$ runs over the spatial dimensions. The scale factor is a power-law controlled by the cosmological parameter $\vep$: $a(\eta) = (\eta/\eta_{0})^{-(1+\varepsilon)}$. 
Setting $\vep$ to specific integer values recovers well studied cosmologies \cite{De:2023xue, Arkani-Hamed:2023kig}. In particular, $0<\vep\ll1$ is expected to capture the essential features of a near de Sitter inflationary cosmology.

Central to our analysis is the wavefunction of the universe $\Psi[\Phi]$. Formally, this is computed as a path integral by integrating over all bulk field configurations $\phi(\vec{x},\eta)$ with non-vanishing Dirichlet boundary condition in the future $\phi(\vec{x},\eta=0)=\Phi(\vec{x})$. It is standard to work in momentum space where the wavefunction has the expansion
\begin{align}
    \Psi[\Phi] &= \int_{\phi(-\infty(1-i\epsilon))}^{\phi(0)=\Phi} \hspace{-3pt} \mathcal{D} \phi \, e^{i S[\phi]} \nonumber \\
    &{\equiv} \exp{\left[i \sum_n \frac{1}{n!} \int  \left( \prod_{i=1}^n \frac{\d^d \vec{k}_i}{(2\pi)^d} \, \Phi(\vec{k}_i)\right) \, \int\left( \prod_{j=1}^\ell \d^d\vec{L}_j \right)
        \psi^{(n,\ell)}(\{\vec{k}_i\})
    \ \delta^{(d)}\left(\sum_{i=1}^n \vec{k}_i\right)\right]},
    \label{eq:Uniwavefunc}
\end{align}
and the standard $i \epsilon$ prescription selects the adiabatic/Bunch-Davies/Hartle-Hawking vacuum at the early-time boundary $\eta \to -\infty$. 
The kernels $\psi^{(n,\ell)}$ are the $n$-point 
\textit{wavefunction coefficients} with $n$ denoting the number of field insertions on the late-time boundary and $\{\vec{k}_i\}$ the corresponding set of all \textit{spatial} momentum.
In our analysis, we will be more interested in the number of vertices (or equivalently sites) in a graph $\G$ rather than the number of external fields on the future boundary. To differentiate this fact, the wavefunction coefficient for the $n$-site, $\ell$-loop graph will be denoted by $\psi_n^{(\ell)}$.
There has been significant progress in the perturbative computations of these wavefunction coefficients in the Bunch-Davies state, both for conformally-coupled scalars as well as massless and light states \cite{ArkaniHamed2017, Benincasa:2019vqr}. 
Also, we note that $\vec{L}_j$ denote the loop momenta and consequently the wavefunction coefficient $\psi_n^{(\ell)}$ corresponds to the \emph{loop integrand} of a given graph.
For a discussion on the associated loop integrals and recent progress, please refer to \cite{Benincasa:2024lxe, Lee:2023jby, Benincasa:2024ptf, Chowdhury:2023arc}.

The FRW wavefunction coefficients $\psi_n^{(\ell)}$ can be constructed out of their (\textit{shifted}) flat-space counterparts $\hat{\Omega}_n^{\elll}$ (as expressed in \eqref{eq:wvfn}), which admit a purely geometric description in terms of the associated cosmological polytope.
We review the cosmological polytope and the corresponding hyperplane arrangement in section \ref{sec:UniversalIntegrand}. 
Then in section \ref{sec:FRWUplift}, we elaborate on how to uplift the flat-space canonical form to FRW spacetimes and the associated twisted cohomology.
Section \ref{sec:relCohom} reviews the the intersection number and dual relative twisted cohomology; our main tools for analyzing the analytic structure of FRW integrals.

\subsection{The universal integrand \label{sec:UniversalIntegrand}}

From a traditional QFT perspective, the FRW wavefunction coefficient $\psi_n^{(\ell)}$ is represented by a Feynman diagram $\G$ with $n$ vertices and $\ell$ loops.  
However, the corresponding flat space coefficient $\hat\Omega_n^{(\ell)}$ from which it emerges can be defined without ever referencing ``Feynman rules'' and has a purely combinatorial-geometric origin as the \emph{canonical form} of the cosmological polytope $\p_\G$ \cite{ArkaniHamed2017}.

The cosmological polytope $\p_\G$ is a \emph{positive geometry} encoding the singularity structure of wavefunction coefficients for conformally coupled scalars in flat spacetimes.
A canonical form is the unique differential form (up to sign) that can be associated to a bounded region $\check{\Delta}$ with logarithmic singularities on its corresponding boundaries $\partial\check{\Delta}$. 
Moreover, the residue of a canonical form with respect to a boundary divisor is the canonical form of that boundary.
The existence of a canonical form whose residues are all canonical forms is a recursive way to define a positive geometry \cite{Arkani-Hamed:2017tmz}.

For a given graph $\G$ (without external legs), there are a total of $2n+\ell-1$ kinematic parameters: $\{X_i\}_{i=1}^{n}$ and $\{Y_i\}_{i=1}^{n+\ell-1}$. The hyperplane polynomials, $S_i$, are generated by the 1-tubings of $\G$. 
A 1-tubing $\tau_i$ consists of two sets: the vertices $\mathcal{V}_i$ encircled by the tubing and the edges $\mathcal{E}_i$ that this tubing crosses. 
To each 1-tubing $\tau_i$ of the Feynman graph, the associated hyperplane polynomial is 
\begin{align} \label{eq:Sdef}
    S_i(\mbf{X},\mbf{Y}) = \sum_{v\in \mathcal{V}_i} X_v + \sum_{e \in \mathcal{E}_i} Y_e
    \,.
\end{align}    
For example, the hyperplane polynomials for the 3-site tree level graph are
\begin{equation}
\label{eq:3siteS}
\begin{aligned}
    S_1 & = 
    \,\includegraphics[align=c,width=4em]{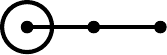}\,
    =  X_1 {+} Y_1~,
    &
    S_2 &=
    \,\includegraphics[align=c,width=4em]{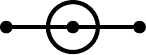}\,
    = X_2 {+} Y_1 {+} Y_2~,
    \\
    S_3 &= 
    \,\includegraphics[align=c,width=4em]{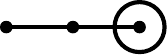}\,
    = X_3 {+} Y_2~, 
    &
    S_4 &{=} 
    \,\includegraphics[align=c,width=4em]{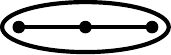}\,
    = X_1 {+} X_2 {+} X_3~,
    \\
    S_5 &= 
    \,\includegraphics[align=c,width=4em]{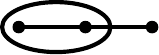}\,
    = X_1 {+} X_2 {+} Y_2~,
    &
    S_6 &= 
    \,\includegraphics[align=c,width=4em]{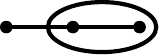}\,
    = X_2 {+} X_3 {+} Y_1~.
\end{aligned}
\end{equation}
This 3-site tree level chain graph will be an important pedagogical example to illustrate our main ideas in section \ref{sec:3SitePed}.

Usually, the simplest way to compute the canonical form is by summing rational functions associated with the set of compatible complete tubings (non-crossing $(2n{+}\ell{-}1)$-tubings) of a given graph $\G$ \cite{ArkaniHamed2017}.
Explicitly, if $\mathsf{T}$ is the set of compatible complete tubings, the canonical form $\Omega^{(\ell)}_n$ and function $\hat{\Omega}^{(\ell)}_n$ are 
\begin{align}\label{eq:canForm}
    \Omega^{(\ell)}_n = 
    \hat{\Omega}^{(\ell)}_n(\mbf{X},\mbf{Y})
    \frac{\d^n\mbf{X} \wedge \d^{n{+}\ell{-}1}\mbf{Y}}{\text{GL}(1)}
    ,\qquad 
    \hat{\Omega}^{(\ell)}_n(\mbf{X},\mbf{Y})
    = \mathcal{N}_n^\elll \sum_{\tau \in \mathsf{T}} 
    \prod_{t\in\tau} \frac{1}{S_t}
    \,,
\end{align}
and $\mathcal{N}_n^\elll = \prod_{i=1}^{n+\ell-1}2Y_i$ is a normalization factor ensuring that the canonical form has unit residues.
Returning to the 3-site example, the canonical function is  
\begin{align}\begin{aligned} \label{eq:3siteCanForm}
    \hat{\Omega}^{(0)}_{\text{3-chain}}(\mbf{X},\mbf{Y})
    &= 4Y_1Y_2
    \left[
        \includegraphics[align=c,width=.17 \textwidth]{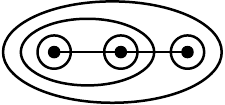}
        {+}
        \includegraphics[align=c,width=.17 \textwidth]{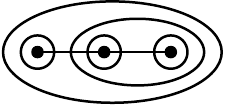}
    \right]
    = \frac{4Y_1Y_2}{S_1 \cdots S_4}
    \left(
        \frac{1}{S_5}
        {+}
        \frac{1}{S_6}
    \right)
    \,
\end{aligned}\end{align}
where, in this context, each 5-tubing corresponds to the inverse of the product of the depicted $S_i$'s.

\subsection{From flat-space to FRW spacetimes \label{sec:FRWUplift}}

Importantly, the flat-space wavefunction coefficients/canonical forms define a  universal integrand for their counterparts in \emph{any} power-law FRW spacetime. 
For any $\vep \notin\mathbb{Q}$, this is obtained by integrating the shifted flat-space wavefunction against a specific kernel $u$ called the \emph{twist},
\begin{align} \label{eq:psiphys}
    \psi_{n}^{(\ell)} = 
    \int_0^\infty 
    u\  \Psi_{n}^{(\ell)}
    \,,
    \quad\text{where}\quad
    \Psi_{n}^{(\ell)} = 
    \hat{\Omega}_n^{(\ell)}(\mbf{x}+\mbf{X},\mbf{Y})\;
    \d^n\mbf{x}
    \,,
\end{align}
Here, $u=\prod_{i=1}^n x_i^{\alpha_i}$ is a multi-valued function ($\alpha_i\notin\mathbb{Z}$) called the \emph{twist} whose vanishing loci defines the twisted hyperplanes $\mathcal{T}_i = \{x_i=0\}$ (coordinate hyperplanes). 
The exponents $\alpha_i$ are related to the underlying power-law cosmology ($\vep$), number of spatial dimensions ($d$) and valency of the $i^\text{th}$ vertex ($p_i$):  \begin{align}
    \alpha_i = d + \vep (d + 1) + \frac{1}{2} p_i (1 + \vep)(1 - d) \notin \mathbb{Z}
    \,.
\end{align} 
The condition $\vep \notin \mathbb{Q}$ guarantees that $\alpha_i \notin \mathbb{Z}$, which is essential for the techniques of twisted cohomology employed in this paper. 
However, it is important to note that one can often specialize to the case $\alpha_i\in\mathbb{Z}$ by an analytic continuation of the case $\alpha_i\notin\mathbb{Z}$. 
In the remainder of this paper, we assume $d=3$ and $p_i=3$ such that $\alpha_i=\vep$ for all vertices and $u=\prod_{i=1}^n x_i^\vep$.%
\footnote{Technically, our cosmological model \eqref{eq:actioninFRW} needs at least quartic interactions (only terms with $p\geq4$ in \eqref{eq:actioninFRW}) to ensure that the kinematic space of the $(\ell\geq1)$-loop 2-site graphs are non-trivial. However, when we treat these integrals in section \ref{sec:goodTubes}, the specific form of the twist is not important.}

As seen above in \eqref{eq:psiphys}, the FRW form $\Psi_{n}^{(\ell)}$ is simply generated from the canonical function $\hat{\Omega}^{(\ell)}_n(\mbf{X},\mbf{Y})$ of the cosmological polytope by a shift in its $\mbf{X}$-variables.
We denote the shifted linear factors appearing in the FRW form $\Psi_{n}^{(\ell)}$ by 
\begin{align}
    B_i(\mbf{x};\mbf{X},\mbf{Y}) = S_i(\mbf{x}+\mbf{X},\mbf{Y})
    \;. 
\end{align}
Then, 
\begin{align}
    \Psi_{n}^{(\ell)} = \d^n\mbf{x}\; 
    \left(\prod_{i=1}^{n+\ell-1} 2 Y_i\right)
    \left( 
        \sum_{\tau \in \mathsf{T}} 
        \prod_{t\in\tau} \frac{1}{B_t}
    \right)
    \;, 
\end{align}
where $\mathsf{T}$ is the set of compatible complete tubings of a given graph $\G$.

Since our objective is understanding the mathematical structure of wavefunction coefficients for arbitrary $\vep$, our primary focus is the hyperplane arrangement generated by both the $B_i$ and $T_i$ not the cosmological polytope $\p_\G$ (generated solely by the $S_i$'s).%
\footnote{We refer to this structure as the ``FRW cosmological polytope'' to remind ourselves that it is defined in $\mbf{x}$-space like the FRW form $\Psi_n^{\elll}$.
In an abuse of notation, we will also refer to the FRW cosmological polytope by $\p_{\G}$.
}
Explicitly, we wish to classify the differential forms on the topological space $M \setminus \B$ where
\begin{align}\begin{aligned}\label{eq:M}
    M {:=} \mathbb{C}^n \setminus \mathcal{T}
    \,,
    \quad
    \mathcal{T} &{:=} \bigcup_{i=1}^n \mathcal{T}_i
    \,,
    \quad 
    \mathcal{T}_i {:=} \{x_i =0\}
    \,,
    \quad
    \B {:=} \bigcup_j \B_j
    \,,
    \quad\text{and}\qquad
    \B_j {:=} \{B_j=0\}
    \,. 
\end{aligned}\end{align}
This requires understanding the associated twisted cohomology $H^n(M\setminus\mathcal{B}; \nabla)$;  the cohomology with respect to the  covariant derivative $\nabla := \d + \omega \wedge$ with flat connection $\omega = \dlog\, u$
\begin{align}\label{eq:twCohom}
    H^n(M\setminus\B;\nabla) := 
    \frac{\text{covariantly closed $n$-forms on $M\setminus \B$}}{\text{covariantly exact $n$-forms on $M\setminus \B$}}
    \;.
\end{align}
It classifies all non-trivial (do not integrate to zero) differential $n$-forms on $M\setminus\B$ that can have poles on the untwisted singular loci $\mathcal{B}$ and twisted loci $\mathcal{T}$ (coordinate axes). 
Explicitly, the twisted cohomology of FRW integrals is populated by forms
\begin{align} \label{eq:vphiEx}
    \vphi \sim
    \frac{P(\mbf{x})\; \d^n\mbf{x}}{\prod_i T_i^{\mu_i} \prod_j B_j^{\nu_j}}
    \in H^n(M\setminus\B;\nabla)
\end{align}
where $\mu_i, \nu_i \in \mathbb{Z}$ and $P(\mbf{x})$ is any polynomial in $\mbf{x}$.
While the above is automatically covariantly closed $(\nabla\vphi=0)$ since it is a holomorphic top-form, it may or may not be covariantly exact. 
Depending on $P$, it is possible for $\vphi$ to be exact $\vphi = \nabla [(n-1)\text{-form}] \simeq 0$ and lie outside of $H^n(M\setminus\B;\nabla)$.

Importantly, the integrand of these twisted integrals (i.e., $u\ \Psi_n^{(\ell)}$ or any other $u\ \vphi$) exhibit very different behavior near the twisted and untwisted singular loci. 
Due to the twist $u$, there exists a choice of $\vep$ such that integral in a local neighborhood of where an $x_i$ vanishes converges; the integral for all other ``non-convergent'' values of $\vep$ are well defined through analytic continuation. 
On the other hand, nothing saves the integral from blowing up whenever a $B_i$ vanishes. 
Therefore, singularities at $B_i=0$ and $x_i=0$ must be managed differently. 
This has already been partially accounted for in the covariant derivative $\nabla$ since the connection $\omega$ has singularities on the twisted hyperplanes. 
From the perspective of contours, $B_i$-residues of the integrand are allowed but $x_i$-residues are not since the coordinate axes are branch surfaces ($u \sim x_i^{\alpha_i}$ with $\alpha_i \notin \mathbb{Z}$).\footnote{For forging connections to cosmological spacetimes one will generally encounter cases where $\alpha_i=\vep \in \mathbb{Z}$. We note that the DEQs (and their solutions) for twisted cosmological integrals can be analytically continued in $\vep$ and one can extract the physical wavefunction by taking appropriate limits.}
This distinction places crucial restrictions on the kind of allowed contours and cuts. 
This will be most obvious in the dual cohomology defined in section \ref{sec:relCohom} where each dual form can be thought of as a coarse approximation to a cut contour. 
This is made precise in appendix \ref{app:dualCanForm} and will play an important role in the development of a diagrammatic coaction method \cite{coaction, Abreu:2019wzk}.

\subsection{The intersection number and relative twisted cohomology  \label{sec:relCohom}}

The \emph{intersection number} $\la \cvphi \vert \vphi \ra$ defines an \emph{inner product} on the space of twisted differential forms (and therefore, cosmological integrals) by pairing an element of FRW cohomology $\vphi \in H^n(M\setminus\mathcal{B};\nabla)$ (recall \eqref{eq:twCohom}) with an element of the associated  dual cohomology $\cvphi \in \c{H}^n$ (defined in \eqref{eq:relCoDef}).
The dual cohomology has more structure than its FRW counterpart since it is essentially the direct sum of the the twisted cohomology of each cut. 
Therefore, it is automatically organized in an ideal way for understanding cuts.
It is also easily linked to the space of contours on the topological space $M\setminus\mathcal{B}$ (see appendix \ref{app:dualCanForm}).

The intersection number is the link between the FRW cohomology and dual cohomology 
\begin{align}
    \la \cvphi \vert \vphi \ra: 
    H^n(M,\mathcal{B}; \c{\nabla})
    \times 
    H^n(M\setminus\mathcal{B};\nabla) 
    \mapsto \mathbb{C}
    \,.
\end{align}
For FRW correlators,  the intersection number is easy to compute; it is simply a weighted residue where the dual forms dictate the exact linear combination (\eqref{eq:cobdAct} and \eqref{eq:CombInt}). 
Moreover, the intersection number facilitates a simple formula for computing the differential equation in a way that emphasizes the role of cuts (demonstrated in section \ref{sec:deq}). 
From this formula, we deduce that an FRW form $\vphi$ couples to $\partial_{X_i} \Psi_n^{(\ell)}$ or $\partial_{Y_i} \Psi_n^{(\ell)}$ if and only if $\vphi$ shares a cut with the FRW form $\Psi_{n}^{(\ell)}$ (also in section \ref{sec:deq}).

The dual cohomology to \eqref{eq:twCohom} is the \emph{relative}%
\footnote{Name for the cohomology of a topological space with boundaries ($\B$ in our case).}
twisted cohomology 
\begin{align} \label{eq:relCoDef}
    \c{H}^n &:= H^n(M,\mathcal{B}; \c{\nabla})
    \subseteq \bigoplus_{p=0}^n \bigoplus_{|J|=p}
    \delta_{J}\left(H^{n-p}(M_J; \c{\nabla}\vert_J)\right)
    \,,
    &
    \c{\nabla}=\d-\omega\wedge
    \,,
\end{align}
where $J$ is a multi-index denoting a cut, $M_J = M \cap \mathcal{B}_J$ is the topological space associated to this cut and $\mathcal{B}_J = \cap_{j\in J} \{B_j=0\}$.
Importantly, dual forms are represented by twisted differential forms localized to the cuts $M_J$.
The localization to the cut $M_J$ is denoted by the coboundary symbol $\delta_J$\footnote{See \cite{De:2023xue,Caron-Huot:2021iev,Caron-Huot:2021xqj} for a more in-depth explanation of the coboundary and its properties. See \cite{hwa1966homology} for a more formal exposition in the context of (untwisted) homology.}
and its argument contains a differential form on $M_J$:
\begin{align}
    \cvphi &= \sum_J \delta_J(\phi_J)
    \,,
    &
    \phi_J & \in H^{n-|J|}(M_J; \c{\nabla}\vert_J)
    \,.
\end{align}
The $\delta_J$ indexes the component of the direct sum corresponding to the $J$-boundary cohomology and keeps track of boundary terms.%
\footnote{While not explicitly used in this work, boundary terms are generated by the derivative in the dual relative twisted cohomology: $\d\delta_J(\phi_J) = (-1)^{|J|} \big[\delta_J(\d\phi_J) + \sum_{i\notin J} \delta_{Ji}(\phi_J\vert_i)\big]$. \label{foot:bdTerms}}
They are also anti-symmetric in their indices $\delta_{ij} = - \delta_{ji}$ and induce a residue/cut inside the intersection number
\begin{equation} \label{eq:cobdAct}
    \la \delta_J(\cphi) \vert \vphi \ra
    = \Big\la \cphi \Big\vert \res_J[\vphi] \Big\ra_J
    =: \la \cphi \vert \phi_J \ra_J
    \,,
\end{equation}
where 
\begin{align}\begin{aligned}
    &\res_J[\bullet] := \res_{j_1 \cdots j_{|J|}}[\bullet] :=
    \res_{\B_{j_{|J|}}\cap\cdots\cap\B_{j_2}\cap \B_{j_1}}
    \circ
    \res_{\B_{j_2}\cap \B_{j_1} } 
    \circ
    \res_{\B_{j_1}}~,
\end{aligned}\end{align}
denotes the iterated residue.

After taking care of the coboundary operators, the remaining intersection number $\la \cphi \vert \phi_J \ra_J$ localizes to the maximal codimension intersection points of the twisted surfaces  on the cut.%
\footnote{\label{foot:CombInt}Explicitly,
\begin{align} \label{eq:CombInt}
    \la \cphi \vert \phi_J \ra_J
    =
        \sum_{\bs{z}^* \in \text{Int}_J} 
        \frac{
            \text{Res}_{\bs{z}^*=\bs{0}}[\cphi]
            \text{Res}_{\bs{z}^*=\bs{0}}[\phi_J]
        }{
            \prod_{i=1}^{n-|J|} \res_{z^*_i=0}[\omega]
        } 
    \,,
\end{align}
and only plays a minor role in the remainder of this paper. 
Here, $\text{Int}_J = \bigcup_{K:|K| = n-|J|}\; (\B_J \cap \mathcal{T}_K)$ and $\mathcal{T}_K := \bigcap_{k\in K} \mathcal{T}_k$. See appendix A of \cite{De:2023xue} for a more detailed discussion.
}
Since $\la \cphi \vert \phi_J \ra_J$ only plays a minor role in the remainder of this article, its definition has been relegated to footnote \ref{foot:CombInt}.
While equation \eqref{eq:cobdAct} (and \eqref{eq:CombInt}) are sufficient for our purposes, it is important to note that they are \emph{only valid for logarithmic differential forms}; there are known corrections to these formulas if the forms have higher order poles \cite{matsumoto1998kforms, Mizera:2019vvs}.

\section{DEQs from residues and finding the physical subspace \label{sec:deq}}

We choose a spanning set  for both dual and FRW forms that makes the cut structure manifest. 
The first step is to determine which cuts have a non-trivial twisted cohomology. 
This is done by computing the Euler characteristic of each cut surface $\chi_J =\chi(M_J)$ and putting $|\chi_J|$-many canonical forms into the coboundary operator
\begin{align} \label{eq:dualBasis}
    \{ \cvphi_a \}_{a=1}^{\sum_J |\chi_J|} = 
    \bigcup_{J} \bigg\{ \delta_J(\Omega_{J,k}) \bigg\}_{k=1}^{|\chi_J|}~,
\end{align}
Importantly, the canonical forms are non-singular at infinity and are always a linear combination of $\dlog T_i\vert_J \wedge \cdots \wedge \dlog T_k\vert_J$.
Dual forms are closely related to cut contours; they are the \emph{dual canonical form} of a cut contour and the intersection number is the leading order term of the cut integral (see appendix \ref{app:dualCanForm}). 

For the FRW cohomology, we choose the set 
\begin{align} \label{eq:frwBasis}
    \{\vphi_a\}_{a=1}^{\sum_J |\chi_J|} = 
    \bigcup_J \left\{ 
    \dlog_J
    \wedge \tilde{\Omega}_{J,k} 
    \right\}_{k=1}^{|\chi_J|}
    \quad\text{where}\quad
    \dlog_J := \bigwedge_{j\in J} \dlog B_j 
    \,,
\end{align}
and $\tilde{\Omega}_{J,k}$ is obtained from $\Omega_{J,k}$ by removing the restriction to the cut $J$ acting on the $T_i$'s (an explicit example of this procedure is given in section \ref{sec:3SitePed} and diagrammatic rules are provided in section \ref{sec:goodTubes}). 
By organizing the bases according to the cuts automatically produces a block diagonal intersection matrix 
\begin{align} \label{eq:blockDiag}
    C_{ab} 
    := \la \cvphi_a \vert \vphi_b \ra
    = \la \delta_I(\Omega_{I,k}) \vert \dlog_J \wedge \tilde{\Omega}_{J,l} \ra 
    = \delta_{IJ}  \la \Omega_{J,k} \vert \Omega_{J,l} \ra_J~. 
\end{align}
where $\Omega_{J,k} = \tilde{\Omega}_{J,k}\vert_J$.
Note that it is easy to translate between these two sets via the replacement $\delta_J(\Omega) \leftrightarrow \dlog_J\wedge\tilde{\Omega}$.%
\footnote{\label{foot:almostDual}%
Technically, the sets \eqref{eq:dualBasis} and \eqref{eq:frwBasis} are not quite dual since the intersection matrix \eqref{eq:blockDiag} is only block diagonal and not full rank. 
However, this technicality will dealt with shortly in a simple way. 
}

The intersection number facilitates a simple formula for the DEQs in terms of residues that highlights the importance of cuts.
Schematically, the connection matrix takes the form%
\footnote{The precise formula is 
$    
A_{ab} = C_{bc}^{-1} \la \cvphi_{c} \vert \nabla_\kin \vphi_a \ra
$. 
It will be used to compute the DEQs in section \ref{sec:3SitePed} as well as appendices \ref{app:4SiteChain} and \ref{app:3Site1Loop}.} 
\begin{align} \label{eq:DEQFromRes}
    A_{ab} \sim
    \Big\la \delta_{J_b}(\cphi_b) \Big\vert \nabla_\kin \vphi_a \Big\ra
    = 
    \Big\la \cphi_b \Big\vert \res_J[\nabla_\kin \vphi_a] \Big\ra_{J_b}
    \to 0 
    \text{ whenever } \res_{J_b}[\nabla_\kin\vphi] = 0
    \,. 
\end{align}
This equation implies that $\vphi_b$ couples to $\vphi_a$ in the DEQ if and only if both $\vphi_a$ and $\vphi_b$ share a residue since%
\footnote{In order to use  equations \eqref{eq:cobdAct} and  \eqref{eq:CombInt}, one needs a representation for  $\nabla_\kin\vphi$ that has at most simple poles.
We provide a simple recipe for such a representative for $\nabla_\kin\vphi$  \emph{without} invoking integration-by-parts in appendix \ref{app:noNewPoles}.
We also derive \eqref{eq:resdphi_iff_resphi} as a consequence.}
\begin{align} \label{eq:resdphi_iff_resphi}
    \res_{J_b}[\nabla_\kin\vphi_a]=0 
    \iff 
    \res_{J_b}[\vphi_a] = 0
    \,.
\end{align}
This implies that the physical subspace is spanned by all FRW forms in \eqref{eq:frwBasis} that have overlapping residues with the physical wavefunction. 
More explicitly, 
\begin{align} \label{eq:physBasis}
    \tcboxmath{
    H^n_\phys \subset 
    \text{Span}
    \left\{
        \dlog_J \wedge \tilde{\Omega}_{J}
    \right\}_{J}
    \quad\text{such that} \quad \res_J[\Psi_{n}^{(\ell)}] \neq 0
    \,.
    }
\end{align}
This is the first main result of this paper. 
Moreover, \eqref{eq:DEQFromRes} and \eqref{eq:resdphi_iff_resphi} imply that $\dlog_J$-forms only couple to themselves and $\dlog_{J \setminus j}$-forms where $j\in J$.

If the FRW hyperplane arrangement was non-degenerate, \eqref{eq:physBasis} would be the whole story.
However, as we will see shortly (section \ref{sec:3SitePed}), even simple cosmological examples are \emph{non-generic} or \emph{degenerate} and we have to work a bit harder to reduce \eqref{eq:physBasis} from a spanning set to a basis. 
We will discuss how to systematically work with this degeneracy in the remainder of this article.

\section{A pedagogical example: the 3-site chain}
\label{sec:3SitePed}

We illustrate how the physical subspace arises by organizing the basis of forms according to whether they have compatible sequential residues with the physical FRW form in the context of the simplest example: the 3-site chain.
Moreover, we highlight how the associated hyperplane arrangement dictates this organization; a ubiquitous feature for all tree-level and loop-level cosmological arrangements. 
We comment on further worked examples (the 4-site chain, 4-site star and 1-loop 3-gon of appendix \ref{sec:examples}) near the end of this section.

\paragraph{The hyperplane polynomials:} 
For the convenience of the reader, we collect the  3-site FRW hyperplane polynomials $B_i$ (shifted versions of \eqref{eq:3siteS}) 
\begin{equation}
\label{eq:3siteB}
\begin{aligned}
    B_1 & = 
    \,\includegraphics[align=c,width=4em]{figs/3site/S1.pdf}\,
    =  x_1 {+} X_1 {+} Y_1~,
    &
    B_2 &=
    \,\includegraphics[align=c,width=4em]{figs/3site/S2.pdf}\,
    = x_2 {+} X_2 {+} Y_1 {+} Y_2~,
    \\
    B_3 &= 
    \,\includegraphics[align=c,width=4em]{figs/3site/S3.pdf}\,
    = x_3 {+} X_3 {+} Y_2~, 
    &
    B_4 &{=} 
    \,\includegraphics[align=c,width=4em]{figs/3site/S4.pdf}\,
    = x_1 {+} x_2 {+} x_3 {+} X_1 {+} X_2 {+} X_3~,
    \\
    B_5 &= 
    \,\includegraphics[align=c,width=4em]{figs/3site/S5.pdf}\,
    = x_1 {+} x_2 {+} X_1 {+} X_2 {+} Y_2~,
    &
    B_6 &= 
    \,\includegraphics[align=c,width=4em]{figs/3site/S6.pdf}\,
    = x_2 {+} x_3 {+} X_2 {+} X_3 {+} Y_1~.
\end{aligned}
\end{equation}

\paragraph{The FRW form:} 
Following the discussion in section \ref{sec:FRWUplift}, the FRW form (shifted version of \eqref{eq:3siteCanForm}) is 
\begin{align} \label{eq:3siteFRWForm}
    \Psi_3^{(0)}= 
    \frac{4Y_1Y_2}{B_1 \cdots B_4}\left(\frac{1}{B_5}+\frac{1}{B_6}\right)
    \d^3\mbf{x}
    \,.
\end{align}
Recalling \eqref{eq:M}, this is a differential form on the space $M\setminus\mathcal{B}$ where 
\begin{align}
    M &:= \mathbb{C}^3\setminus\left\{x_1 x_2 x_3 = 0\right\}
    \,,
    & 
    \mathcal{B} &:= \bigcup_{i=1}^6 \mathcal{B}_i 
    \,,
    &
    \mathcal{B}_i &:= \{B_i=0\}
    \,.
\end{align}
In the following, we will also be interested in the space $M$ after taking a sequential residue $\res_J$. 
We denote these topological spaces by $M_J := M \cap \mathcal{B}_J$ where $\B_J = \cap_{j\in J} \mathcal{B}_j$.

\paragraph{A linear relation among hyperplane polynomials:} 
The associated hyperplane arrangement is highly non-generic or degenerate due to a linear relation between the FRW hyperplane polynomials of \eqref{eq:3siteB}.%
\footnote{By non-generic or degenerate, we mean that there exists a collection of $|J|$ hyperplanes $\{\B_{j\in J}\}$ that intersect in codimension-$m$ where $m<|J|$.}
A graphical representation for the 3-site hyperplane polynomial relation is 
\begin{align}\begin{aligned} 
\label{eq:3-sitelinrelnew}
    \includegraphics[align=c, width=.2\textwidth]{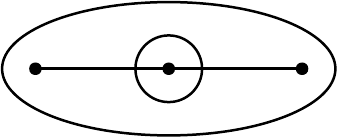}
    \bigg\vert_+
    = \includegraphics[align=c, width=.2\textwidth]{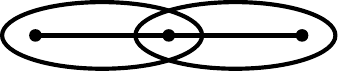}
    \bigg\vert_+
    \iff 
    B_2 + B_4 = B_5 + B_6
    \,, 
\end{aligned}\end{align}
where $\vert_+$ instructs us to add the $B_i$ associated to the 2-tubings. 
This relation implies that the four hyperplanes $\{\B_4,\B_5,\B_6,\B_2\}$ intersect in codimension-3 instead of codimension-4.
Therefore, the geometry of the following 3-cuts is not unique: $M_{4,5,2}=M_{4,6,2}=M_{4,5,6}=M_{5,6,2}$.
This will have important consequences for the space of residue operators and differential forms.

\paragraph{Cuts and residues:}
Following the discussion in section \ref{sec:relCohom} (in particular \eqref{eq:relCoDef} and \eqref{eq:dualBasis}),  it is clear that the dual cohomology is built by constructing the cohomology of each cut. 
Excluding the 0-cut, which is $M$ itself (recall \eqref{eq:M}), the 3-site chain has a total of 37 cuts:
\begin{itemize}
    
\item There are 6 possible 1-cuts (the space $M$ after taking the residues on each of the six $B_i$'s in \eqref{eq:3siteB})
\begin{align}
    \left\{
        {\color{gray}M_1}
        ,
        {\color{gray}M_2}
        ,
        {\color{gray}M_3}
        ,
        {\color{gray}M_5}
        ,
        {\color{gray}M_6}
        ,
        M_4
    \right\}
    \,. 
\end{align}
These are 2-dimensional topological spaces (codimension-1). 
As we will see, the {\color{gray}gray $M_J$} will turn out to have trivial cohomology.

\item There are $\binom{6}{2}=15$ possible 2-cuts, each of which are 1-dimensional (codimension-2)
\begin{align}\begin{aligned}
    &\left\{
        {\color{gray}M_{1,2}}
        ,
        {\color{gray}M_{1,3}}
        ,
        {\color{gray}M_{5,1}}
        ,
        {\color{gray}M_{2,3}}
        ,
        {\color{gray}M_{5,2}}
        ,
        {\color{gray}M_{6,2}}
        ,
        {\color{gray}M_{6,3}}
        ,
        \right.\\&\qquad\qquad\left.
        M_{4,1},M_{6,1},M_{4,2},M_{4,3},M_{5,3},
        M_{4,5},M_{4,6},M_{5,6}
    \right\}
    \,.
\end{aligned}\end{align}

\item Lastly, there are sixteen 3-cuts
\begin{align}
    \Big\{&
        M_{1,2,3},M_{4,1,2},M_{6,1,2},M_{4,1,3},
        M_{5,1,3},M_{6,1,3},M_{4,5,1},M_{5,6,1},
        \nn\\&\qquad
        M_{4,2,3},M_{5,2,3},M_{4,6,3},M_{5,6,3},
        M_{4,5,2},M_{4,6,2},M_{5,6,2},M_{4,5,6}
    \Big\}
    \,.
\end{align}
These are 0-dimensional topological spaces consisting of only a point. 
Note that a fully generic arrangement would have produced $\binom{6}{3}=20$ codimension-3 boundaries. 
In our case, four are missing because some cuts do not exist: $M_{4,5,3} = M_{4,6,1} = M_{5,1,2}$ = $M_{6,2,3} = \varnothing$. 
That is, the associated hyperplanes to these cuts do not intersect.

\end{itemize}

\begin{tcolorbox}[breakable, title=\hypertarget{box:cutOrd}{Cut ordering}]
The cut multi-index $J$ in $\delta_J$, $\res_J$ and $M_J$ is \emph{always} ordered such that the associated 1-tubings with the most vertices comes first. 
For tubings of the same size, we proceed in lexicographical order.
Symbolically, if $\vert\tau_i\vert = \vert\mathcal{V}_i\vert$ is the number of vertices inside the tubing $\tau_i$, we require $\vert\tau_{j_i}\vert \geq \vert\tau_{j_{i+1}}\vert$ for $J=\{j_1, j_2, \cdots\}$. 
\\[1em]
While ad hoc now, the physical reason will become clear in section \ref{sec:goodTubes}. 
Mathematically, an order is necessary to resolve the mentioned degeneracies (more in section \ref{sec:linrel}).\footnotemark
\end{tcolorbox}
\footnotetext{We thank Cl\'ement Dupont and Lecheng Ren for important discussions regarding this point.}

\paragraph{Building a basis from the geometry of cuts:}
To generate elements of the dual cohomology, we search for cuts with bounded chambers. 
The 0-cut, $M$, has no bounded chambers and therefore does not generate a potential basis element. 
This is a general feature of all FRW arrangements; the 0-cut will never generate an interesting form. 

Of all the 1-cuts, only the total energy cut $M_4$ has a bounded chamber. 
For example, 
\begin{align}
    M_5= \includegraphics[align=c,width=.3\textwidth]{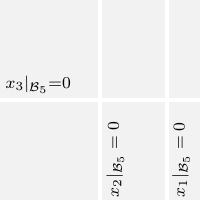}\,,
    \qquad
    M_4= \includegraphics[align=c,width=.3\textwidth]{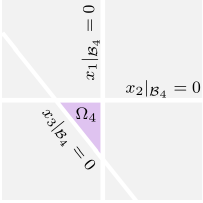}
    \label{eq:trivandnontriv1-cut}
    \,,
\end{align}
where the white lines are removed from $\mathbb{C}^2$. 
Evidently, there is no bounded chamber in $M_5$.
To see this, solve $B_5=0$ by eliminating $x_2$. 
Then, $\mbf{x}\vert_{\B_5}=\left(x_1, -(x_1+X_1+X_2+Y_2), x_3\right)$. 
Clearly, the first two components define parallel lines as depicted above. 
However, there is a single bounded chamber in $M_4$. 
This can be seen by solving the condition $B_4=0$ and eliminating $x_3$. Then,  $\mbf{x}\vert_{\B_4} = \left(x_1,x_2,-(x_1{+}x_3{+}X_1{+}X_2{+}X_3)\right)$ generates the plot above. 
The canonical form associated to this bounded chamber is 
\begin{align} \label{eq:3site1bd}
    \cvphi_1 = \delta_{4}({\Omega}_4) 
    \quad \text{where} \quad
    {\Omega}_4 = \frac{\vep^2}{3} \dlog \frac{x_1}{x_2}\Big\vert_{\B_4} \wedge \dlog \frac{x_2}{x_3}\Big\vert_{\B_4}
    \,, 
\end{align}
The powers of $\vep$ in the normalization ensure $\vep$-form DEQs. 
The overall rational number is convenient. 

The remaining dual forms are generated in the same way and can be found in \cite{De:2023xue}. 
The 2-boundary dual forms are generated by the line segments between the removed twisted points on the 2-cuts
\begin{align}\begin{aligned}\label{eq:3site2bd}
    \cvphi_2 &= \frac{\vep}{2} \delta_{4,1}
    \left(\dlog \frac{x_{2}}{x_{3}} \Big\vert_{\B_{41}} \right)
    ,
    &
    \cvphi_3 &= \frac{\vep}{2} \delta_{6,1}
    \left(\dlog\frac{x_{2}}{x_{3}}\Big\vert_{\B_{61}}\right)
    ,
    &
    \cvphi_4 &= \frac{\vep}{2} \delta_{4,3}
    \left(\dlog\frac{x_{1}}{x_{3}}\Big\vert_{\B_{43}}\right)
    ,
    \\
    \cvphi_5 &= \frac{\vep}{2} \delta_{5,3}
    \left(\dlog\frac{x_{1}}{x_{2}}\Big\vert_{\B_{53}}\right)
    ,
    &
    \cvphi_6 &= \frac{\vep}{2} \delta_{4,5}
    \left(\dlog \frac{x_{1}}{x_{2}} \Big\vert_{\B_{45}} \right)
    ,
    &
    \cvphi_7 &= \frac{\vep}{2} \delta_{4,6}
    \left(\dlog\frac{x_{1}}{x_{2}} \Big\vert_{\B_{46}} \right)
    ,
    \\
    \cvphi_8 &= \frac{\vep}{2} \delta_{4,2}
    \left(\dlog\frac{x_{2}}{x_{3}}\Big\vert_{\B_{42}}\right)
    ,
    &
    \cvphi_9 &= \frac{\vep}{2} \delta_{5,6}
    \left(\dlog \frac{x_{1}}{x_{2}} \Big\vert_{\B_{56}}\right)
    ,
    &
    \cvphi_{10} &= \frac{\vep}{2} \delta_{5,6}
    \left(\dlog \frac{x_{2}}{x_{3}}\Big\vert_{\B_{56}}\right)
    .    
\end{aligned}\end{align}
Note that the cut $M_{56}$ has two canonical forms since the restriction of all coordinate axes to the cut form two line segments
\begin{align}
    M_{56} = \includegraphics[align=c,width=.3\textwidth]{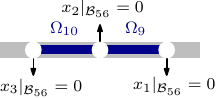}
    \;.
\end{align}
The 3-boundary forms are 
\begin{align}\begin{aligned}\label{eq:3site3bd}
    \cvphi_{11} &= \delta_{1,2,3}(1)
    &
    \cvphi_{12} &= 
    \delta_{6,1,2}(1)
    \,,
    &
    \cvphi_{13} &= 
    \delta_{4,1,3}(1)
    \,,
    &
    \cvphi_{14} &= 
    \delta_{5,1,3}(1)
    \,,
    \\
    \cvphi_{15} &= 
    \delta_{6,1,3}(1)
    \,,
    &
    \cvphi_{16} &= 
    \delta_{4,5,1}(1)
    \,,
    &
    \cvphi_{17} &= 
    \delta_{5,2,3}(1)
    \,,
    &
    \cvphi_{18} &= 
    \delta_{4,6,3}(1)
    \,,
    \\
    \cvphi_{19} &= 
    \delta_{4,5,2}(1)
    \,,
    &
    \cvphi_{20} &= 
    \delta_{4,6,2}(1)
    \,,
    &
    \cvphi_{21} &= 
    \delta_{5,6,2}(1)
    \,,
    &
    \cvphi_{22} &= 
    \delta_{4,5,6}(1)
    \,,
    \\
    \cvphi_{23} &= 
    \delta_{4,1,2}(1)
    \,,
    &
    \cvphi_{24} &= 
    \delta_{5,6,1}(1)
    \,,
    &
    \cvphi_{25} &= 
    \delta_{4,2,3}(1)
    \,,
    &
    \cvphi_{26} &= 
    \delta_{5,6,3}(1)
    \,.
\end{aligned}\end{align}
Note that the argument of each $\delta_J$ is simply the  the 0-form 1; the canonical form for a point. 

\paragraph{The {\color{Black}physical}, {\color{Black}unphysical}  and {\color{Black}degenerate/mixed} cuts:} 
The non-trivial dual forms corresponding to the 1-, 2- and 3-cuts given by \eqref{eq:3site1bd}, \eqref{eq:3site2bd} and \eqref{eq:3site3bd} will now be color-coded according to whether they are {\color{BrickRed}physical}, {\color{MidnightBlue}unphysical} or {\color{Orange}degenerate/mixed}. 

The {\color{BrickRed}physical} dual forms induce residues that do not annihilate $\Psi_n^{(\ell)}$ in the intersection number
\begin{align}
    &\res_{\color{BrickRed}J}[\Psi_3^{(0)}] \neq 0
    \,,
\end{align}
where 
\begin{align}\begin{aligned}
    \color{BrickRed}
    J \in\Big\{
    &\color{BrickRed}
    \{1\}, \{4,1\}, \{6,1\}, \{4,3\}, \{5,3\},
    \{4,5\}, \{4,6\}, \{1,2,3\}, 
    \\&\qquad
    {\color{BrickRed}
    \{6,1,2\}, \{4,1,3\},\{5,1,3\}, \{6,1,3\}, 
    \{4,5,1\}, \{5,2,3\}, \{4,6,3\}
    \Big\}
    }
    \,.
\end{aligned}\end{align}
These generate the physical FRW forms in \eqref{eq:3site1cut}, \eqref{eq:3site2cut} and \eqref{eq:3site3cut} via the replacement $\delta_J(\Omega) \to \dlog_J\wedge\tilde{\Omega}$. 

The {\color{MidnightBlue}unphysical} dual forms induce residues that annihilate $\Psi_n^{(\ell)}$ in the intersection number 
\begin{align}
    \res_{\color{MidnightBlue}J}[\Psi_3^{(0)}] = 0
    \,,
    \label{eq:3sitebadboundaries}
\end{align}
where 
\begin{align}
    \color{MidnightBlue}
    J \in\Big\{
    &\color{MidnightBlue}
    \{4,2\}, \{5,6\}, 
    \{4,1,2\}, \{5,6,1\}, \{4,2,3\}, \{5,6,3\} 
    \Big\}
    \,.
\end{align}
These generate unphysical FRW forms.
Importantly, note that the unphysical dual forms correspond to sequential residues that set either side of the linear relation \eqref{eq:3-sitelinrelnew} to zero. 
The unphysical residues correspond to Steinmann relations in the cosmological context \cite{Benincasa:2020aoj, Benincasa:2021qcb}.

The {\color{Orange}degenerate/mixed} class of dual forms correspond to residues that may or may not annihilate $\Psi_n^{(\ell)}$
\begin{align}\begin{aligned} \label{eq:3siteUnphysJ}
    &\res_{\color{Orange}4,5,2}[\Psi_3^{(0)}] 
    = -\res_{\color{Orange}4,6,2}[\Psi_3^{(0)}] 
    = \res_{\color{Orange}4,5,6}[\Psi_3^{(0)}] 
    = 1
    \\
    &\res_{\color{Orange}5,6,2}[\Psi_3^{(0)}] = 0
    \,.
\end{aligned}\end{align}
As discussed earlier all four of these cuts correspond to the same loci $\color{Orange} M_{452} = M_{462} = M_{562} = M_{456}$ and are therefore called degenerate. 
Additionally, the associated residues operators induced by the $\delta_{\color{Orange}J}$ satisfy linear relations. 
For example,
\begin{align} \begin{aligned}\label{eq:3siteDegen}
    \res_{\color{Orange} 4,5,2} = \res_{\color{Orange} 4,5,6}
    \,,
    \quad\text{and}\quad
    \res_{\color{Orange} 4,6,2} = \res_{\color{Orange} 4,6,5}
    \,.
\end{aligned}\end{align}
From the point of view of direct computation, it takes additional work to understand which FRW forms generated by dual forms on the degenerate cut should be associated to the physical sector. 
A cut is degenerate whenever all but one hyperplane in a linear relation appears in a sequential residue. 
In this example, it happens whenever a cut involves three of the hyperplane polynomials that appear in the linear relation \eqref{eq:3-sitelinrelnew}.

This classification of cuts with a non-trivial cohomology into {\color{BrickRed}physical}, {\color{MidnightBlue}unphysical} and {\color{Orange}degenerate/mixed} 
directly follows from the linear relation(s) of the hyperplane arrangement.
Simple rules, that circumvent direct computation, for classifying the non-trivial cuts are provided in section \ref{sec:linrel} and \ref{sec:goodTubes}. 
Further rules for generating the exact physical differential forms are also provided.

\paragraph{The FRW forms:} 
A spanning set of FRW forms is obtained from the dual forms by the rule $\delta_J(\Omega_J) \to \dlog_J\wedge\tilde{\Omega}_J$ and guarantees that the intersection matrix is block diagonal on the cuts.
We also drop any normalization associated to the dual forms in this procedure. 

There is one \textcolor{BrickRed}{physical} FRW form associated to the only non-trivial 1-cut from \eqref{eq:3site1bd}
\begin{align}
    \color{BrickRed}\vphi_1 &= \dlog_{\color{BrickRed}4} \wedge 
    \dlog\frac{x_{1}}{x_{2}} \wedge \dlog \frac{x_{2}}{x_{3}}
    \,.
    \label{eq:3site1cut}
\end{align}
The nine dual forms \eqref{eq:3site2bd} generate the \textcolor{BrickRed}{physical} $\vphi_{2-7}$ and \textcolor{MidnightBlue}{unphysical} \color{black} $\vphi_{8-10}$ FRW forms
\begin{align}\begin{aligned}
    \color{BrickRed}\vphi_2 &= 
    \dlog_{\color{BrickRed}4,1} \wedge 
    \dlog \frac{x_{2}}{x_{3}}
    \,,
    &
    \color{BrickRed}\vphi_3 &= 
    \dlog_{\color{BrickRed}6,1} \wedge 
    \dlog \frac{x_{2}}{x_{3}}
    \,,
    &
    \color{BrickRed}\vphi_4 &= 
    \dlog_{\color{BrickRed}4,3} \wedge 
    \dlog\frac{x_{1}}{x_{3}}
    \,,
    \\
    \color{BrickRed}\vphi_5 &= 
    \dlog_{\color{BrickRed}5,3} \wedge 
    \dlog \frac{x_{1}}{x_{2}}
    \,,
    &
    \color{BrickRed}\vphi_6 &= 
    \dlog_{\color{BrickRed}4,5} \wedge 
    \dlog \frac{x_{1}}{x_{2}}
    \,,
    &
    \color{BrickRed}\vphi_7 &= 
    \dlog_{\color{BrickRed}4,6} \wedge 
    \dlog \frac{x_{1}}{x_{2}}
    \,,
    \\
    \color{MidnightBlue}\vphi_8 &= 
    \dlog_{\color{MidnightBlue}4,2} \wedge 
    \dlog\frac{x_{2}}{x_{3}}
    \,,
    &
    \color{MidnightBlue}\vphi_9 &= 
    \dlog_{\color{MidnightBlue}5,6} \wedge 
    \dlog \frac{x_{1}}{x_{2}}
    \,,
    &
    \color{MidnightBlue}\vphi_{10} &= 
    \dlog_{\color{MidnightBlue}5,6} \wedge 
    \dlog \frac{x_{2}}{x_{3}}
    \,.    
    \label{eq:3site2cut}
\end{aligned}\end{align}
The ten dual forms \eqref{eq:3site3bd} generate the \textcolor{BrickRed}{physical} $\vphi_{11-18}$,  \textcolor{MidnightBlue}{unphysical} $\vphi_{23-26}$ and \textcolor{orange}{degenerate} $\vphi_{19-22}$ FRW forms
\begin{align}\begin{aligned}
    \color{BrickRed}\vphi_{11} &= 
    \dlog_{\color{BrickRed}1,2,3}
    \,,
    &
    \color{BrickRed}\vphi_{12} &= 
    \dlog_{\color{BrickRed}6,1,2}
    \,,
    &
    \color{BrickRed}\vphi_{13} &= 
    \dlog_{\color{BrickRed}4,1,3}
    \,,
    &
    \color{BrickRed}\vphi_{14} &= 
    \dlog_{\color{BrickRed}5,1,3}
    \,,
    \\
    \color{BrickRed}\vphi_{15} &= 
    \dlog_{\color{BrickRed}6,1,3}
    \,,
    &
    \color{BrickRed}\vphi_{16} &= 
    \dlog_{\color{BrickRed}4,5,1}
    \,,
    &
    \color{BrickRed}\vphi_{17} &= 
    \dlog_{\color{BrickRed}5,2,3}
    \,,
    &
    \color{BrickRed}\vphi_{18} &= 
    \dlog_{\color{BrickRed}4,6,3}
    \,,
    \\
    \color{Orange}\vphi_{19} &= 
    \dlog_{\color{Orange}4,5,2}
    \,,
    &
    \color{Orange}\vphi_{20} &= 
    \dlog_{\color{Orange}4,6,2}
    \,,
    &
    \color{Orange}\vphi_{21} &= 
    \dlog_{\color{Orange}5,6,2}
    \,,
    &
    \color{Orange}\vphi_{22} &= 
    \dlog_{\color{Orange}4,5,6}
    \,,
    \\
    \color{MidnightBlue}\vphi_{23} &= 
    \dlog_{\color{MidnightBlue}4,1,2}
    \,,
    &
    \color{MidnightBlue}\vphi_{24} &= 
    \dlog_{\color{MidnightBlue}5,6,1}
    \,,
    &
    \color{MidnightBlue}\vphi_{25} &= 
    \dlog_{\color{MidnightBlue}4,2,3}
    \,,
    &
    \color{MidnightBlue}\vphi_{26} &= 
    \dlog_{\color{MidnightBlue}5,6,3}
    \,.
    \label{eq:3site3cut}
\end{aligned}\end{align}

The only remaining task is to understand how to deal with the {\color{Orange}degenerate} forms.

\paragraph{Disentangling the degenerate cuts via direct computation:}
Since we will soon describe how to find the physical subspace without performing any direct computation, we focus on conceptual details; some explicit computations can be found in appendix \ref{sec:examples}. 

As mentioned above, the intersection matrix is block diagonal with one block coming from the unphysical two-dimensional cohomology on the 56-cut and the other from the degenerate 452-, 462-, 562- and 456-cuts
\begin{align} \label{eq:3siteocCmat}
    \mat{C} = \includegraphics[align=c, width=.3\textwidth]{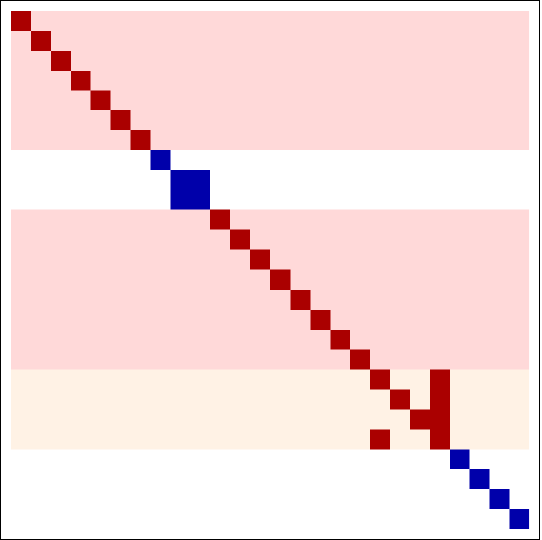}
    \qquad\text{where}\qquad
    \mat{C}_\text{degen} = 
    \begin{pmatrix}
        1 & 0 & 0 & \phantom{-}1
        \\
        0 & 1 & 0 & -1
        \\
        0 & 0 & 1 & \phantom{-}1
        \\
        1 & 0 & 0 & \phantom{-}1
    \end{pmatrix}
    \,.
\end{align}
Each row of this matrix corresponds to a dual form and therefore to a cut and are color coded according to their classification.
In particular, the {\color{Orange}degenerate} rows in {\color{Orange}orange} correspond to residues that localize to the same loci. 
Yet, some annihilate $\Psi^{(0)}_3$ and some do not (c.f., \eqref{eq:3siteDegen}). 
Therefore, it is not clear if they should be classified as physical or unphysical.

Also, the rank of the $4\times4$ degenerate block $\mat{C}_\text{degen}$ is only three. 
This is a consequence of a linear relation among $\dlog$-forms and a generic feature of degenerate hyperplane arrangements (see section \ref{sec:dlogRel}). 
Therefore, we have to remove one element in this $4\times4$ block from our spanning sets to get a basis. 
We remove the elements associated to the 456-cut (elements $\cvphi_{22}$ and $\vphi_{22}$) to form the bases $\bs{\cvphi}^\prime$ and $\bs{\vphi}^\prime$. 
The corresponding intersection matrix $\mat{C}^\prime$ is full rank.

At this stage, the number of linearly independent elements in cohomology that share cuts with $\Psi_3^{(0)}$ are the number of {\color{BrickRed}red} elements along the diagonal of $\mat{C}^\prime$: 18. 
To reduce this more, we need to decompose the degenerate space into a component along $\Psi_3^{(0)}$ and its orthogonal complement. 
This is accomplished by constructing a gauge transformations $\mat{\c{U}}$ and $\mat{U} = {( \mat{\c{U}} \cdot \mat{C}^\prime)^{-1}}^\top$ detailed in appendix \ref{app:3siteDegenBlock}. 
Applying these gauge transformations we obtain the bases 
$\bs{\cvphi}^{\prime\prime} = \mat{\c{U}}\cdot \bs{\cvphi}^\prime$
and 
$\bs{\vphi}^{\prime\prime} = \mat{U}\cdot\bs{\vphi}^\prime$. 
Since these bases have unit intersection matrix $\mat{C}^{\prime\prime}=\mathds{1}$, they are truly dual making it easy to map between the FRW and dual cohomologies (c.f., footnote \ref{foot:almostDual}).

The new basis $\bs{\cvphi}^{\prime\prime}$ contains only 16 dual forms that correspond to residues that do not annihilate $\Psi_3^{(0)}$; the corresponding FRW forms share sequential residues with $\Psi_3^{(0)}$ and make up the physical subspace. 
Computing the $\bs{\vphi}^{\prime\prime}$-DEQs, $A^{\prime\prime}_{ab} = (C_{bc}^{\prime\prime})^{-1} \la \cvphi^{\prime\prime}_{c} \vert \nabla_\kin \vphi^{\prime\prime}_a \ra = \la \cvphi^{\prime\prime}_{b} \vert \nabla_\kin \vphi^{\prime\prime}_a \ra$, we find that the 16-dimensional physical subspace closes and is integrable 
\begin{align}
   \nabla_\text{kin}\,  \includegraphics[align=c, height=12em]{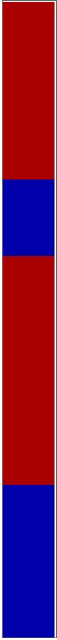}
    = \includegraphics[align=c, height=12em]{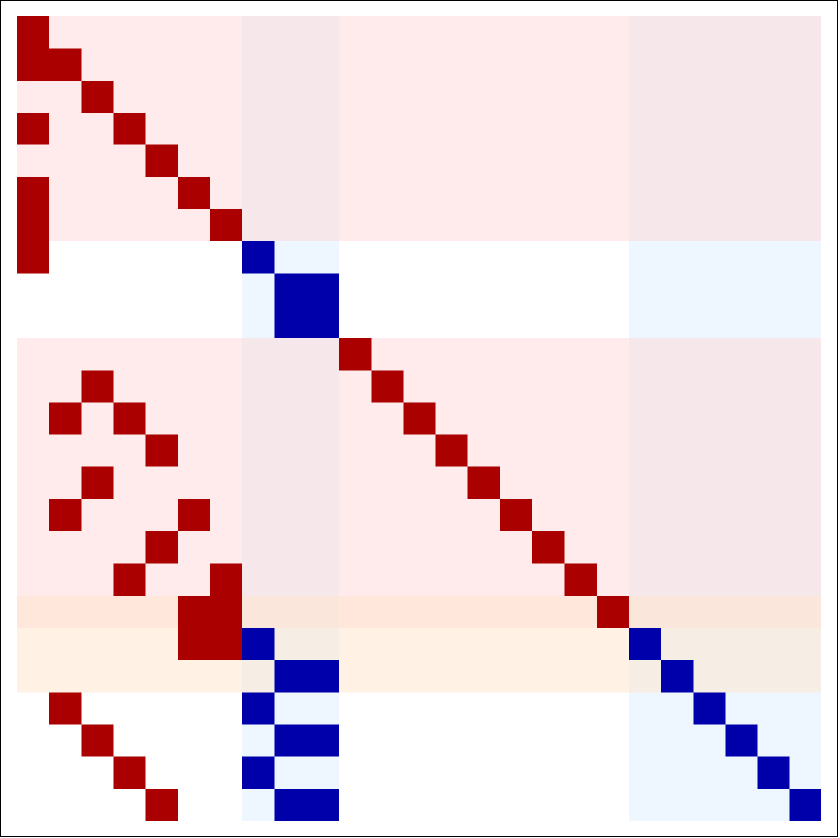}
    \cdot 
    \includegraphics[align=c, height=12em]{figs/3site/RedBlueBasis.pdf}
    \,.
\end{align}
Clearly the derivative of any {\color{BrickRed}physical} FRW form only talks to other {\color{BrickRed}physical} FRW forms.
Explicitly, the physical subspace contains all forms in {\color{BrickRed}red} and one linear combination of the  {\color{Orange}orange} forms of the degenerate cut
\begin{align}\begin{aligned}
	\{\vphi_{\phys,i}\}_{i=1}^{16} 
	= \{\vphi_1, \dots \vphi_7, \vphi_{11}, \dots \vphi_{18}, (\vphi_{19} {-} \vphi_{20})/2 \}
\label{eq:3-sitephysicalforms}
	\,.
\end{aligned}\end{align}
From the perspective of direct calculation, the linear combination $(\vphi_{19} - \vphi_{20})/2$ in \eqref{eq:3-sitephysicalforms} comes from the gauge transformation matrix $\mat{U}$.
\texttt{Mathematica} files with the physical basis, gauge transformations and DEQs can be found in appendix \ref{app:3siteDegenBlock} and at the \texttt{github} repository \github. 

Of course, we need to decompose $\Psi_3^{(0)}$ into the physical basis in order to use the DEQs for the physical quantity of interest. 
One can either use partial fractions (see appendix \ref{app:partialFrac} for a diagrammatic approach) or the intersection number: 
\begin{align}
    \Psi_3^{(0)} = (0,0,0,0,0,0,0,1,-1,-1,1,1,-1,-1,1,2)\cdot\bs{\vphi}_\text{phys}
    \,.
\end{align}

\paragraph{Further examples and loops:}
At no step in any part of our algorithm has the specification to tree-level cosmological correlators been needed; everything generalizes straightforwardly to loop-level. 
We demonstrate the splitting of the full cohomology into physical and unphysical subspaces via direct calculation for the 
4-site chain (appendix \ref{app:4SiteChain}), 
4-site star (appendix \ref{app:4SiteStar}) 
and 1-loop 3-gon (appendix \ref{app:3Site1Loop}). 
Using the intersection matrix, we demonstrate that the full cohomologies with dimensions 213, 312 and 99 reduce to a physical basis with 64, 64 and 50 elements in agreement with both the kinematic flow algorithm \cite{Arkani-Hamed:2023bsv, Arkani-Hamed:2023kig, Baumann:2024mvm, Hang:2024xas} and those obtained from a time integral perspective \cite{He:2024olr}. 
The DEQs for the 4-site chain and 1-loop 3-gon  can be found in the \texttt{Mathematica} files at the \texttt{github} repository \github.

\paragraph{Disentangling the degenerate cuts avoiding direct computation:}
Before moving on, we preview how to find the physical subspace and avoid direct calculation.
A systematic algorithm is developed in  sections \ref{sec:linrel} and \ref{sec:goodTubes}.

To every $\delta_J$ or equivalently $\res_J$, we can associate a cut tubing. 
For example, the total energy residue can be represented diagrammatically by the dashed tubing
\begin{align}
    \delta_4 \leftrightarrow \res_{4} := \includegraphics[align=c,scale=0.67]{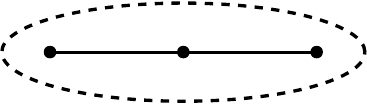}
    \,,
\end{align}
Next, note that \emph{only} cut tubings with \emph{no crossings} appear in the physical basis \eqref{eq:3-sitephysicalforms}. 
This is because of the two following facts:
\begin{itemize}
    \item[i)] The set of compatible complete tubings of a graph $\G$ only involves non-crossed tubings (see \eqref{eq:3siteCanForm}, \eqref{eq:4chainTri} and \eqref{eq:3gonTri} for examples).
    \item[ii)] Due to our  \hyperlink{box:cutOrd}{ordering}, the only residues with crossed cut tubings that do not annihilate $\Psi_n^\elll$ are degenerate. 
    Moreover, the degenerate crossed cut tubings that do not annihilate $\Psi_n^\elll$ are equivalent to a non-crossed cut tubings.
    For example, in the context of the 3-site example,  $\res_{4,5,6}=\res_{4,5,2}$
    \begin{align}
        \includegraphics[align=c,width=.2\textwidth]{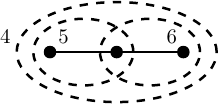}
        =
        \includegraphics[align=c,width=.2\textwidth]{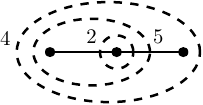}
        \,.
    \end{align}
    The generalization of such residue relations is presented in section \ref{sec:resRel}.%
\end{itemize} 
\begin{tcolorbox}[breakable, title = \hypertarget{box:noCrossCuts}{No crossed cut tubings}]
The physical subspace only contains forms associated to non-crossed tubings:
\begin{align}
    \includegraphics[align=c,width=.2\textwidth]{figs/Res452.pdf}
    \quad \checkmark
    \,,
    \qquad
    \includegraphics[align=c,width=.2\textwidth]{figs/Res456.pdf}
    \quad \tikzxmark
    \,.
    \nn
\end{align}
\end{tcolorbox}

Each degenerate block also only contributes one form to the physical subspace since we decompose the degenerate block into a component along $\Psi_3^{(0)}$ and its orthogonal complement.
The $\dlog_J$-part of the physical form associated to a degenerate cut is a linear combination of $\dlog_J$-forms associated to the non-crossing cut tubings of the degenerate cut. 
Specifically, there is a relative sign between all pairs of $\dlog_J$'s that only differ by tubings that cross. 
Indeed, $\vphi_{\text{phys},16} \propto \vphi_{19} {-} \vphi_{20} = \dlog_{4,5,2}-\dlog_{4,6,2}$ satisfies this rule; the $\dlog$'s share all but one factor and these unshared $B_i$ have tubings that cross.

\section{Linear relations and degenerate arrangements}
\label{sec:linrel}

The classification of cuts/residues with non-trivial cohomologies into physical, unphysical and degenerate/mixed cuts is controlled entirely by the system of linear relations associated to the hyperplane arrangement of the graph $\G$. 
Importantly, these hyperplane arrangements (section \ref{sec:FRWUplift}) are non-generic in the sense that there exists codimension $m$ loci where more than $m$ hyperplanes intersect.
This non-genericity in cosmological hyperplane arrangements is entirely encoded in the linear relations among the corresponding hyperplane polynomials $B_i$.
As a consequence of this degeneracy, some sequential residues do not necessarily anti-commute and satisfy linear relations (recall \eqref{eq:3siteUnphysJ} and \eqref{eq:3siteDegen}).

The linear relations among the $B_i$ (equivalently the $S_i$) are generated by  2-tubings that intersect.\footnote{This is not always a minimal presentation; the set of all pairs of crossed tubings often produces an over complete set of relations.} 
For the 3-site chain, there is just one relation
\begin{align}\begin{aligned} \label{eq:3-sitelinrel}
    \includegraphics[align=c, width=.2\textwidth]{figs/3siteSrelLHS.pdf}
    \bigg\vert_+
    =~ \includegraphics[align=c, width=.2\textwidth]{figs/3siteSrelRHS.pdf}
    \bigg\vert_+
    \iff 
    B_2 + B_4 = B_5 + B_6
    \,, 
\end{aligned}\end{align}
where $\vert_+$ instructs us to add the $B_i$ associated to the 2-tubings. 
Via equation \eqref{eq:Sdef}, this graphical rule is easy to verify since the crossed 2-tubings encircle the same vertices and cross the same edges the same number of times.
Moreover, this graphical rule holds beyond tree-level graphs and generates all possible linear relations.
Indeed, it is easy to check the following identity for the 1-loop 3-gon graph below
\begin{align}
    \left.
    \includegraphics[align=c, width=.15\textwidth]{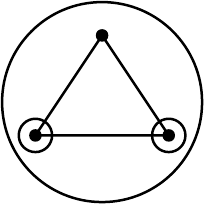}
    \right\vert_+
    ~~~~=~~~~
    \left.
    \includegraphics[align=c, width=.15\textwidth]{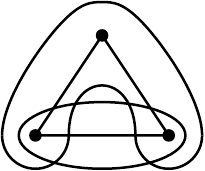}
    \right\vert_+
    \,.
\end{align}
We also note that not all linear relations between hyperplane polynomials will generate interesting consequences for the physical basis; some only effect the unphysical subspace.

To keep track of the linear relations and deduce their consequences, we introduce the notion of a minimally dependent indexing set $I$. 
A minimally dependent indexing set 
\begin{align} \label{eq:minDepSet}
    I = \{I_\text{LHS} \vert I_\text{RHS}\}
    \,,
\end{align} 
is a collection of two sets indexing the hyperplane polynomials $B_i$ that appear on the LHS and RHS of linear relations. 
For example, in this notation, equation \eqref{eq:3-sitelinrel} is equivalent to $I=\{2, 4 | 5, 6\}$ or with some abuse of notation $I=\{B_2, B_4 | B_5, B_6\}$. 
Furthermore, note that $I_\text{RHS}$ is always a set of two 1-tubings that cross.

The set $I$ is minimally dependent in the sense that removing any single element, say $B_j$, results in a linearly independent set of $B_i$\footnote{In some of the relations that follow, we suppress the $B$'s and denote a hyperplane polynomial $B_i$ simply by its label $i$.}
\begin{align} \label{eq:minDep}
    \sum_{i \in I}  
    \underset{\neq 0}{\underbracket{\ \alpha_{i}\ }}
    B_{i} {=} 0
    \quad\text{but}\quad 
    \sum_{{i} \in I \setminus {i_j}} \!\! \alpha_i B_i {=} 0 {\iff} \alpha_i {=} 0
    \,.
\end{align}
Every minimally dependent set $I$ generates a relation between sequential residues (section \ref{sec:resRel}) as well as a relation among $\dlog$-forms (section \ref{sec:dlogRel}). 
After identifying a degenerate boundary, we also provide a simple prescription for constructing the physical form without computing the intersection matrix in section \ref{ssec:picstoforms}.

\subsection{Consequences for residue operations \label{sec:resRel}}

As a consequence of the linear relations among hyperplane polynomials, iterated residues do not necessarily anti-commute under ordering exchanges and satisfy relations. To see why iterated residues do not necessarily anti-commute, consider the following example of a toy (non-cosmological) arrangement below where the lines $\B_1$, $\B_2$ and $\B_3$ all intersect at the same point (in {\color{BrickRed} red})
\begin{center}
    \includegraphics[width=.42\textwidth]{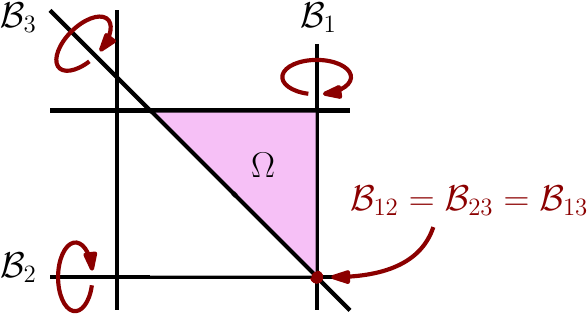}
\end{center}
While the canonical form $\color{Plum} \Omega$ is singular at the point $\mathcal{B}_{12}=\mathcal{B}_{23}=\mathcal{B}_{31}$, its residue at this point depends on the order. 
The $13$-residues of $\color{Plum} \Omega$ behave normally
\begin{align}
    \res_{13}[{\color{Plum} \Omega}]
    &= -\res_{31}[{\color{Plum} \Omega}]
    \,.
\end{align}
On the other hand, one ordering of the $23$- and $12$-residues vanish
\begin{align}
    \res_{21}[{\color{Plum} \Omega}] = \res_{23}[{\color{Plum} \Omega}] 
    = 0
    \,. 
\end{align}
The other ordering is equal to an ordering of the $13$-residues
\begin{align}
    \res_{32}[{\color{Plum} \Omega}] 
    &= \res_{31}[{\color{Plum} \Omega}] 
    \neq 0
    \,,
    \\
    \res_{12}[{\color{Plum} \Omega}] 
    &= \res_{13}[{\color{Plum} \Omega}] 
    \neq 0
    \,.
\end{align}
For this reason, we need to fix an ordering when taking residues. 
As mentioned in section \ref{sec:3SitePed}, we always order residues such that the $B_i$ associated to the tubing that encircles the most vertices always comes first. 
Diagrammatically, this means that we work outward-in for residue tubings
\begin{align}
    \res_{452} := 
    \includegraphics[align=c,width=.2\textwidth]{figs/Res452.pdf}
    \neq \res_{425} \neq \res_{542} \neq \cdots \,.
\end{align}
For tubings of the same size, we assign lexicographical order
\begin{align}
    \res_{413}:=
    \includegraphics[align=c,width=.2\textwidth]{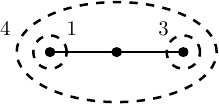}
    \,.
\end{align}
For now, this ordering is a mathematical necessity. 
However, in section \ref{sec:goodTubes}, we will show that our ordering choice aligns with physical intuition. 

Every minimally dependent set, $I$, of hyperplane polynomials \eqref{eq:minDep} seeds a residue relation 
\begin{align}  
    \res_{i_j}\circ
    \res_{\sigma(I\setminus \{i_j,i_k\})}
    {=} \res_{i_k}\circ
    \res_{\sigma(I\setminus \{i_j,i_k\})}
    \,,
    \label{eq:degenboundequation}
\end{align}
where $i_j,i_k \in I$ and $\sigma \in S_{|I|-2}$ is a permutation acting on $I\setminus \{i_j,i_k\}$. 
To see why this is the case, note that when restricted to the cut $I\setminus \{i_j,i_k\}$, the linear relation between the hyperplane polynomials now becomes $\alpha_{i_j} B_{i_j}\vert_{I\setminus \{i_j,i_k\}} + \alpha_{i_k} B_{i_k}\vert_{I\setminus \{i_j,i_k\}} = 0$. 
Since $B_{i_j}\vert_{I\setminus \{i_j,i_k\}}$ and $B_{i_k}\vert_{I\setminus \{i_j,i_k\}}$ are proportional, the final residues in the above equation are equivalent.
Note that this relation is insensitive to the actual values of the coefficients $\alpha_i$ appearing in the linear relations \eqref{eq:minDep}.

\subsection{Consequences for differential forms \label{sec:dlogRel}}

Similarly, linear relations among the hyperplane polynomials generate linear relations among differential forms. 
Like before, we identify a minimally dependent set of hyperplane polynomials $I$. 
After fixing an order for $I$, the following identity holds
\begin{align} \label{eq:OrlikSolomon}
    \sum_{j=1}^{|I|} (-1)^{j-1} 
    \left(\bigwedge_{k \in I\setminus i_j} \dlog B_k \right) = 0
    \,,
\end{align}
where $i_j$ is the $j^\text{th}$ element of $I$.
For example, it is easy to verify the relation 
\begin{align}
    \dlog_{456} - \dlog_{452} + \dlog_{462} - \dlog_{562} 
    = 0
    \,,
\end{align}
in terms of the 3-site example (section \ref{sec:3SitePed}), which can be deduced from the nullspace of the intersection matrix \eqref{eq:3siteocCmat}. 
Such relations are known in the mathematical literature as Orlik-Solomon relations \cite{dupont2015orliksolomonmodelhypersurfacearrangements, orlik1980combinatorics}.\footnote{We thank Cl\'ement Dupont for bringing the Orlik-Solomon algebra to our attention.}
In fact, the Orlik-Solomon algebra generates all linear relations of degenerate hyperplane arrangements circumventing the need to use blow-ups. 
Furthermore, a nice presentation for the Gr\"obner basis of monomials in $\dlog$'s is known. 

Equation \eqref{eq:OrlikSolomon}, also gives a relations between the $(|I|-2)$-residues of $(|I|-1)$-forms. 
Let $I$ be a minimally dependent set and $j,k\in\{1,\dots, |I|\}$. 
Then, 
\begin{align} \label{eq:Im2ResRel}
    \res_{I \setminus \{i_j i_k\}} 
    \dlog_{I \setminus i_j}
    = (-1)^{j-k}
    \res_{I \setminus \{i_j i_k\}} 
    \dlog_{I \setminus i_k}
    .
\end{align}
The sign in \eqref{eq:Im2ResRel} comes from the fact that we are comparing the residues of different $\dlog$-forms.

\section{Cut tubings -- a basis for all trees and loops} \label{sec:goodTubes}

In this section, we demonstrate how to construct a basis of physical forms for \emph{any} $n$-site, $\ell$-loop FRW wavefunction coefficient $\psi_{n}^{(\ell)}$ via simple universal graphical rules that systematize the construction of \textit{residue} or equivalently \textit{cut tubings}. 

As mentioned in section \ref{sec:3SitePed}, an $\ell$-loop cut tubing, $\C_{n_i,B_i}^{(\ell)}$, is denoted by a dashed tube (in comparison to the solid tubings $\tau_i$ of \eqref{eq:3siteS}) and defines an instruction to compute a residue on the hyperplane polynomial $B_i=0$.
The label $n_i$ in $\C_{n_i,B_i}^{(\ell)}$ denotes the number of vertices enclosed by the tubing.
For example, a residue on the total energy pole $B_{T=4}=0$ in the tree-level 3-site graph is represented by the following cut tubing:
\begin{equation}
   \C_{3,B_4}^{(0)} := \includegraphics[align=c,scale=0.67]{figs/residue_tubings/3-sitetotenergytubing.pdf} ~:= \res_{\B_4}  \Psi_3^{(0)}~.
   \label{eq:residuetubing}
\end{equation}
While primarily a bookkeeping tool to enumerate the basis elements of any given graph $\G$, these cut tubings carry both physical and mathematical information:
\begin{enumerate}
    \item From a mathematical perspective, they correspond to all physical boundaries $M_J$ that have a one-dimensional twisted cohomology and count each degenerate cut exactly once.
    
    \item From a physics perspective, they correspond to all  ways to factorize the shifted flat-space wavefunction coefficient $\Psi_n^\elll$ solely into product(s) of lower-point flat-space scattering amplitudes.

\end{enumerate}
We elaborate on these connections as we present the universal rules for generating the physical cut tubings, which remarkably hold true for \textit{all} tree and loop level wavefunction coefficients.

In section \ref{ssec:cuttubingevolution}, we present a set of simple universal graphical rules (with detailed examples) associated to cut tubings that arise from the associated  system of linear relations as well as from cohomological arguments. 
In section \ref{ssec:basisoftubings}, we will see how these simple rules systematize the construction of the space of all tubings corresponding to each physical cut $M_J$ (i.e., the space of physical 1-tubings, 2-tubings, and so on), thereby enumerating the physical subspace of forms that couple to $\Psi_n^\elll$ in the DEQs. 
We end by explicitly implementing these universal rules to generate the physical basis for the tree-level 3-site graph in section \ref{ssec:3sitefromtubings} and the 2-loop 2-site sunset graph in section \ref{ssec:sunset}. 
In section \ref{ssec:cutstrat}, we present some remarkable patterns that our analysis reveals.

\subsection{Universal selection rules for physical cut tubings}
\label{ssec:cuttubingevolution}

The linear relations of section \ref{sec:linrel} constrain the space of physical cuts. 
In particular, they constrain how cut tubings can be nested. 
In subsection \ref{sec:nesting}, we present the \textit{good cut} and \textit{degenerate cut} conditions that govern the mechanism for evolving any cut tubing (i.e., going from a $J$-cut to $(J{+}1)$-cut) in a sequentially nested fashion.
Following this, we present an additional set of rules in subsection \ref{sec:beyondNesting} motivated by cohomological arguments, which accounts for cuts beyond this scope of sequential nesting.

\subsubsection{The good cut and degenerate cut conditions}
\label{sec:nesting}

For a given graph $\G$, the residue of the wavefunction coefficient $\Psi_n^{(\ell)}$  on its total energy pole $B_T=0$ is called the \emph{(true) scattering facet} of $\p_{\mathcal{G}}$ \cite{ArkaniHamed2017}. At tree level, we have
\begin{equation}
   \C_{n,B_T}^{(0)} := \res_{\B_{T}} \Psi_n^{(0)} := ~\includegraphics[align=c,scale=0.57]{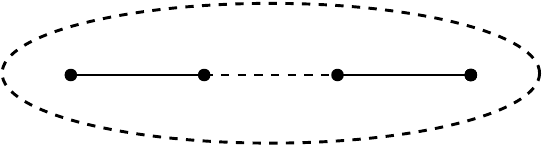} ~.
\end{equation}
Subsequent cut tubings contained inside the true facet $\C_{n_i,B_i}^{(0)} \subset \C_{n,B_T}^{(0)}$ behave as scattering facets associated to a lower $n_i$-point wavefunction coefficient $\Psi_{n_i<n}^{(0)}$. 
This extends to any loop and number of sites and means that a  $\C^{(\ell^{\prime})}_{n_i,B_i} (\subset \C^{(\ell)}_{n,B_T})$ evolves like an $(n_i<n)$-site ($\ell^\prime <\ell$)-level cut tubing regardless of what other cut tubings surround $\C^{(\ell^{\prime})}_{n_i,B_i}$.
This observation implies that $\C^{(\ell^{\prime})}_{n_i,B_i}$ only directly constrains whether a (dashed) tubing placed wholly inside of $B_i$ lead to a physical cut.
In particular, this means that the evolution of a cut tubing---what further cut tubings can be drawn inside it---does not depend on the details of the larger graph outside of it. 
In this way, information from lower-point and lower-loop cuts is recycled!
This structure is reflected in the good-cut and degenerate-cut conditions discussed below.

\begin{tcolorbox}[breakable,title=The good cut condition]
Suppose we want to evolve the cut tubing  $\C^{(\ell)}_{n_i,B_i}$ by nesting the cut tubing $\C^{(\ell^{\prime})}_{n_j,B_j}$: $\C^{(\ell^{\prime})}_{n_j,B_j} \subset \C^{(\ell)}_{n_i,B_i}$. 
We first identify all minimal dependent sets \eqref{eq:minDepSet} where $\tau_i$ (corresponding to $B_i$) is the largest tubing\footnotemark
\begin{equation}
    I^{(i)}_{\alpha} = \{
    I^{(i)}_{\alpha,\text{LHS}}
    \vert 
    I^{(i)}_{\alpha,\text{RHS}}
    \}\,.
\end{equation}
The residue tubings that can be paired with $\C_{n_i,B_i}^{(\ell)}$ to form physical cuts \textit{do not} set all elements of $I^{(i)}_{\alpha,\text{LHS}}$ to zero (also no element of $I^{(i)}_{\alpha,\text{RHS}}$ should be set to zero). 
This is the \textit{good cut condition} associated to the cut tubing $\C_{n_i,B_i}^{(\ell)}$.
It constrains the space of physical cuts with $|I^{(i)}_{\alpha,\text{LHS}}|-1$ further cuts inside $\C_{n_i,B_i}^{(\ell)}$. 
\end{tcolorbox}
\footnotetext{Note that $\tau_i$ is always an element of $I^{(i)}_{\alpha,\text{LHS}}$.}

Sequential residues that set all elements in $I^{(i)}_{\alpha,\text{LHS}}$ (and none in $I^{(i)}_{\alpha,\text{RHS}}$) to zero annihilate the physical form $\Psi_n^{(\ell)}$.
Therefore, the good cut condition tells us how to avoid these unphysical boundaries and generate physical cuts.%
\footnote{%
Recall that all elements of $I^{(i)}_{\alpha,\text{RHS}}$ contain crossed tubings and correspond to sequential residues/cuts that are either unphysical or degenerate. 
For degenerate cuts, these crossed tubings can always be uncrossed using the linear relation of section \ref{sec:resRel}.  
Hence such crossed residue tubings do not appear in our analysis. 
See the discussion around the \hyperlink{box:noCrossCuts}{no crossed cuts box} for more. 
} 

For example, the tree-level 3-site chain graph has a single minimal dependent set $I^{(T)}_1=\{B_2,B_{T=4}|B_5, B_6\}$ (depicted in \eqref{eq:3-sitelinrel}) that constrains the space of physical 2-cuts. 
The double residue $\res_{42}[\Psi_{3}^{(0)}]$ annihilates $\Psi_3^{(0)}$ (as shown in \eqref{eq:3sitebadboundaries}) implying that $\C^{(0)}_{1,B_2} \subset \C^{(0)}_{3,B_4}$ is an unphysical 2-cut
\begin{equation}
  \C_{1,B_2}^{(0)} \subset \C_{3,B_{T=4}}^{(0)}=\includegraphics[align=c,scale=0.67]{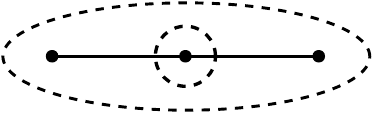} = 0~.
  \label{eq:badboundary3-site}
\end{equation}
Thus, the good cut condition for the true facet $C_{3,B_{T=4}}^{(0)}$ implies that pairing any of the cut tubings $\C_{1,B_1}^{(0)}, \C_{1,B_3}^{(0)}, \C_{2,B_5}^{(0)}, \C_{2,B_6}^{(0)}$ with it corresponds to physical 2-cuts (see \eqref{eq:2tubingsSF}).
In general, $|I_{\alpha,\text{LHS}}|=2$ for all tree-level graphs. 
Therefore, the good cut condition only restricts the next cut $\C_{n_j,B_j}^{(0)} \subset \C_{n_i,B_i}^{(0)}$ such that $B_j \notin I^{(i)}_{\alpha,\text{LHS}}$.

For loop-level graphs, $|I_{\alpha,\text{LHS}}|\geq2$ and the next cut $\C_{n_j,B_j}^{(\ell^{\prime})} \subset \C_{n_i,B_i}^{(\ell)}$ is unconstrained whenever $|I_{\alpha,\text{LHS}}|>2$. 
For example, let's understand how the minimal dependent set $I_1^{(T)}=\{B_1, B_2, B_{T=5}|B_6, B_{14}\}$ of the the 1-loop 4-gon box graph affects the evolution of its true facet.
While the space of 2-tubings involving $\C_{4,B_{T=5}}^{(1)}$ is unconstrained, the 3-element set $I^{(T)}_{1,\text{LHS}}$ constrains the space of physical 3-cuts according to the good cut condition of the true facet. 
Explicitly, the the following 3-cut is unphysical  
\begin{equation}
(\C_{1,B_2}^{(0)} \times \C_{1,B_1}^{(0)}) \subset \C_{4,B_{T=5}}^{(1)} =  \includegraphics[align=c,scale=0.7]{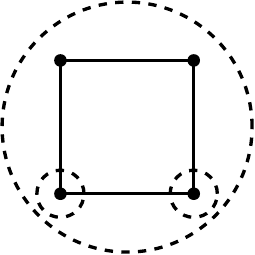} = 0~. 
\end{equation}

\begin{tcolorbox}[breakable, 
title=The degenerate cut condition]
Suppose that we want to evolve an $m=|I^{(i)}_{\alpha,\text{LHS}}|$ cut 
\begin{align}
    \C^{(\ell_{j_{m-1}})}_{n_{j_{m-1}}, B_{j_{m-1}}} 
    \subset \cdots \subset 
    \C^{(\ell_{j_{1}})}_{n_{j_{1}}, B_{j_{1}}} 
    \subset 
    \C_{n_i,B_i}^{(\ell)}
    \,,
\end{align} 
(that could be embedded inside another cut) into an $(m+1)$-tubing.  
The next tubing is placed in accordance with the good cut condition of the most recent cut $\C^{(\ell_{j_{m-1}})}_{n_{j_{m-1}}, B_{j_{m-1}}}$
\begin{align} \label{eq:mp1cut}
    \C^{(\ell_{j_{m}})}_{n_{j_{m}}, B_{j_{m}}} 
    \subset
    \C^{(\ell_{j_{m-1}})}_{n_{j_{m-1}}, B_{j_{m-1}}} 
    \subset \cdots \subset 
    \C^{(\ell_{j_{1}})}_{n_{j_{1}}, B_{j_{1}}} 
    \subset 
    \C_{n_i,B_i}^{(\ell)}
    \,.
\end{align}
Then, the resulting $(m+1)$-cut tubing is degenerate, in the sense of section \ref{sec:linrel}, if all elements in $I^{(i)}_{\alpha,\text{LHS}}$ appear in the sequence \eqref{eq:mp1cut} (the good cut condition ensures one cut corresponding to an element of $I^{(i)}_{\alpha,\text{LHS}}$ is present in the sequence \eqref{eq:mp1cut}).
This is the \emph{degenerate cut} condition of $\C_{n_i,B_i}^{(\ell)}$ or more precisely $I^{(i)}_{\alpha}$.
\\[1em]
Physical degenerate cuts need to be identified and counted \textit{exactly} once since they form a single degenerate sub-block in the DEQ matrix $\mat{A}$ and only contribute a single differential form to the physical basis. 
\end{tcolorbox}

To illustrate this condition in action, we provide a short discussion with selected examples below:
\begin{itemize}

\item Consider a 4-term minimal dependent set $I^{(j)}_{\alpha}=\{B_i, B_j| B_k, B_l\}$. While the 2-tubing $\C_{n_i,B_i}^{(\ell_i)} \subset \C_{n_j,B_j}^{(\ell_j)}$ is an unphysical 2-cut, $\C_{n_k,B_k}^{(\ell_k)} \subset \C_{n_j,B_j}^{(\ell_j)}$ is a physical 2-cut consistent with the good cut condition of $B_j$. This 2-tubing can now be followed by a residue tubing $\C_{n_i,B_i}^{(\ell_i)}$ (with $\tau_i$ entirely inside $\tau_j$) since this is consistent with the good cut evolution of $\C_{n_j,B_j}^{(\ell_j)}$. However, this 3-tubing satisfies the degenerate cut condition of $\C_{n_i,B_i}^{(\ell_i)}$ associated to $I^{(j)}_{\alpha}$ and must be classified as degenerate. 
This example is essentially a diagrammatic restatement of the residue relation (\ref{eq:degenboundequation}) discussed in section \ref{sec:linrel}.

The first example of such a degenerate cut occurs in the tree-level 3-site graph when the physical 2-tubings $\C_{2,B_5}^{(0)} \subset \C_{3,B_{T=4}}^{(0)}$ or $\C_{2,B_6}^{(0)} \subset \C_{3,B_{T=4}}^{(0)}$ are evolved into a 3-tubing and is a consequence of the minimal dependent set $I^{(T)}_1=\{B_2,B_{T=4}|B_5,B_6\}$. 
As discussed above, the cuts $\C_{2,B_{5}}^{(0)}$  and $\C_{2,B_{6}}^{(0)}$ effectively evolve as 2-site graphs, producing the 3-cuts
    \begin{equation} \label{eq:degen3cuts}
        \includegraphics[align=c,scale=0.67]{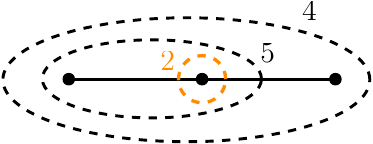} ~=~ \includegraphics[align=c,scale=0.67]{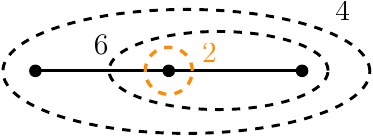}~.
    \end{equation}
However, the new cut tubing in {\color{Orange}orange} saturates $I_1^{(T)}$ and satisfies the degenerate cut condition of the true scattering facet $\C_{3,B_{T=4}}^{(0)}$.
Therefore, these 3-cuts must be identified.
    
\item A similar identification of a degenerate cut can be made for 1-loop graphs where one encounters 5-term minimal dependent sets. For example, for the case of the 1-loop 4-site box graph, we encounter a degenerate cut as a consequence of the minimal dependent set $I_1^{(T)}=\{B_1,B_3,B_{T=5}|B_{10},B_{12}\}$. 
Below, the 4th cut tubing $\C_{1,B_3}^{(0)}$ saturates $I_1^{(T)}$ and is identified:
    \begin{equation}
       \includegraphics[align=c,scale=0.67]{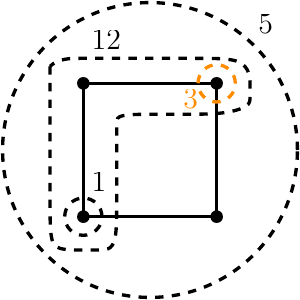} ~=~ \includegraphics[align=c,scale=0.67]{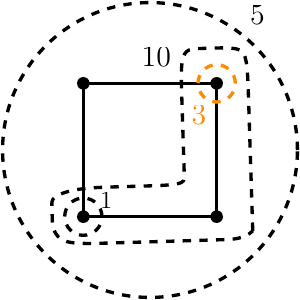}~.
    \end{equation}

\item The further evolution of physical degenerate cuts occurs in the usual fashion; adding a new tubing inside the most recent tubing in accordance with its good cut condition and the degenerate cut conditions (of previous tubings).

\item One often encounters scenarios where on the locus of a degenerate cut arising from some system, say $I^{(i)}$, is equivalent to the locus of another degenerate cut arising from a separate system $I^{(j)}$. In such a case, one simply identifies all $(m{+}1)$-cuts where the latest red tubing is common to all of them. 
For example, the first instance where one has to account for this is in the case of the tree-level 4-site chain graph due to the minimal dependent sets $I_1^{(9)}=\{B_2, B_9|B_6,B_7\}$ and $I^{(T)}_1=\{B_{T=5}, B_7 | B_9, B_{10}\}$:
\begin{equation}
    \includegraphics[align=c,scale=0.42]{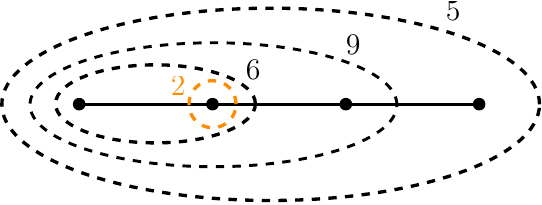} ~=~ \includegraphics[align=c,scale=0.42]{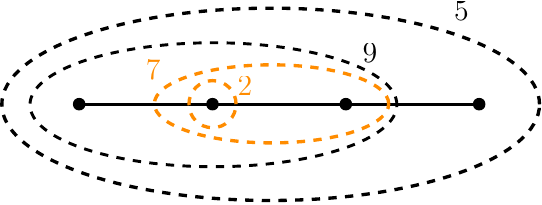}~=~ \includegraphics[align=c,scale=0.42]{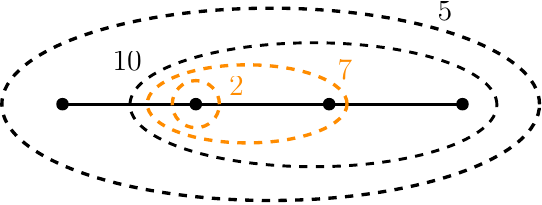}~. 
\end{equation}
Importantly, we note that such degenerate cuts are different from the following degenerate cut, which arises from a distinct system $I_2^{(T)}=\{B_2,B_{T=5}|B_6,B_{10}\}$:
\begin{equation}
    \includegraphics[align=c,scale=0.5]{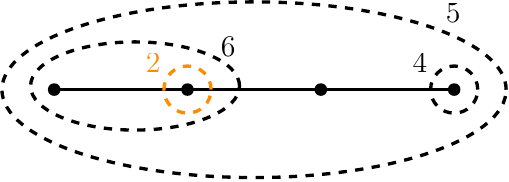} ~=~ \includegraphics[align=c,scale=0.5]{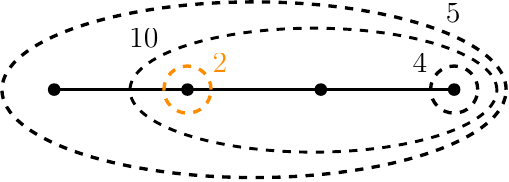}~.
\end{equation}  
\end{itemize}

\subsubsection{The uni-dimensional cohomology condition(s)}
\label{sec:beyondNesting}

The above good and degenerate cut conditions associated to any cut tubing are simple manifestations of the linear relations of the associated hyperplane arrangement and constrain sequentially nested cut tubings.
In addition to these, we need one more set of conditions---the \emph{uni-dimensional cohomology} condition(s)---that goes beyond nested evolution, to generate all physical cuts.
This graphical rule tells us how to avoid cuts with trivial cohomology (do not contribute forms to the basis).

\begin{tcolorbox}[breakable, title=The uni-dimensional cohomology condition(s)]
The uni-dimensional cohomology condition(s) dictates that:
\begin{enumerate}

    \item An $m$ cut tubing of an $n$-site $\ell$-loop graph $\G$ must enclose all the sites/vertices of $\G$ to yield a physical $m$-cut. 
    
    \item Inside each physical cut tube \emph{containing more than a single vertex}, further good cut tubes must leave \emph{at least} one unenclosed vertex.

\end{enumerate}
\end{tcolorbox}

Below, we provide the reasoning behind these conditions as well as a short discussion of selected examples. 
\begin{itemize}
    \item To have a non-trivial cohomology, the cut must contain a bounded chamber with boundaries made from the coordinate axes restricted to this cut. 
    For a 1-cut of $\G$, this implies that all physical 1-cut tubings are associated to hyperplane polynomials $B_i$ that contain the factor $\sum_{i=1}^n (x_i + X_i)$ in them.
    This ensures that one twisted hyperplane on the cut intersects the remaining other twisted hyperplanes creating a bounded chamber (see the right figure in \eqref{eq:trivandnontriv1-cut} corresponding to the 3-site tree graph for an example). 
    
    For tree-level wavefunction coefficients $\Psi_n^{(0)}$, this condition implies that the only allowed 1-tubing with a one-dimensional cohomology is the true scattering facet $\C_{n,B_T}^{(0)}$:
    \begin{equation}
       \C_{n,B_T}^{(0)} :=~\includegraphics[align=c,scale=0.55]{figs/residue_tubings/SFn.pdf}~.
       \label{eq:tree1-tubing}
    \end{equation}
    It is imperative to note that 1-tubings of tree graphs that do not correspond to the true facet tubing $\C_{n,B_T}^{(0)}$ violate this condition despite corresponding to non-vanishing residues of $\Psi_n^{(0)}$. This is because such 1-cuts $\C_{n_i, B_i \neq B_T}^{(0)}$ do not have a bounded chamber and therefore have trivial cohomology (the arrangement of twisted hyperplanes on the cut contains a pair of parallel hyperplanes; see left figure in \eqref{eq:trivandnontriv1-cut} corresponding to the 3-site tree graph for an example). 
    Such cuts correspond to unphysical cut tubings. 
    
    For example, the 1-cut $\C_{n,B_5}^{(0)}$ in the 3-site graph that encloses only two vertices out of the possible three
    \begin{equation}
       \C_{n,B_5}^{(0)} := \res_{\B_5} \Psi_3^{(0)}:=~\includegraphics[align=c,scale=0.55]{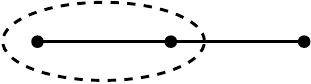}~ \neq 0~, \qquad \text{(invalid tubing)}~
       \label{eq:tree1-tubingB5}
    \end{equation}
    yields a non-zero residue; however, such 1-cuts have \textit{trivial cohomology} and are therefore \textit{inconsistent} with the condition above. 
    From a cuts perspective, such residues split the graph into a flat-space amplitude times a (shifted) lower-point wavefunction coefficient, as discussed in \cite{Baumann:2021fxj}. 
    
    For loop integrands, in addition to the true facet $\C_{n,B_T}^{(\ell)}$, there are further 1-cut tubings corresponding to hyperplanes $B_{{T}^{\prime}}$ that enclose all the sites of the graph and have a one-dimensional cohomology. For example, the $n$-gon has a total of $n{+}1$ physical 1-cuts. In the case of the 3-gon (triangle) graph, these four tubings are
    \begin{equation}
       \includegraphics[align=c,scale=0.72]{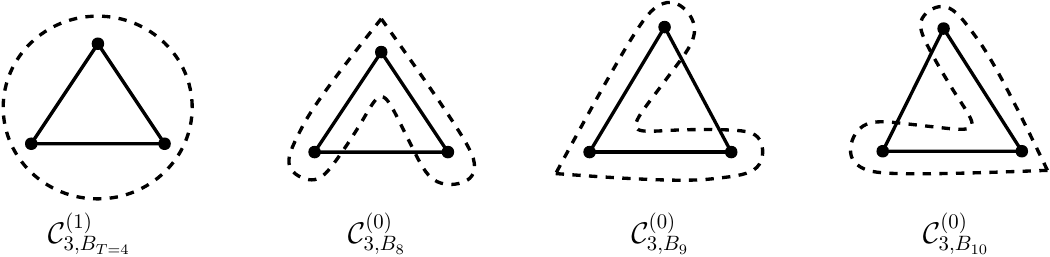}~.
    \label{eq:1-tubingstriangle}
    \end{equation}
    Note that in addition to the true scattering facet tubing $\C_{3,B_{T=4}}^{(1)}$ of the 3-gon, the three other 1-tubings corresponding to the hyperplane polynomials $B_{T^{\prime}=8,9,10} = \sum_{i=1}^3 (x_i+X_i) + 2 Y_{1,2,3}$, enclose all the three vertices of the triangle graph and are therefore consistent with the uni-dimensional cohomology condition. 
    
    For higher cuts, to be consistent with the first uni-dimensional cohomology condition, one has to partition the set of $n$-vertices of an $\ell$-loop graph into non-empty, non-overlapping subsets to give a physical product of residue tubings $\C_{n_1,B_1}^{(\ell_1)} \times \C_{n_2,B_2}^{(\ell_2)} \times \dots \times \C_{n_m,B_m}^{(\ell_m)}$ such that $n_1+n_2+ \dots + n_m=n$, which means no vertices are left unenclosed, thus ensuring a one-dimensional cohomology. An example of such a 2-tubing of the 4-site star graph is 
        \begin{equation}
        \C_{3,B_{10}}^{(0)} \times \C_{1,B_2}^{(0)}~ =  \includegraphics[align=c,scale=0.52]{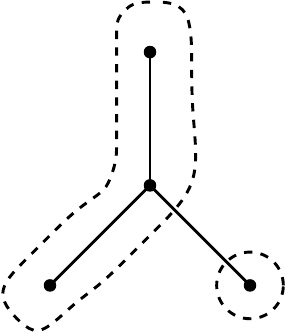}\,.
        \label{eq:2tubingstargraph}
        \end{equation}
    An example of a 4-cut of the double box consistent with this condition is
    \begin{equation}
       \C_{2,B_{12}}^{(0)} \times \C_{2,B_{10}}^{(0)} \times \C_{1,B_2}^{(0)} \times \C_{1,B_1}^{(0)} ~ = \includegraphics[align=c,scale=0.66]{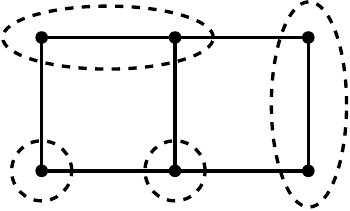}~.
    \end{equation}

\item Given any cut tubing, the subsequent cut tubings (consistent with its good cut condition) contained inside it must leave \emph{at least} one vertex unenclosed to guarantee a one-dimensional cohomology and a non-trivial residue for the corresponding cut. 
When  all vertices inside a cut tubing are enclosed again, the linear system cannot be solved and the corresponding cut/residue does not exist.

For example, one can check this condition explicitly in the case for the 3-site tree graph. The following 3-cuts do not exist
\begin{equation}
    \includegraphics[align=c,scale=0.64]{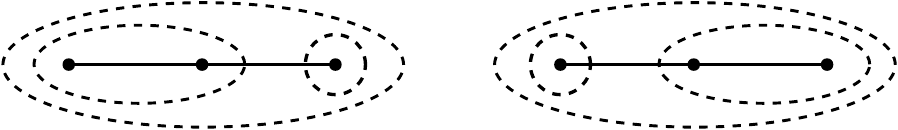}  \qquad \text{(invalid tubings)} \nonumber
\end{equation}    
while, on the other hand,  
\begin{equation}
    \includegraphics[align=c,scale=0.67]{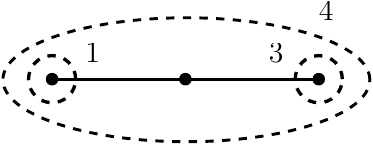}~.
\label{eq:3-site3-cutallowed}
\end{equation}
is a physical 3-cut. 
It abides by both the good cut condition and the uni-dimensional cohomology condition.

In general, the $n_i$-vertices inside any cut tubing $\C_{n_i,B_i}^{\elll}$ can be partitioned into non-empty, non-overlapping subsets to give a physical product of residue tubings $\C_{n_1,B_1}^{(\ell_1)} \times \C_{n_2,B_2}^{(\ell_2)} \times \dots \times \C_{n_m,B_m}^{(\ell_m)}$ such that $n_1+n_2+ \dots + n_m < n_i$. 
This leaves at least one vertex unenclosed and satisfies the second uni-dimensional cohomology condition.
\end{itemize}

The set of three conditions presented above --- the good cut condition, the degenerate cut condition and the uni-dimensional cohomology condition --- universally determine the set of all possible physical cuts contributing to the space of physical forms in the DEQs of \textit{any} given $\Psi_n^{\elll}$, as we demonstrate in the following.

\subsection{Enumerating physical cut tubings}
\label{ssec:basisoftubings}

Equipped with a universal set of rules for identifying physical cut tubings, we elucidate how to systematically construct all physical cuts  for \textit{any} $n$-site, $\ell$-loop graph. 
Moreover, we will demonstrate how these conditions (that identify the space of physical boundaries) are in correspondence with all possible ways to factorize $\Psi_n^{\elll}$ \textit{solely} into product(s) of flat-space scattering amplitudes.

\begin{center}
    \underline{1-cut tubings}
\end{center}
As discussed above, for tree-level wavefunction coefficients $\Psi_n^{\elll}$, the only allowed 1-tubing consistent with the uni-dimensional cohomology condition is the true scattering facet $\C_{n,B_T}^{(0)}$, which yields the corresponding flat-space amplitude $A_n$:\footnote{In order to demonstrate factorization properties, we will now restrict to a $\phi^3$ vertex for simplicity. However, our analysis holds true for arbitrary polynomial interactions.}
\begin{equation}
   \C_{n,B_T}^{(0)} = \includegraphics[align=c,scale=0.55]{figs/residue_tubings/SFn.pdf} \leftrightarrow \includegraphics[align=c,scale=0.53]{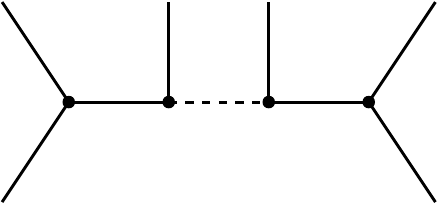}~.
\end{equation}

At loop level there exist additional 1-tubings that have one-dimensional cohomology as illustrated in \eqref{eq:1-tubingstriangle} for the 1-loop 3-gon. 
Note that the three 1-cuts $\C_{3,B_{i=8,9,10}}^{(0)}$ carry a superscript label $(0)$ in comparison to the true facet 1-cut $\C_{3,B_T=5}^{(1)}$. 
The difference between these tubings is that the total energy pole $B_{T=5}$ is involved in 5-term minimal dependent sets $I^{(T)}_{\alpha}$ (or more specifically, with a 3-term $I^{(T)}_{\alpha,\text{LHS}}$) whereas the hyperplane polynomials $B_{i=8,9,10}$ appear in 4-term sets (with a 2-term $I^{(i)}_{\beta,\text{LHS}}$) meaning that such 1-cuts effectively evolve as tree-level 3-site 1-cut tubings. This is a generic feature for all $n$-gon graphs and is an explicit example of how lower-site lower-loop data gets gets recycled for the $n$-site $\ell$-loop graph under study!

\begin{center}
    \underline{2-cut tubings}
\end{center}
Consistent with the rules presented in section \ref{ssec:cuttubingevolution}, there are two possible ways to construct physical 2-tubings that generates the entire space of physical 2-cuts contributing to the DEQ basis:
\begin{itemize}
    \item \textit{Evolution of the 1-cut(s)}: 
    For tree level graphs, the true scattering facet $\C_{n,B_T}^{(0)}$ is evolved into physical 2-tubings by inserting a cut tubing inside $\C_{n,B_T}^{(0)}$ that is consistent with its good cut condition, as described in section \ref{ssec:cuttubingevolution}. 
    These 2-cuts factorize the tree-level 1-cut $\C_{n,B_T}^{(0)}$ or, equivalently the flat-space amplitude $A_n$, into a product of lower-point amplitudes. 
   
    For example in the 3-site tree graph, two possible physical 2-cuts evolved according to the good cut condition of the true facet factorizes $A_5^{(0)}$ as follows:
    \begin{align}
    \C_{1,B_3}^{(0)} \subset \C_{3,B_{T=4}}^{(0)} := \includegraphics[align=c,scale=0.6]{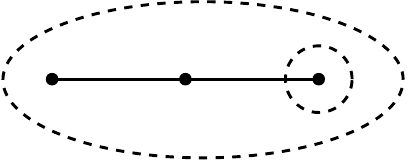} &\leftrightarrow \includegraphics[align=c,scale=0.4]{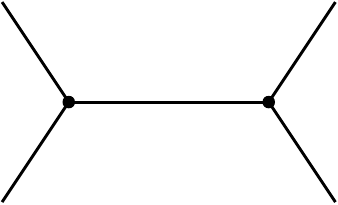} \times \includegraphics[align=c,scale=0.4]{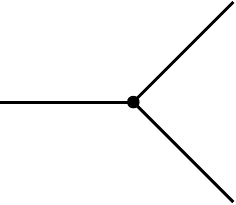}~, \\
    \C_{2,B_5}^{(0)} \subset \C_{3,B_{T=4}}^{(0)} := \includegraphics[align=c,scale=0.6]{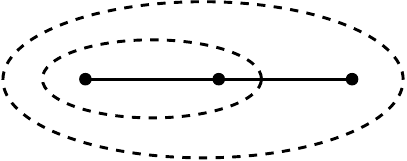} &\leftrightarrow \includegraphics[align=c,scale=0.4]{figs/residue_tubings/A4.pdf} \times \includegraphics[align=c,scale=0.4]{figs/residue_tubings/A3.pdf}~.
    \label{example:2-tubingsSF}
    \end{align}
    It is important to note that while the above 2-cuts result in a the same product of flat-space amplitudes $A_4 \times A_3$, they are distinct cut tubings and hence generate distinct physical forms. 
    This is a manifestation of the fact that the flat-space Mandelstam invariants are obtained from a product of two hyperplane polynomials $B_i, B_j$ (on the residue of the total energy pole) appearing in the wavefunction coefficient. An explicit residue computation for the tubings above in (\ref{example:2-tubingsSF}) gives a result proportional to $g \times \frac{-g^2}{s_{12}}$, where $g$ is the three-point coupling and $s_{12} = -(Y_1-X_1)(Y_1+X_1)$ is the flat-space Mandelstam invariant.
    
    For 1-loop $n$-gons, the 1-tubings $\C_{n,B_i}^{(0)}$ that are not the true facet evolve according to their good cut condition, which is analogous to an $n$-site tree-level 1-tubing. 
    The true facet $\C_{n,B_T}^{(1)}$ is evolved without any restriction; any  cut tubing can be placed inside it. 
    This discussion readily generalizes to generic $\ell$-loop graphs.

    \item \textit{Product of lower-point facets}: 
    As discussed in section \ref{sec:beyondNesting}, a physical 2-cut consistent with the uni-dimensional cohomology condition involves partitioning the set of $n$-vertices of an $\ell$-loop graph into two non-empty, non-overlapping subsets to give a product of two residue tubings $\C_{n_1,B_i}^{(\ell_1)} \times \C_{n_2,B_j}^{(\ell_2)}$ such that $n_1+n_2=n$.
    An example of such a 2-tubing for the tree-level 4-site star graph was given in \eqref{eq:2tubingstargraph} and corresponds to the  factorization:
    \begin{equation}
    \C_{3,B_{10}}^{(0)} \times \C_{1,B_2}^{(0)}~ =  \includegraphics[align=c,scale=0.5]{figs/residue_tubings/2-tubingstarexample.pdf} \leftrightarrow \includegraphics[align=c,scale=0.4]{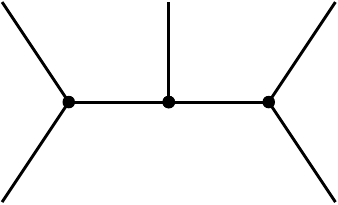} \times \includegraphics[align=c,scale=0.4]{figs/residue_tubings/A3.pdf}~. 
    \end{equation}
    Some examples of such 2-cuts in the 3-gon (triangle) and 4-gon (box) graphs are:
    \begin{equation}
        \includegraphics[align=c,scale=0.6]{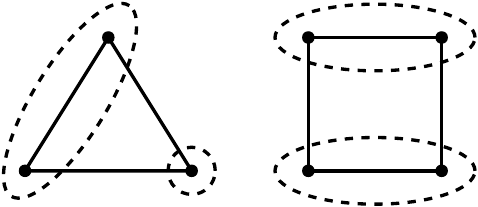}~.
    \end{equation}    
    
    It is important to note that each cut tubing in these 2-cuts will evolve as independent 1-cut tubings, in accordance with its own set of good and degenerate cut conditions. 
    What guarantees this is the existence of linear relations where $B_i, B_j$ now act as the ``true'' facets of lower-point functions with $n_1$ and $n_2$ vertices. 
    
    For example, in the 4-site star example presented above, the hyperplane $B_{10}$ is the ``true facet'' in the minimal dependent set $I^{(10)}_1=\{B_4,B_{10}|B_6,B_8\}$ represented graphically by
    \begin{equation}
    \left.\includegraphics[align=c,scale=0.5]{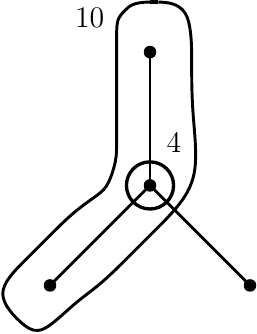} \right|_+
    ~=~ 
    \left.\includegraphics[align=c,scale=0.5]{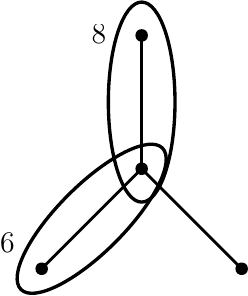} \right|_+
    ~.
    \end{equation}
    As a consequence, the 1-cut tubing $\C_{3,B_{10}}^{(0)}$ evolves as a 3-site 1-tubing.

    The same story extends to product 2-tubings of the $n$-gon. 
    For example, in the 4-gon 2-cut presented above, the two 1-cuts will each evolve as the 1-cut of the 2-site tree-level graph! 
    
    One can similarly work out higher loop examples. For example, in the double-box 6-site 2-loop graph, we have the following possible physical 2-cut:
    \begin{equation}
            \includegraphics[align=c,scale=0.67]{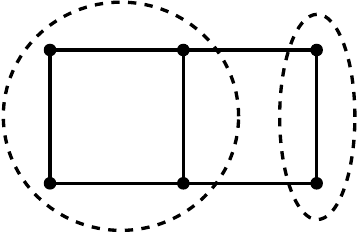}~.
        \end{equation} 
    While the first 1-cut evolves as that of a 4-gon box, the second 1-cut evolves as the 1-cut of the 2-site tree-level graph. This allows us to effectively reuse lower-site, lower-loop data for complicated loop graphs under study!

   \end{itemize}

\begin{center}
    \underline{3-cut tubings}
\end{center}

There are now three possible ways to construct physical 3-tubings that generates the entire space of physical 3-cut forms, while accounting for the appearance of degenerate cuts:
\begin{itemize}
    \item \textit{Evolution of the 1-cut(s)}: At tree level, the evolution of the physical 2-tubings involving the true facet $\C_{n,B_T}^{(0)}$ occurs via the two possible options. 
   
    i) Evolving the latest cut tubing consistent with its good cut condition. In such a scenario, one has to also account for the degenerate cut condition of  section \ref{sec:nesting}. An example of a non-degenerate and degenerate 3-cut for the tree-level 4-site star graph are:
        \begin{align}\begin{aligned}
            \C_{1,B_1}^{(0)} \subset \C_{3,B_9}^{(0)} \subset \C_{4,B_{T=5}}^{(0)} &= \includegraphics[align=c,scale=0.45]{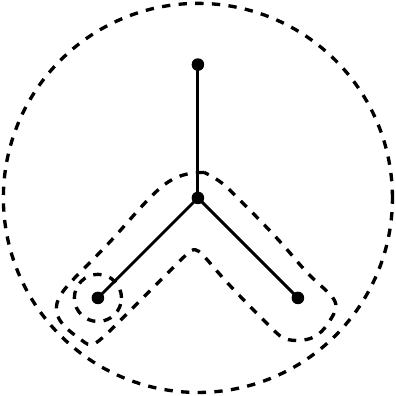} ~~, 
            \\
            \C_{2,B_6}^{(0)} \subset \C_{3,B_9}^{(0)} \subset \C_{4,B_{T=5}}^{(0)} &= \includegraphics[align=c,scale=0.45]{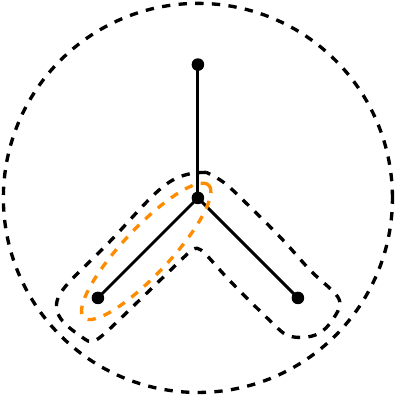}~.
        \end{aligned}\end{align}
        
        ii) Introducing an additional good cut tubing inside the true facet consistent with the second uni-dimensional cohomology condition, i.e., $(\C_{n_1,B_i}^{(0)} \times \C_{n_2,B_j}^{(0)}) \subset \C_{n,B_T}^{(0)} $ such that $n_1 + n_2 < n$.
        From a factorization perspective, this ensures that we avoid cutting an internal edge of the flat amplitude twice simultaneously. 
        
        An example of such a 3-tubing for the 3-site graph is \eqref{eq:3-site3-cutallowed}, 
        This is also the maximal number of tubings one can draw for such a graph and decomposes the flat-space amplitude into its three 3-pt building blocks $A_3^{(0)}$:
        \begin{equation}
            \includegraphics[align=c,scale=0.68]{figs/residue_tubings/4,1,3-3tubing-3-site.pdf} \leftrightarrow \includegraphics[align=c,scale=0.54]{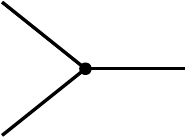} \times \includegraphics[align=c,scale=0.49]{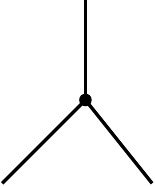} \times \includegraphics[align=c,scale=0.47]{figs/residue_tubings/A3.pdf}~.
            \label{eq:3site3cuteallowedcuts}
        \end{equation}

\item \textit{Evolution of the lower-point facets}: As discussed, the evolution of the 2-cut $\C_{n_1,B_i}^{(\ell_1)} \times \C_{n_2,B_j}^{(\ell_2)}$ proceed without extra subtleties except that any evolution of the individual 1-cuts has to be consistent with its good cut conditions. An example of such a 3-tubing for the 4-gon is shown as follows:
\begin{equation}
(\C_{1,B_{3}}^{(0)} \subset \C_{3,B_{12}}^{(0)}) \times \C_{1,B_2}^{(0)}:= \includegraphics[align=c,scale=0.75]{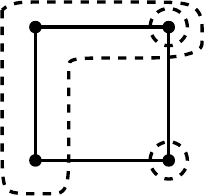}
\,.
\end{equation}
    \item \textit{Product of lower-point facets}: As before, one can construct physical 3-cuts consistent with the uni-dimensional cohomology condition, which involves a simple partition of the $n$-vertices into three non-empty, non-overlapping subsets to give a product of three 1-cuts $\C_{n_1,B_i}^{(\ell)} \times \C_{n_2,B_j}^{(\ell)} \times \C_{n_3,B_k}^{(\ell)}$ such that $n_1+n_2+n_3=n$. An example of such a 3-tubing for the 4-site star tree graph is:
    \begin{equation}
    \C_{2,B_8}^{(0)} \times \C_{1,B_1}^{(0)} \times \C_{1,B_2}^{(0)} ~:= \includegraphics[align=c,scale=0.55]{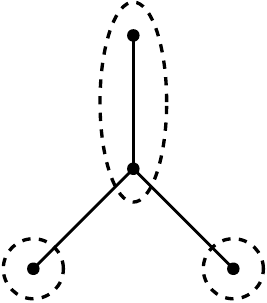} \leftrightarrow \includegraphics[align=c,scale=0.4]{figs/residue_tubings/A4.pdf} \times \includegraphics[align=c,scale=0.45]{figs/residue_tubings/A32.pdf} \times \includegraphics[align=c,scale=0.4]{figs/residue_tubings/A3.pdf}~.
    \end{equation}
\end{itemize}

\begin{center}
    \underline{$n$-cut tubings}
\end{center}

The procedure to generate the $n$-cut tubings for any $n$-site, $\ell$-loop graph proceeds in a similar fashion --- evolution and construction of each cut tubing such that it is consistent with its \textit{good cut}, \textit{degenerate cut} and the \textit{uni-dimensional cohomology} conditions at each step. For a generic graph, the  maximal number of residue operations one can perform is $n$, which from a cuts perspective corresponds to decomposing the flat-space amplitude $A_n$ into its $n$ 3-point building blocks. The process ends when one has generated all physical boundaries contributing to 1-, 2-, $\dots$, $n$-cuts, which in turn enumerate all the physical forms contributing to the physical subspace of the DEQs!

\subsection{From cut tubings to physical FRW forms \label{ssec:picstoforms}}

This space of physical cut tubings can be translated to the corresponding physical FRW forms that constitute physical DEQ basis. 
Not only do these cut tubings enumerate the space of physical cuts and therefore, the $\dlog_J$ part of a basis form (c.f., \eqref{eq:frwBasis}), they also determine $\tilde{\Omega}_J$ (and consequently $\Omega_J$), giving the full FRW form $\vphi_a$. 
We provide the rule to generate non-degenerate physical forms before providing the rule for degenerate forms.

\paragraph{Non-degenerate physical FRW forms:}
Enumerating the physical cut tubings via the rules presented in sections \ref{ssec:cuttubingevolution} and \ref{ssec:basisoftubings} is equivalent to listing all good $\delta_J$'s.
For non-degenerate cuts, each $\delta_J$ is dual via the intersection number to $\dlog_J$.
The remaining part of the FRW form is linked to the the canonical form of the cut $M_J$ (recall each physical cut has a 1-dimensional cohomology and unique canonical form). 

To determine $\tilde{\Omega}_J$ (or equivalently $\Omega_J = \tilde{\Omega}\vert_J$) from the cut tubings, one links each pair of vertices $\{i,j\}$ without crossing a cut tubing.%
\footnote{The variable $x_i$ is fully determined if it is the only unenclosed vertex inside another cut tubing. Therefore, it cannot participate in the $\dlog$-form contributing to $\tilde{\Omega}_J$.}
Each linked pair of vertices corresponds to the $\dlog$-form: $\dlog\frac{x_i}{x_j}$. 
Wedging together all pairs produces $\tilde{\Omega}_J$. 
We illustrate this procedure using selected examples from the physical basis of the 4-site chain (see appendix \ref{app:4SiteChain})
\begin{align}
    \includegraphics[align=c,scale=1]{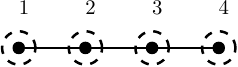}
    &= \vphi_{38}^\text{4-chain} = \dlog_{1,2,3,4}
    \,,
     \\
    \includegraphics[align=c,scale=1]{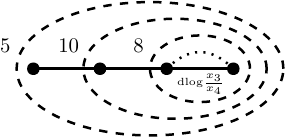}
    &= \vphi_{34}^\text{4-chain} = \dlog_{5,10,8}
    \wedge\dlog\frac{x_3}{x_4} 
    \,,
    \\
    \includegraphics[align=c,scale=1]{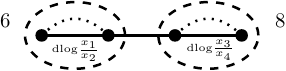}
    &= \vphi_{10}^\text{4-chain} = \dlog_{6,8}
    \wedge\dlog\frac{x_1}{x_2} 
    \wedge\dlog\frac{x_3}{x_4} 
    \,,
    \\
    \includegraphics[align=c,scale=1]{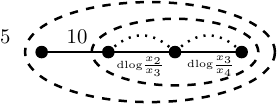}
    &= \vphi_7^\text{4-chain} = \dlog_{5,10}
    \wedge\dlog\frac{x_2}{x_3} 
    \wedge\dlog\frac{x_3}{x_4} 
    \,,
    \\
    \includegraphics[align=c,scale=1]{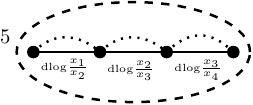}
    &= \vphi_1^\text{4-chain} = \dlog_{5}
    \wedge\dlog\frac{x_1}{x_2} 
    \wedge\dlog\frac{x_2}{x_3} 
    \wedge\dlog\frac{x_3}{x_4} 
    \,,
\end{align}
and the 1-loop 3-gon (see appendix \ref{app:3Site1Loop})
\begin{align}
    \includegraphics[align=c,scale=.75]{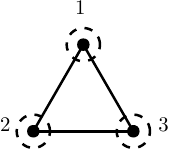}
    &= \vphi_{24}^\text{3-gon} = \dlog_{1,2,3}
    \,,
\end{align}
\begin{align}
     \includegraphics[align=c,scale=.75]{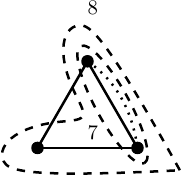}
    &= \vphi_{24}^\text{3-gon} = \dlog_{8,7}
    \wedge\dlog\frac{x_1}{x_3}  
    \,,
\\
    \includegraphics[align=c,scale=.75]{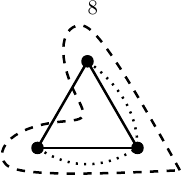}
    &= \vphi_2^\text{3-gon} = \dlog_{8}
    \wedge\dlog\frac{x_1}{x_2} 
    \wedge\dlog\frac{x_2}{x_3} 
    \,,
\\
    \includegraphics[align=c,scale=.75]{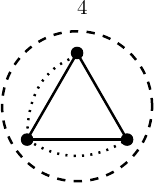}
    &= \vphi_1^\text{3-gon} = \dlog_{4}
    \wedge\dlog\frac{x_1}{x_2} 
    \wedge\dlog\frac{x_2}{x_3} 
    \,.
\end{align}
Note that the indices in the $\dlog_J$ have been ordered according to our \hyperlink{box:cutOrd}{cut ordering rules}: larger tubes come before smaller tubes.

\paragraph{Degenerate physical FRW forms: 
}

As discussed previously, there are \hyperlink{box:noCrossCuts}{no crossed cut tubings} in the physical subspace. 
The physical forms associated to degenerate cuts always come with a specific linear combination of $\dlog_J$'s where each $J$ is associated to a non-crossing tubing of the degenerate cut. 
For example, in the 3-site chain we have 
\begin{align}\label{eq:3sitePhiPhys16}
    \vphi_{\text{phys},16} 
    = \frac{\vphi_{19}-\vphi_{20}}{2} 
    = \dlog B_4 \wedge \dlog \frac{B_5}{B_6}
    \wedge \dlog B_2
    \,.
\end{align}
We conjecture that the physical linear combination of $\dlog$'s associated to any degenerate cut is obtained by first identifying all non-crossed cut tubings associated to this degenerate cut. 
Then, the corresponding $\dlog$'s are added with coefficients $\pm1$ such that the $B_i$ corresponding to crossed tubings appear in ratios like in \eqref{eq:3sitePhiPhys16} whenever possible.

To further illustrate this point, consider the degenerate 4-cut involving $B_5,B_9,B_{10},B_7$ and $B_3$. 
There are three non-crossing cut tubings associated to this cut
\begin{equation}
    \includegraphics[align=c,scale=0.42]{figs/residue_tubings/5962-4site-degen.pdf},\, 
    \includegraphics[align=c,scale=0.42]{figs/residue_tubings/5972-4site-degen.pdf},\,
    \includegraphics[align=c,scale=0.42]{figs/residue_tubings/51072-4site-degen.pdf}
    \,. 
\end{equation}
From the discussion above, we want the crossing tubings to appear as ratios in the physical form 
\begin{align}
    \vphi_{63} &=
        \dlog_{5,9,6,2} 
        - \dlog_{5,9,7,2}
        + \dlog_{5,10,7,2}
    \,,
\end{align}
as given in \eqref{eq:degen4cuts4sitechain} (produced via the direct calculation method outlined in section \ref{sec:3SitePed}).  
Indeed this is easy to check
\begin{align}\begin{aligned}
    \vphi_{63} &= 
    \dlog B_5 \wedge \dlog \frac{B_6}{B_{7}}
    \wedge \dlog B_7 \wedge \dlog B_3 
    + \dlog_{5,10,7,3}
    \\
    &= \dlog_{5,9,6,2} 
    + \dlog B_5 \wedge \dlog \frac{B_{10}}{B_9}
    \wedge \dlog B_7 \wedge \dlog B_2
\end{aligned}\end{align}
where the tubings $\{\tau_6,\tau_{7}\}$ and $\{\tau_{9},\tau_{10}\}$ cross.
However, the pair $\{\tau_{10},\tau_{6}\}$ also cross but cannot be put into a ratio because they differ by more than one factor of $B_i$. 

In general, the physical $\dlog$ form for a degenerate $k$-cut is 
\begin{align} \label{eq:physDegen}
    \sum_{a=1}^r 
    \underset{\pm1}{\underbracket{
    \left(\frac{\bigwedge_{i\in J_a^{(k)}} \d B_i}{\bigwedge_{j\in J_1^{(k)}} \d B_j} \right)
    }}
    \dlog_{J_a^{(k)}}
    \,,
\end{align}
where $J_a^{(k)}$ is a set indexing the $B_i$ generated by the non-crossing $k$-tubings associated to the degenerate $k$-cut: $\{\tau^{(k)}_a\}$ with $a=1, \dots, r\geq 2$. 
The only freedom is the overall sign which is fixed by a choice of $\tau_1^{(k)}$ or equivalently $J_1^{(k)}$.

To see why this formula must be correct, recall that all FRW forms are generated by the set of compatible complete tubings of $\G$, which by definition, do not contain crossed tubings.
Additionally, note that each term in this set comes with coefficient $+1$.
Up to a sign, the non-crossing residues of $\Psi^\elll_n$ corresponding to a degenerate cut must be the same. 
Because of this, the relative sign of the residues and corresponding combination of $\dlog$'s that appear in $\Psi^\elll_n$ comes solely from the relative sign of the differentials $\bigwedge_{j\in J_a^{(k)}} \d B_j$'s.

From the formula \eqref{eq:physDegen}, it is easy to see that the $B_i$ that cross must come with signs such that they form ratios in the $\dlog$-form. 
Choose two terms corresponding to sets $J_a$ and $J_b$ that share $k-1$ elements
\begin{align}
    J_a^{(k)} = \{(J_a^{(k)} \cap J_b^{(k)}), p\}
    \quad\text{and}\quad
    J_b ^{(k)}= \{(J_a^{(k)} \cap J_b^{(k)}), q\}
    \,.
\end{align}
The tubings $\tau_p$ and $\tau_q$ necessarily cross which means that they appear on the same side of a linear relation with a ``$+$''-sign. 
Thus, $B_q = -B_p + \cdots$ and therefore there is a relative sign between the differentials $\wedge_{i\in J_a^{(k)}} \d B_i$ and $\wedge_{j\in J_1^{(k)}} \d B_j$. 
When combined, we find a $\dlog$-form with the ratio $B_p/B_q$. 

An alternative construction of the physical form associated to a degenerate boundary using a diagrammatic method for partial fractions can be found in appendix \ref{app:partialFrac}. 

Since all degenerate cuts correspond to the same topological space, the associated $\tilde{\Omega}_{J^{(k)}}$ are the same and generated in the same way as for non-degenerate cuts. 
For example, 
\begin{align}
    \includegraphics[align=c,scale=.7]{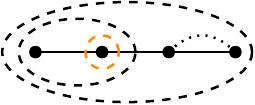}
    {-} \includegraphics[align=c,scale=.7]{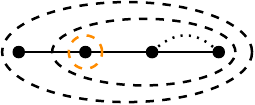}
    &= \vphi_{35}^\text{4-chain}
    = \left(\dlog_{5,6,2}{-}\dlog_{5,10,2}\right)
    \wedge\dlog\frac{x_3}{x_4} 
    \,,
    \\
     \includegraphics[align=c,scale=.8]{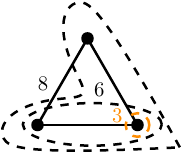}
    {-} \includegraphics[align=c,scale=.8]{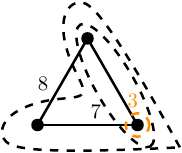}
    &= \vphi_{48}^\text{3-gon}
    = \dlog_{8,6,3}-\dlog_{8,7,3}
    \,, 
\end{align}
where $\vphi^\text{4-chain}_{35}$ and $\vphi^\text{3-gon}_{48}$ can be found in appendix \ref{app:4SiteChain} and \ref{app:3Site1Loop}.

\subsection{Example: the physical basis for the 3-site tree graph from cut tubings}
\label{ssec:3sitefromtubings}
We now demonstrate the rules governing the evolution and construction of cut tubings to enumerate the physical boundaries contributing to the DEQ system with explicit examples. 
We begin with the 3-site tree graph and show that our prescription involving drawing consistent residue tubings land us on the 16 physical forms listed in (\ref{eq:3-sitephysicalforms}). For the hyperplane polynomials $B_i$ and the associated linear relation system $I$ for the 3-site tree graph, please refer to \eqref{eq:3siteB} and \eqref{eq:3-sitelinrelnew} respectively. 

\begin{center}
    \underline{1-cut tubing}
\end{center}
As discussed in section \ref{ssec:basisoftubings}, the only allowed 1-tubing consistent with the uni-dimensional cohomology condition is the true scattering facet of the 3-site tree graph $\C_{3,B_{T=4}}^{(0)}$. This corresponds to the residue that yields the 5-pt flat-space tree amplitude $A_5$:
\begin{equation}
        \includegraphics[align=c,scale=0.65]{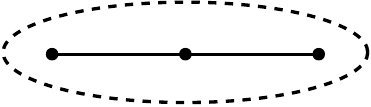}~ \leftrightarrow \includegraphics[align=c,scale=0.52]{figs/residue_tubings/A5.pdf}
        \,,
        \label{eq:diag1-cut}
\end{equation}  
as well as the physical form $\color{BrickRed}\vphi_1$ given in \eqref{eq:3site1cut}.

\begin{center}
    \underline{2-cut tubings}
\end{center}
Now, there are two possible ways to construct physical 2-tubings, as discussed in section \ref{ssec:basisoftubings}:
\begin{itemize}
\item[i)] Evolve the scattering facet tubing \eqref{eq:diag1-cut} consistent with its good cut condition. Following the discussion leading up to \eqref{eq:badboundary3-site}, we obtain four such good 2-tubings that evolve from the scattering facet which generate the physical forms $\color{BrickRed}\vphi_2, \color{BrickRed}\vphi_4, \color{BrickRed}\vphi_6, \color{BrickRed}\vphi_4$ listed in \eqref{eq:3site2cut}:
\begin{equation}
        \includegraphics[align=c,scale=0.6]{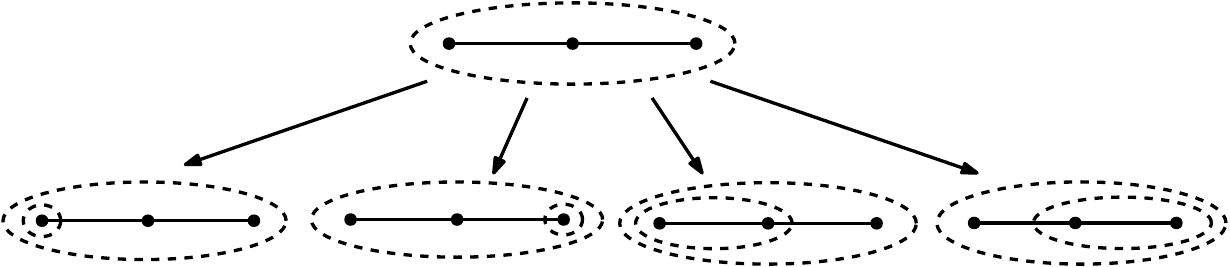}          \label{eq:2tubingsSF}
    \end{equation}     
Each of these 2-cuts above factorize the 5-pt flat amplitude $A_5$ into the product $A_4 \times A_3$, as was shown in \eqref{example:2-tubingsSF} for two such 2-cuts. 
\item[ii)] Construct physical 2-tubings by decomposing the graph into a product of two lower-point residue tubings. For the 3-site graph, we have two such possibilities:
\begin{align}
\begin{aligned}
    \includegraphics[align=c,scale=0.67]{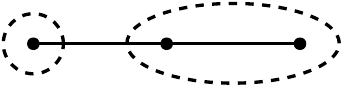} &\leftrightarrow \includegraphics[align=c,scale=0.52]{figs/residue_tubings/A31.pdf} \times \includegraphics[align=c,scale=0.42]{figs/residue_tubings/A4.pdf}~, \\
   \includegraphics[align=c,scale=0.67]{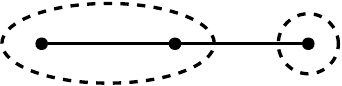} &\leftrightarrow \includegraphics[align=c,scale=0.42]{figs/residue_tubings/A4.pdf} \times \includegraphics[align=c,scale=0.4]{figs/residue_tubings/A3.pdf}~.
    \label{eq:2tubingsnotSF}
    \end{aligned} 
\end{align}
These 2-cuts correspond to the physical forms $\color{BrickRed}\vphi_3$ and $\color{BrickRed}\vphi_5$ in \eqref{eq:3site2cut}.

\end{itemize}

\begin{center}
    \underline{3-cut tubings}
\end{center}
The construction of 3-cut tubings proceeds in three different ways, as outlined in section \ref{ssec:basisoftubings}:  
\begin{itemize}

\item[i)] Continue evolving the 2-tubings that originated from the scattering facet \eqref{eq:2tubingsSF}, except that now we have to be consistent with the appropriate degenerate cut condition. 
Additionally, we can construct a 3-cut consistent with the uni-dimensional cohomology condition, as shown in \eqref{eq:3site3cuteallowedcuts}.
Thus, we get three non-degenerate physical 3-tubings that correspond to the forms $\color{BrickRed}\vphi_{13}, \color{BrickRed}\vphi_{16}, \color{BrickRed}\vphi_{18}$ in \eqref{eq:3site3cut}. 
The degenerate 3-cuts \eqref{eq:degen3cuts} are identified and counted precisely \textit{once}; they correspond to the physical form $({\color{Orange}\vphi_{19}}-{\color{Orange}\vphi_{20}})/2$.
The full evolution of the scattering facet tubing into 3-cuts is depicted below:
\begin{equation}
        \includegraphics[align=c,scale=0.6]{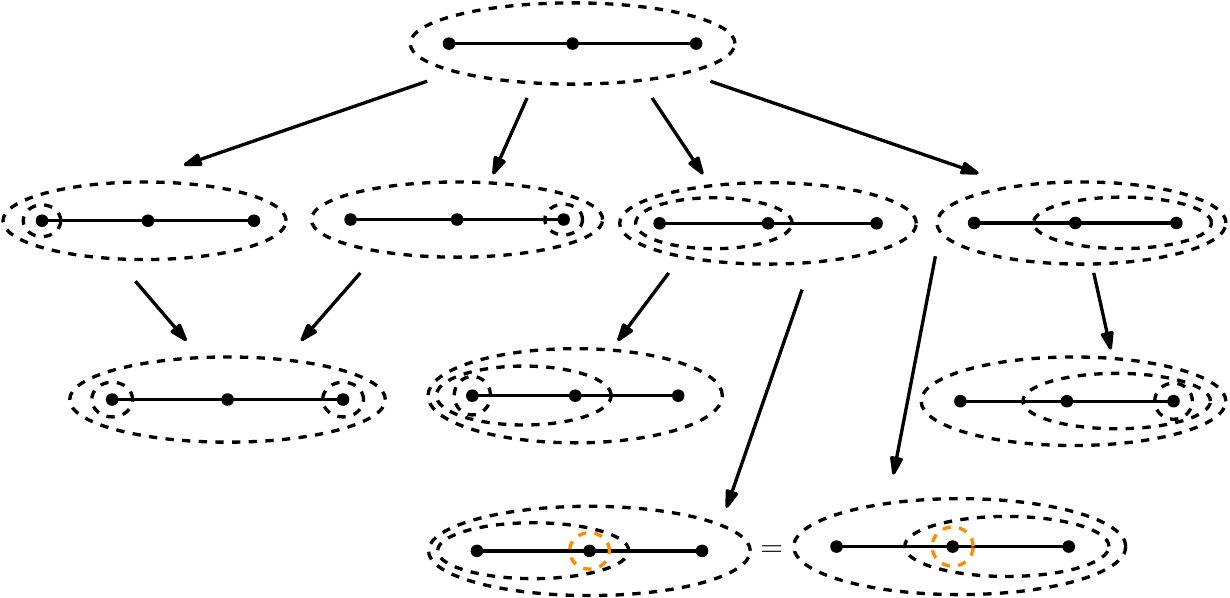}  
    \end{equation} 
All the 3-cuts above correspond to the product $A_3 \times A_3 \times A_3$, as shown explicitly for the 3-cut in \eqref{eq:3site3cuteallowedcuts}.

\item[ii)] The second set of physical 3-tubings are obtained by evolving the 2-tubings constructed in \eqref{eq:2tubingsnotSF}. We note the tubings $\C_{2,B_5}^{(0)}$ and $\C_{2,B_6}^{(0)}$ in \eqref{eq:2tubingsnotSF} are effectively scattering facets of a 2-site tree graph, which can be evolved without restriction (as there are no associated minimal dependent sets $I$ associated with $\Psi_2^{(0)}$):
\begin{equation}
        \includegraphics[align=c,scale=0.65]{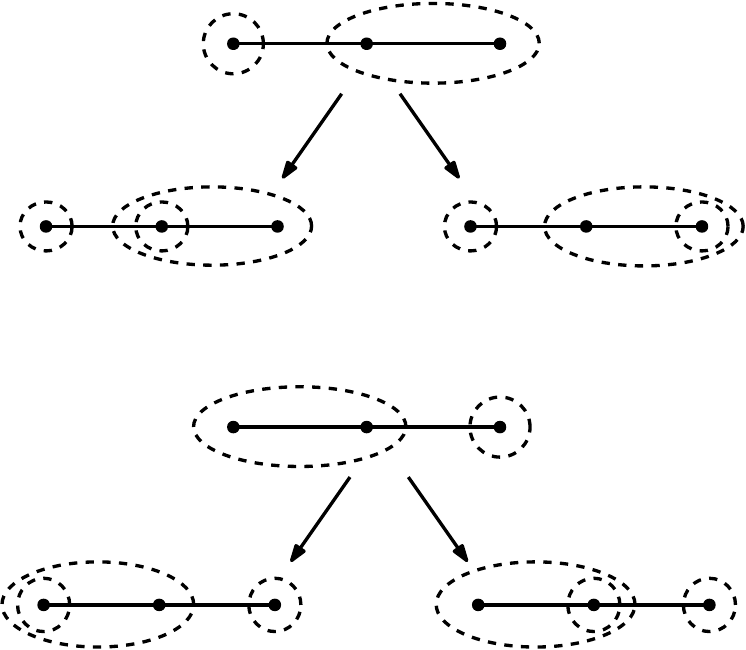} 
    \end{equation} 
These 3-cuts correspond to the physical forms $\color{BrickRed}\vphi_{12}, \color{BrickRed}\vphi_{15}, \color{BrickRed}\vphi_{14}, \color{BrickRed}\vphi_{17}$ in \eqref{eq:3site3cut} and also decompose the flat amplitude $A_5$ into its three building blocks $A_3$.

\item[iii)] The remaining physical 3-cut:
\begin{equation}
        \includegraphics[align=c,scale=0.65]{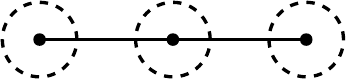} ~ \leftrightarrow \includegraphics[align=c,scale=0.5]{figs/residue_tubings/A31.pdf} \times \includegraphics[align=c,scale=0.5]{figs/residue_tubings/A32.pdf} \times \includegraphics[align=c,scale=0.4]{figs/residue_tubings/A3.pdf}
        \,,
\label{eq:3-site3-tubing123}
    \end{equation}
generates the remaining physical 3-form $\color{BrickRed}\vphi_{11}$ in \eqref{eq:3site3cut}. 
\end{itemize}

For the 3-site tree graph, the maximal number of residues one can take is three and hence the flow ends with drawing all consistent 1-, 2- and 3-tubings, which are in \emph{one-to-one} correspondence with all possible ways to factorize the 5-pt flat amplitude down to its 3-point building blocks, as demonstrated above. 
Therefore, these simple rules for evolving cut tubings generate the 16 physical forms in the DEQs of the 3-site tree-level wavefunction coefficient!

\subsection{Example: the physical basis for 2-site 2-loop sunset graph from cut tubings}
\label{ssec:sunset}

Having demonstrated the universal rules for residue tubings for the simplest 3-site tree-level graph, we show these rules in action for the 2-site 2-loop sunset graph. 

\paragraph{Hyperplane polynomials:} The FRW hyperplane polynomials $B_i$ associated with the sunset graph are
\begin{align}
    B_1 & {=} 
    \includegraphics[align=c,width=4em]{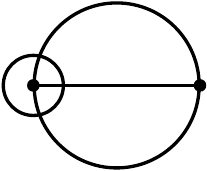}
    {=}  x_1 {+} X_1 {+} Y_1 {+} Y_2{+} Y_3
    ,
    &
    B_2 &{=}
    \includegraphics[align=c,width=4em]{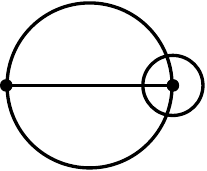}
    {=} x_2 {+} X_2 {+} Y_1 {+} Y_2{+} Y_3
    ,
    \nn\\
    B_{T{=}3} &{=} 
    \includegraphics[align=c,width=4em]{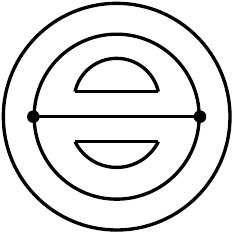}
    {=} x_1 {+} x_2 {+} X_1 {+} X_2
    , 
    &
    B_4 &{=} 
    \includegraphics[align=c,width=4em]{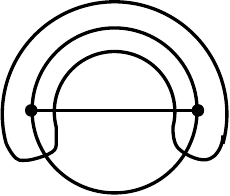}
    {=} x_1 {+} x_2 {+} X_1 {+} X_2{+} 2Y_2 {+} 2Y_3
    ,
    \nn
\end{align}
\begin{align}
    B_5 &{=} 
    \includegraphics[align=c,width=4em]{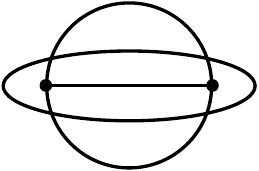}
    {=} x_1 {+} x_2 {+} X_1 {+} X_2{+} 2Y_1 {+} 2Y_3
    ,
    &
    B_6 &{=} 
    \includegraphics[align=c,width=4em]{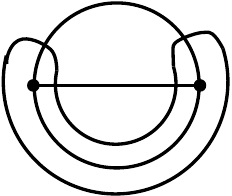}
    {=} x_1 {+} x_2 {+} X_1 {+} X_2{+} 2Y_1 {+} 2Y_2
    ,
    \nn\\
    B_7 &{=} 
    \includegraphics[align=c,width=4em]{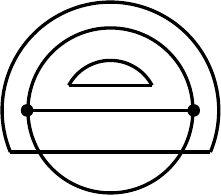}
    {=} x_1 {+} x_2 {+} X_1 {+} X_2{+} 2Y_3
    , 
    &
    B_8 &{=} 
    \includegraphics[align=c,width=4em]{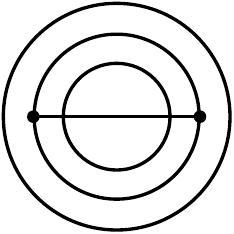}
    {=} x_1 {+} x_2 {+} X_1 {+} X_2{+} 2Y_2
    ,
    \nn\\
    B_9 &{=} 
    \includegraphics[align=c,width=4em]{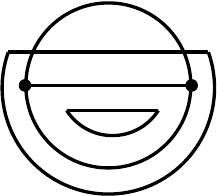}
    {=} x_1 {+} x_2 {+} X_1 {+} X_2{+} 2Y_1
    .
    \label{eq:sunsetB}
\end{align}

\paragraph{Linear relations:} There exists several linear relations involving the total energy hyperplane $B_{T=3}$. These include three 4-term relations
giving us the minimal dependent sets
\begin{equation}
    I_1^{(T)} =\{B_{T=3},B_4|B_7,B_8\}~~,~~I_2^{(T)} =\{B_{T=3},B_5|B_7,B_9\}~~,~~I_3^{(T)}=\{B_{T=3},B_6|B_8,B_9\}~.
    \label{eq:mindepsetssunset}
\end{equation}
There also exists linear relations involving the other hyperplane polynomials, which all give 5-term minimal dependent sets. For example, a set involving $B_9$ gives us
\begin{equation}
    I^{(9)}=\{B_1, B_2, B_9|B_5, B_6\}~. 
\end{equation}
However, such 5-term sets constrain the space of physical 3-cuts as per the good cut condition; hence, they will not affect our analysis for the sunset graph, where the maximal number of residues one can perform is two. 

We can now construct the physical 1- and 2-cuts that contribute to the physical basis of the sunset graph.

\begin{center}
    \underline{1-cut tubings}
\end{center}
There are seven 1-tubings that are consistent with the uni-dimensional cohomology condition:
\begin{equation}
        \includegraphics[align=c,scale=0.64]{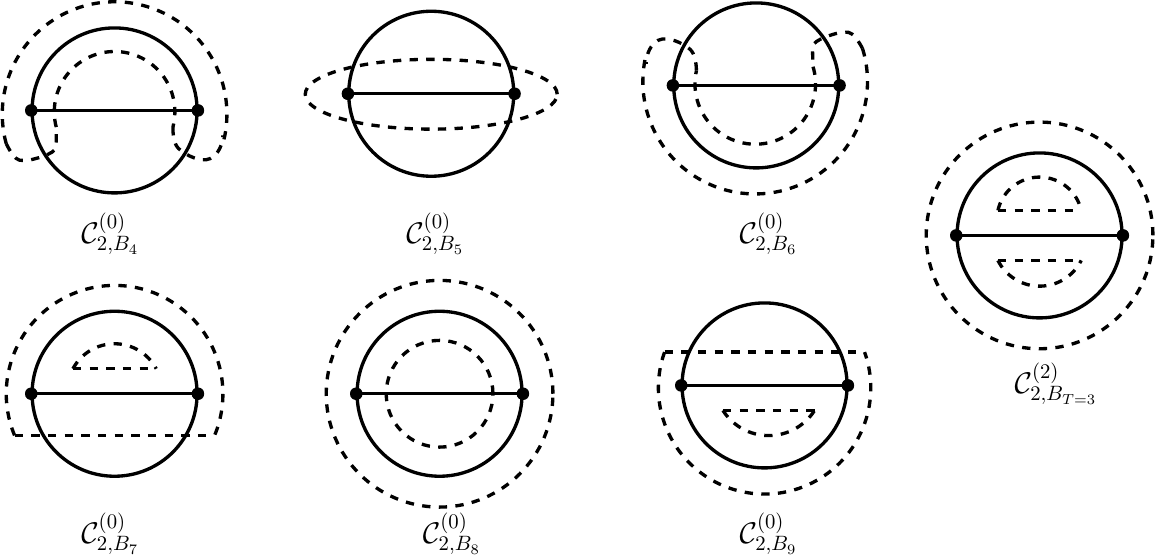}
    \end{equation}
\label{eq:sunset1-tubings}
We note that in addition to the true scattering facet $\C_{2,B_T=3}^{(2)}$, there are six further physical 1-tubings.

\begin{center}
    \underline{2-cut tubings}
\end{center}
Now, the true scattering facet $\C_{2,B_T=3}^{(2)}$ is evolved
to 2-tubings consistent with its good cut condition resulting from the sets $I_{\alpha}^{(T)}$ listed in \eqref{eq:mindepsetssunset}. However, coupled with the uni-dimensional cohomology condition, we only get the following two physical 2-cuts arising from the true facet:
\begin{equation}
        \includegraphics[align=c,scale=0.55]{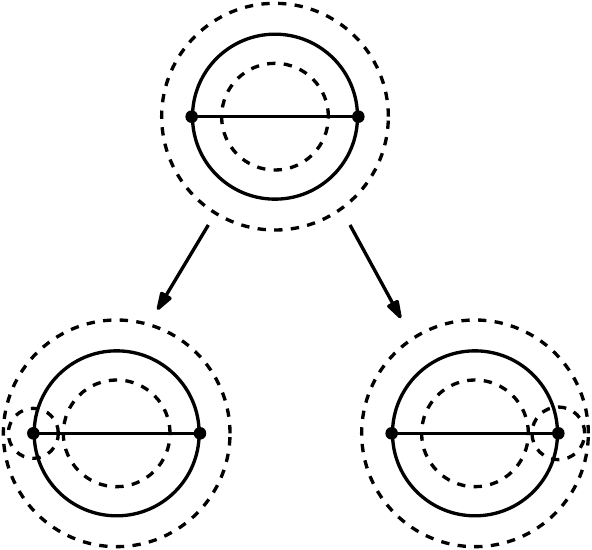}
        \label{eq:sf2-tubingsunset}
    \end{equation}
Each of the remaining six 1-tubings are evolved to give two physical 2-cuts, consistent with the uni-dimensional cohomology condition:
\begin{equation}
        \includegraphics[align=c,scale=0.47]{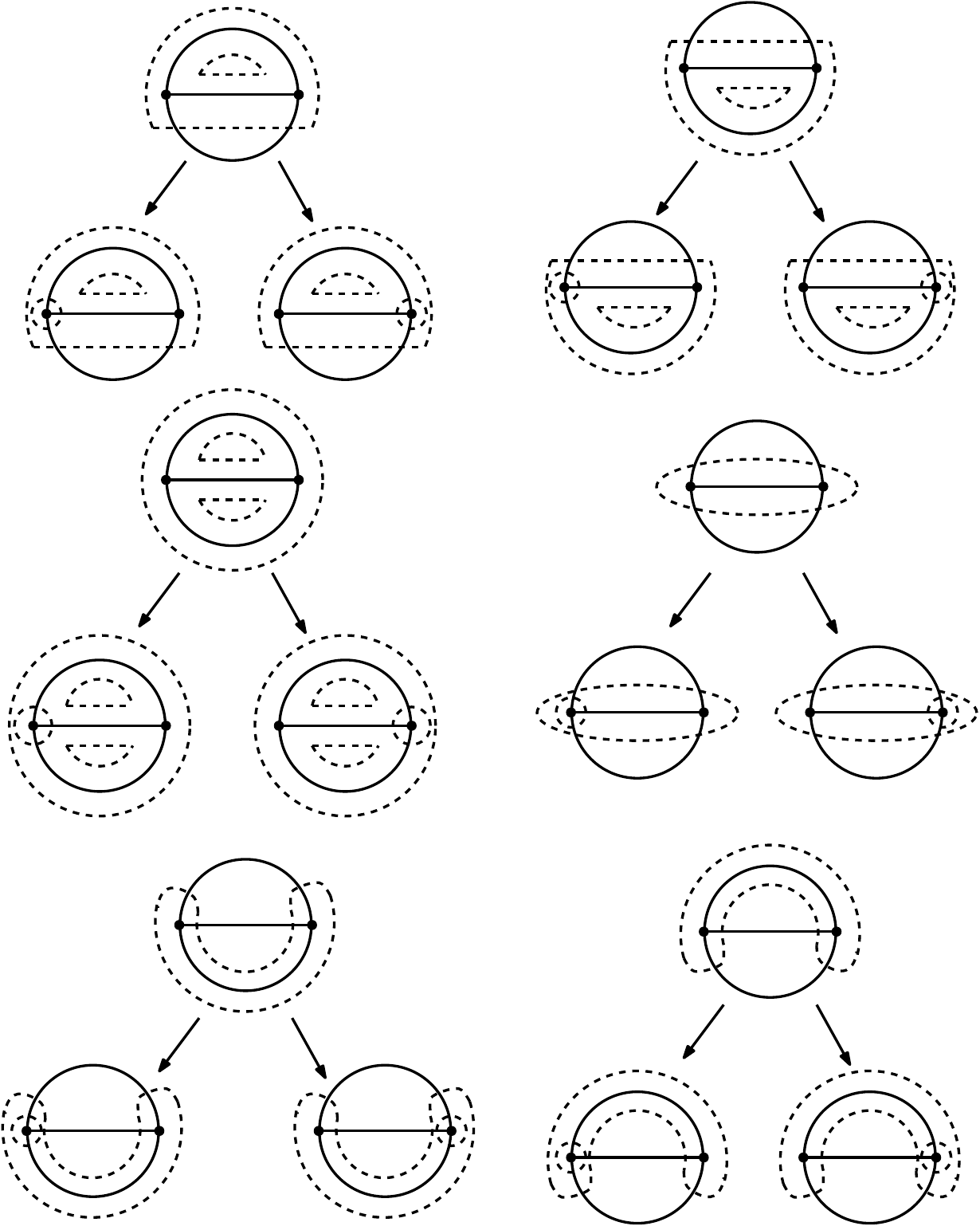} 
    \label{eq:2-tubingsunset}
\end{equation}
    
The remaining 2-tubing is simply given by the product of two 1-tubings enclosing each of the vertices:
\begin{equation}
        \includegraphics[align=c,scale=0.55]{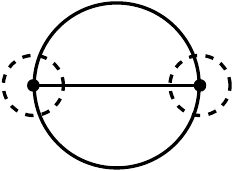}~. 
    \label{eq:rem2-tubingsunset}
\end{equation}
Counting all the seven 1-cut tubings in \eqref{eq:sunset1-tubings} along with the fifteen 2-cut tubings in \eqref{eq:sf2-tubingsunset}, \eqref{eq:2-tubingsunset} and \eqref{eq:rem2-tubingsunset}, yields the 22 physical forms for the sunset graph. 

More generally, it is interesting to note that for a 2-site $\ell$-loop graph, the uni-dimensional cohomology condition implies all its 1-cuts behave as those corresponding to a 2-site tree level graph.
This, along with accounting for the single 2-cut of the form \eqref{eq:rem2-tubingsunset}, allows us to predict the number of the physical forms contributing to the DEQs of the 2-site $\ell$-loop wavefunction coefficient $\psi_2^{(\ell)}$:
\begin{equation}\label{eq:2siteAnyLoop}
    \underbrace{(2^{\ell+1}-1)}_{\# ~\text{1-cuts}} + \underbrace{(2^{\ell+2}-1)}_{\# ~\text{2-cuts}} = 2(3 \cdot 2^{\ell}-1)\,.
\end{equation}
This matches the counting of the physical forms for the bubble ($\ell=1$) and the sunset ($\ell=2$) graphs using the rules of kinematic flow \cite{Hang:2024xas, Baumann:2024mvm}. 
Moreover, at $\ell$-loop order, \eqref{eq:2siteAnyLoop} matches the counting presented in \cite{He:2024olr} obtained by adding the number of functions at each level in their corresponding DEQ system.

\subsection{Cut stratification for trees and loops}
\label{ssec:cutstrat}
We end this section on residue tubings with comments on certain surprising patterns associated to the space of physical boundaries/cuts for tree and loop graphs. At tree-level, the number of physical cut tubings/forms is $4^{n-1}$ and splits according to the codimension of the $b$-cuts 
\begin{equation} \label{eq:treeCutStrat}
     4^{n-1} = \sum_{b=1}^n \binom{n-1}{b-1} 3^{b-1}\;,
\end{equation}
which can be organized into a Pascal triangle:
\begin{center}
\begin{tabular}{c|cccccccc}
     & 1-cuts
     & 2-cuts
     & 3-cuts
     & 4-cuts
     & 5-cuts
     & 6-cuts
     & $\cdots$
     \\
     \hline
     & $3^0$
     & $3^1$
     & $3^2$
     & $3^3$
     & $3^4$
     & $3^5$ 
     & $\cdots$
     \\
     \hline
     $n=2$
     & 1
     & 1
     \\
    $n=3$
     & 1
     & 2
     & 1
     \\
    $n=4$
     & 1
     & 3
     & 3
     & 1
     \\
    $n=5$
     & 1
     & 4
     & 6
     & 4
     & 1
     \\
    $n=6$
     & 1
     & 5
     & 10
     & 10
     & 5
     & 1
     \\
     $\vdots$ 
     &$\vdots$
     &$\vdots$
     &$\vdots$
     &$\vdots$
     &$\vdots$
     &$\vdots$
\end{tabular}
\end{center}
Importantly, we note that this stratification is topology \textit{independent}!

The number of physical forms for $n$-gons at 1-loop generated by this flow of residue tubings matches the number ($=4^n-2(2^n-1)$) predicted in \cite{He:2024olr}. 
While this data does not form a Pascal triangle structure as in the case of tree graphs, its cut stratification is given by
\begin{center}
\begin{tabular}{c|ccccc}
     & 1-cuts 
     & 2-cuts
     & 3-cuts
     & 4-cuts
     & $\cdots$
     \\
     \hline
     $n=2$ (bubble)
     & 3
     & 7
     \\
    $n=3$ (triangle)
     & 4
     & 21
     & 25
     \\
    $n=4$ (box)
     & 5
     & 42
     & 100
     & 79
     \\
     $\vdots$ 
     & $\vdots$
     &$\vdots$
     &$\vdots$
     &$\vdots$
\end{tabular}
\end{center}

\section{Discussion and Outlook \label{sec:conclusion}}

The FRW wavefunction coefficient is expressed as a twisted integral whose integrand is associated to a particular hyperplane arrangement (geometry). 
Recently, the kinematic flow algorithm \cite{Arkani-Hamed:2023bsv, Arkani-Hamed:2023kig, Hang:2024xas, Baumann:2024mvm} demonstrated that the twisted cohomology of FRW cosmological integrands vastly over counts the number of integrands needed to define DEQs for FRW wavefunction coefficients. 
The minimal space of integrands that couple to the wavefunction coefficient is called the physical subspace and its characterization is imperative for understanding how to build DEQs for loop integrated wavefunction coefficients. 

In this article, we demonstrate that the physical subspace is intimately tied to the physical (non-trivial) cuts/residues of the FRW form $\Psi_n^\elll$ (integrand of the FRW wavefunction coefficient). 
Using the intersection number (an inner product on the space of integrands) and exploiting extra structure in the dual relative twisted cohomology (automatically organized by cuts), we show that the physical subspace is spanned by the forms that share cuts/residues with $\Psi_n^\elll$ via equations \eqref{eq:DEQFromRes} and \eqref{eq:physBasis}. 

We compute the DEQs for the full twisted cohomology of the 3-site chain (section \ref{sec:3SitePed}), 4-site chain (appendix \ref{app:4SiteChain}) and 1-loop 3-gon (appendix \ref{app:3Site1Loop}) graphs using the intersection number. 
Then, we demonstrate how to construct the physical subspace and check that this subspace has an integrable DEQs that closes.
\texttt{Mathematica} files for these examples can be found at the github repository \github.

We provide geometric and graphical rules for generating the differential forms that make up the physical subspace. 
Importantly, one has to disentangle differential forms associated to so-called degenerate cuts. 
Degenerate cuts are codimension-$m$ surfaces that correspond to the intersection of more than $m$ (untwisted) hyperplanes ($\B_i=\{B_i=0\}$). 
As a consequence, transposing the order of two resides in a sequential residue does not necessarily (anti-) commute. 
Moreover, there are relations among sequential resides and differential forms. 
We show that each degenerate cut contributes \emph{only} one differential form to the physical subspace and provide a simple rule to construct this form. 

Degenerate cuts are a consequence of linear relations satisfied by the polynomials that generate the cosmological hyperplane arrangement. 
In section \ref{sec:linrel}, we show how all linear relations of a cosmological hyperplane arrangement are a result of simple graphical rules. 
Since each hyperplane is generated by a graphical 1-tubing of the Feynman graph, linear relations are the result of uncrossing a crossed 2-tubing. 

Amazingly, these linear relations classify all physical, unphysical and degenerate cuts as well as differential forms. 
In section \ref{sec:goodTubes}, we develop graphical rules for enumerating all physical cuts and identifying degenerate cuts.  
We also provide rules that map the diagrammatics to explicit forms. 
The good cut, degenerate cut and uni-dimensional cohomology conditions enumerate the physical subspace for any graph at any loop order (at the level of loop integrands).
At tree-level, the physical subspace has a nice interpretation: the physical subspace is all forms associated to cuts that factorize the FRW form into exclusively flat space amplitudes. 
A completely satisfactory explanation of this fact is missing and left for future work. 

Our results for the size of the physical subspace is in perfect agreement with other methods (kinematic flow \cite{Arkani-Hamed:2023bsv, Arkani-Hamed:2023kig, Hang:2024xas, Baumann:2024mvm}). 
Moreover, we find that there is a simple Pascal's triangle structure to how many of the $4^{n-1}$ physical forms are associated with each $b$-cut
(see \eqref{eq:treeCutStrat}). 
For the $\ell$-loop 2-site graph, we verified that the size of the physical subspace is $2(3 \cdot 2^{\ell}-1)$ and break down how it is organized on each cut in equation \eqref{eq:2siteAnyLoop}. 

The flow of cut tubings---how to evolve a physical $m$-cut to a physical $(m+1)$-cut---also predict which physical dual forms couple to each other in the DEQs. 
This is a consequence of the structure of dual cohomology; specifically, how the derivative produces boundary terms (see footnote \ref{foot:bdTerms}). 
Then, since the dual and FRW DEQs are related by a sign and transposition, one simply reverses the arrows to get the flow of FRW forms. 
This effectively predicts all the zeros of the DEQs and provides a possible explanation for the mechanism responsible for the structure observed in the kinematic flow algorithm. 
Graphical rules for determining the entries of the connection matrix akin to the kinematic flow is currently under investigation.

\section*{Acknowledgments}
SD and AP are grateful to Anastasia Volovich and Marcus Spradlin for valuable feedback throughout the project. 
We also thank Cl\'ement Dupont, Andrew McLeod, Lecheng Ren, He-Chen Weng and Karen Yeats for engaging discussions that helped shape our final results. 
This work was supported in part by the US Department of Energy under contract DESC0010010 Task F (SD, AP).

\appendix

\section{From cut contours to the dual canonical form \label{app:dualCanForm}}

In this appendix, we provide more intuition for the nature of dual forms by relating them to more familiar objects: cut contours. 
The ideas presented here are also important for the development of a diagrammatic coaction and are part of some work in progress \cite{coaction}.

For a positive geometry, the {\color{BrickRed}canonical form} maps a dual contour $\check{\Delta}$ (bounded chamber with boundaries on either $\B_i$ or $\mathcal{T}_i$) to an FRW form $\Omega[\check{\Delta}]$ with logarithmic singularities on $\partial\check{\Delta}$.
There also exists a dual version of this map, which maps a cut contour $\Delta_J \wedge \circlearrowleft_J$
(where $\Delta_J$ is a bounded chamber on the cut $M_J$ with boundaries $\mathcal{T}_i \cap \B_J$) to an FRW form via the {\color{MidnightBlue} blue arrow(s)}
\begin{equation}\label{eq:contourToForm}
\begin{tikzcd}
    \mathrm{dual\; forms\;}  
    \arrow[r, dashed, MidnightBlue, "\la\bullet\vert\bullet\ra = \mathds{1}"]
    & \mathrm{FRW\; forms\;} 
    \\
    \mathrm{dual\; contours\;} 
    \arrow[ur, BrickRed, swap, "\Omega" near end]
    & \mathrm{FRW\; contours\;} 
    \arrow[ul, dashed, crossing over, MidnightBlue, "\check{\Omega}" near end]
    \arrow[u, MidnightBlue]
\end{tikzcd}\,.
\end{equation}
In this way, our basis of FRW forms inherits the essential properties of cut contours: we map contours on the space of interest to forms on the space of interest. 
This should be contrasted with the usual positive geometry workflow where the canonical form maps contours belonging to an axillary space to forms on the space of interest.

The first (diagonal) {\color{MidnightBlue} blue dashed arrow} in \eqref{eq:contourToForm} represents the {\color{MidnightBlue} dual canonical form} associated to a cut contour 
\begin{align} \label{eq:gammaToForm}
    \gamma = \Delta_J \wedge \circlearrowleft_J
    \quad
    \mapsto
    \quad
    {\color{MidnightBlue}\check{\Omega}}[\Delta_J \wedge \circlearrowleft_J] 
    := \delta_J(\Omega[\Delta_J])
    \;. 
\end{align}
This is exactly how we build up our basis of dual forms in \eqref{eq:dualBasis}. 
The second (horizontal) {\color{MidnightBlue} blue dashed arrow} in \eqref{eq:contourToForm}, represents choosing a basis of FRW forms that diagonalizes the intersection matrix. 
The following replacement rule 
\begin{align}
    \delta_J(\Omega[\Delta_J])
    \quad
    \mapsto
    \quad
    \dlog{[\textstyle J]} \wedge \tilde{\Omega}[\Delta_J]
    \;. 
\end{align}
generates a basis of FRW forms that almost diagonalizes the intersection matrix as demonstrated by \eqref{eq:blockDiag}.
Recall that $\tilde{\Omega}[\Delta_J]$ is obtained from $\Omega[\Delta_J]$ by removing the restriction to the cut $J$ acting on the $T_i$'s (an explicit example of this procedure is given in section \ref{sec:3SitePed}).
A simple recipe for diagonalizing the degenerate cuts in physical subspace is provided in section \ref{sec:linrel}. 

Intuitively, the action of a dual canonical form $\cvphi = \delta_J(\Omega[\Delta_{J}])$ on an FRW form $\vphi$ via the intersection number should be thought of as an approximation to the the integral $\int_{\gamma} u\, \vphi$ where $\gamma=\Delta_{J} \wedge \circlearrowleft_J$ is the cut contour that generates the dual canonical form $\cvphi$. 
That is, 
\begin{align}
    \int_{\Delta_{J} \wedge \circlearrowleft_J} 
    (u \times  \bullet)
    \approx
    \la \delta_J(\Omega_{J}[\Delta_J]) \vert \bullet \ra
    \;.
\end{align}
Formally, this is because the dual canonical form is a realization of Poincar\'e duality:
$H_n(M\setminus \B;\mathcal{L}_{u}) \simeq \check{H}^n(M\setminus\B;\nabla) = H^n(M,\B;\check{\nabla})$.%
\footnote{In general, Poincar\'e duality is an isomorphism between the $p^\text{th}$ homology and the $(2n-p)^\text{th}$ cohomology where $n$ is the complex dimension of the topological space ($M\setminus\B$ or $(M,\B)$ in our case). For $p=n$, this is an isomorphism between the $n^\text{th}$ homology and cohomology as stated in the main text.}\fnsep%
\footnote{The symbol $\mathcal{L}_u$ is the local system to associated to the twist. Intuitively, this is extra information in the contour specifying which branch of the multi-valued function $u$ should be used when integrating. It is only here for completeness and can be ignored for the purposes of this work.}\fnsep%
\footnote{The existence of the dual canonical form should not be surprising since  $H_n(M\setminus \B;\mathcal{L}_{u})$ has an analogous decomposition to \eqref{eq:relCoDef} (see \cite{hwa1966homology, pham2011singularities} for such a decomposition in the untwisted case).}
Since we diagonalize the intersection matrix of the physical subspace, this implies, that our physical FRW forms have a diagonal pairing with physical cut contours to leading order
\begin{align} \label{eq:cosmologicalPerscriptiveUnitarity}
    \int_{\gamma_{\text{phys},a}}
    u\, \vphi_{\text{phys},b} 
    \approx \delta_{ab}
    \,. 
\end{align}
Like physical dual forms, physical contours do not contain residues that annihilate the physical FRW form. 
The problems with degenerate residues of cosmological arrangements is dealt with in the same way as dual forms (section \ref{sec:linrel}). 

Equation \eqref{eq:cosmologicalPerscriptiveUnitarity} can also be thought of as a generalization of prescriptive unitarity for integrals with multi-valued integrands. 
Usually, prescriptive unitarity instructs one to find a basis of forms that is diagonal with respect to cut contours \cite{Bourjaily:2017wjl}.  
However, due to the twist $u$, the best one can do is diagonalize the leading order part of this matrix.
Moreover, achieving this using the intersection number is much easier that actually carrying out the integrations and expanding in $\vep$.

\section{No new poles\label{app:noNewPoles}}

In this appendix, we derive a simple representation for $\nabla_\kin \vphi$ that has at most simple poles. 
Such a representation is needed to use the simplified formulas \eqref{eq:cobdAct} and  \eqref{eq:CombInt} for computing the intersection number. 
Then, we show that $\res_{J_b}[\nabla_\kin\vphi_a]=0 \iff \res_{J_b}[\vphi_a] = 0$ from which we can deduce equation \eqref{eq:physBasis}. 

To this end, it is easiest to work on on the full space of kinematic and integration variables: $\mathcal{M} \setminus \mathcal{B}$.
Here, we set $\mathcal{M} = \mathbb{C}^{3n+\ell-1} \setminus \mathcal{T}$ where there are $3n-1$ total variables: $n$ integration variables $x_i$, $n$ kinematic variables $X_i$ and an additional $n+\ell-1$ kinematic variables $Y_i$.
We also interpret the vanishing loci $\{T_i=0\}$ and $\{B_i=0\}$ on $\mathbb{C}^{3n+\ell-1}$ instead of just the integration space $M$.

Then all $\dlog$-forms on the integration space $M$ can be canonically uplifted to the full space $\mathcal{M}$ simply by swapping all instances of $\dlog$ for $\dtot\log = \dlog + \d_\kin\log$.
We also upgrade the covariant derivative $\nabla \to \nabla_\dtot = \nabla + \nabla_\kin = \dtot + \dtot\log u \wedge $.
Denoting the uplift of $\vphi$ by $\Phi$, the part of $\nabla_\dtot \Phi$ that is a 1-form on kinematic space is cohomologous to $\nabla_\kin\vphi$ in the usual sense. 
Thus, we have a  closed form $\dlog$ representation of $\nabla_\kin\phi$
\begin{align} \label{eq:nicedphi}
    \nabla_\kin\vphi \simeq 
    \vep\ \dtot\log(T_1\cdots T_n) \wedge \Phi
    \vert_{\text{(kin }p>1\text{)-forms}\to0}
\end{align}
since $\Phi$ only has simple poles and $\D\Phi=0$.
The projection $\text{(kin }p>1\text{)-forms}\to0$ is equivalent to adding up all possible ways one can assign one $\dtot \to \d_\kin$ with all others $\dtot\to\d$.
Having a simple representation for $\nabla_\kin\vphi$ with only simple poles allows for the use of formulas \eqref{eq:cobdAct} and \eqref{eq:CombInt}. 

Then, the claim that ${\res_K[\vphi]=0}$ $\implies$ ${\res_K[\nabla_\kin\vphi] = 0}$ follows straightforwardly from \eqref{eq:nicedphi}. 
Explicitly, start with the following representation for the physical differential form
\begin{align}
    \Psi_n^{(\ell)} = \sum_{J: |J|=n} \psi_J\ \dlog_J
\end{align}
for some coefficients $\psi_J\in\mathbb{Q}$.
Since $\Psi_n^{(\ell)}$ does not have any poles on the twisted hyperplanes, only $\dlog_J$ forms need to be included in the above decomposition. 
Next, we make the replacement $\dlog \to \dtot\log$ to uplift $\Psi_n^\elll$ and compute its total covariant derivative
\begin{align}
    \label{eq:FRWuplift}
    \Psi_n^\elll &\to \sum_{J: |J|=n} \psi_J\ \dtot\log_J
    \\
    \label{eq:dFRWuplift}
    \nabla_\dtot \eqref{eq:FRWuplift}
    &= \vep \sum_{J: |J|=n} \psi_J\ \dtot\log(T_1 \cdots T_n) \wedge \dtot\log_J
\end{align}
Since the residue $\res_K$ is simply the coefficient of $\dtot\log_K$, we have 
\begin{align}\label{eq:resK}
    \res_K[\eqref{eq:FRWuplift}] 
    &= \sum_{J \supset K} 
        (-1)^{J \setminus K} \psi_J 
        \bigwedge_{j\in J\setminus K} \dtot\log B_j\vert_K
\end{align}
and 
\begin{align} \label{eq:dresK}
    \res_K[\eqref{eq:dFRWuplift}]
    {=}
    \vep\ (-1)^{|K|} 
    \dtot\log(T_1 {\cdots} T_n \vert_K)
    {\wedge} \eqref{eq:resK}
\end{align}
Therefore, 
\begin{align}
    \res_K [\Psi_n^{(\ell)}] &= \eqref{eq:resK}\vert_{\text{(kin }p>1\text{)-forms}\to0} = 0
    \,,
    \\
    &\Downarrow\nn
    \\
    \res_K[\nabla_\kin\Psi_n^{(\ell)}] &= \eqref{eq:dresK}\vert_{\text{(kin }p>1\text{)-forms}\to0} = 0
    \,,
\end{align}
as claimed. 
The same manipulations work for all other basis elements, not just $\Psi_n^\ell$.

\section{Further examples (including loops) via direct computation \label{sec:examples}}

In this appendix, we elaborate on how to construct the gauge transformation that splits the twisted cohomology into a physical and unphysical subspace (appendix \ref{app:3siteDegenBlock}). 
Then we apply the methods of section \ref{sec:3SitePed} to the 4-site chain (appendix \ref{app:4SiteChain}), the 4-site star (appendix \ref{app:4SiteStar}) and the 1-loop 3-gon (appendix \ref{app:3Site1Loop}) graphs. 
Using the intersection matrix, we demonstrate that the corresponding full cohomologies with dimensions 213, 312 and 99 reduce to a physical basis with 64, 64 and 50 elements, respectively, in agreement with both the kinematic flow algorithm \cite{Arkani-Hamed:2023bsv, Arkani-Hamed:2023kig, Hang:2024xas, Baumann:2024mvm} and those obtained from a time integral perspective \cite{He:2024olr}. 
The DEQs for the 4-site chain and the 1-loop 3-gon  can be found in the \texttt{Mathematica} files at the \texttt{github} repository \github.

\subsection{Explicit decomposition of the 3-site chain degenerate block \label{app:3siteDegenBlock}}

To decompose the degenerate block in the 3-site chain example, we compute the vector 
\begin{align}\begin{aligned}
    \vec{w}_{4562} &:= \left(
        \res_{452}[\Psi_3^{(0)}], 
        \res_{462}[\Psi_3^{(0)}], 
        \res_{562}[\Psi_3^{(0)}]
    \right)
    = (1,-1,0)
    \,,
\end{aligned}\end{align}
and its nullspace, $\mat{W}_{4562} := \text{Null}(\vec{w}_{4562})$. 
Next, we build the matrix
\begin{align} \label{eq:gauge}
    \mat{\c{U}} = 
    \begin{pmatrix}
        \mathds{1}_{18\times18} &  & 
        \\
         & (\vec{w}_{4562})_{1\times3} & 
        \\
         & (\mat{W}_{4562})_{2\times3} & 
        \\
         &  & \mathds{1}_{4\times 4}
    \end{pmatrix}
    = \begin{pmatrix}
        \mathds{1}_{18\times18} & & & 
        \\
         & 1 & -1 & 0
        \\
         & 0 & 0 & 1
         \\
         & 1 & 1 & 0
        \\
         &  & & & \mathds{1}_{4\times 4}
    \end{pmatrix}
    \,.
\end{align}
The new basis $\bs{\cvphi}^{\prime\prime} = \mat{\c{U}}\cdot \bs{\cvphi}^\prime$ contains 16 dual forms whose induced residues do not annihilate $\Psi_3^{(0)}$.
A basis of FRW forms with the analogous property is given by the basis $\bs{\vphi}^{\prime\prime} = \mat{U}\cdot\bs{\vphi}^\prime$ where 
\begin{align}
    \mat{U} = {( \mat{\c{U}} \cdot \mat{C}^\prime)^{-1}}^\top
    = \begin{pmatrix}
        \mathds{1}_{8\times8} & & & & &
        \\
        & \mat{\mathcal{U}}_{2\times2} & & & & 
        \\
        & & \mathds{1}_{8\times8}
        \\
        & & & & \frac{1}{2} & -\frac{1}{2} & 0 &
        \\
        & & & & 0 & 0 & 1 &
        \\
        & & & & \frac{1}{2} & \frac{1}{2} & 0 &
        \\
        & & & &  &  &  & \mathds{1}_{4\times4} 
    \end{pmatrix}
    \,,
\end{align}
and $\mat{\mathcal{U}}_{2\times2}$ is a matrix corresponding to the unphysical 56-cut with a 2-dimensional cohomology; its exact form is not needed.

The forms of \eqref{eq:3-sitephysicalforms} make up the physical subspace. Their corresponding differential equation is 
\begin{align}
    \mat{A}_\phys = \mat{A}_\phys^\text{diag} + \mat{A}_\phys^\text{off-diag} 
    \,,
\end{align}
where 
\begin{align}
    \mat{A}_\phys^\text{diag} =- \vep\,  \text{diag}\bigg(&\scriptscriptstyle
        \left\llbracket{S_4^3}\right\rrbracket              
        \,,
        \left\llbracket{{S_1^+} (S_6^-)^2}\right\rrbracket           
        \,,
        \left\llbracket{{S_1^+} (S_6^+)^2}\right\rrbracket          
        \,,
        \left\llbracket{{S_3^+} (S_5^-)^2}\right\rrbracket        
        \,,
        \left\llbracket{{S_3^+} (S_5^+)^2}\right\rrbracket       
        \,,
        \left\llbracket{-}{{S_3^-} (S_5^+)^2}\right\rrbracket     
        \,,
        \nn\\&\quad\scriptscriptstyle
        \left\llbracket{-}{{S_1^-} (S_6^+)^2}\right\rrbracket  
        \,,
        \left\llbracket{{S_1^+} {S_2^{++}} {S_3^+}}\right\rrbracket
        \,,
        \left\llbracket{{S_1^+} {S_2^{++}} {S_3^-}}\right\rrbracket 
        \,,
        \left\llbracket{-}{{S_1^+} {S_2^{--}} {S_3^+}}\right\rrbracket 
        \,,     
        \left\llbracket{-}{{S_1^+} {S_2^{-+}} {S_3^+}}\right\rrbracket           
        \,,
        \nn\\&\quad{\scriptscriptstyle
        \left\llbracket{{S_1^+} {S_2^{+-}} {S_3^+}}\right\rrbracket         
        \,,
        \left\llbracket{-}{{S_1^+} {S_2^{-+}} {S_3^-}}\right\rrbracket       
        \,,
        \left\llbracket{-}{{S_1^-} {S_2^{++}} {S_3^+}}\right\rrbracket     
        \,,
        \left\llbracket{-}{{S_1^-} {S_2^{+-}} {S_3^+}}\right\rrbracket   
        \,,
        \left\llbracket{-}{{S_1^-} {S_2^{++}} {S_3^-}}\right\rrbracket 
    }\bigg) \,,
\end{align}
and 
\begin{align}
&\mat{A}_\phys^\text{off-diag} = \vep
\left({\tiny 
\begin{array}{cccccccc}
 0  & 0 & 0 & 0 & 0 & 0 & 0 & 0\cdots 
 \\
 \frac{2}{3} \left\llbracket{-}\frac{{S_1^+}}{{S_6^-}}\right\rrbracket & 0 & 0 & 0 & 0 & 0 & 0 & 0\cdots
 \\
 0 & 0 & 0 & 0 & 0 & 0 & 0 & 0 \cdots
 \\
 \frac{2}{3} \left\llbracket{-}\frac{{S_3^+}}{{S_5^-}}\right\rrbracket & 0 & 0 & 0 & 0 & 0 & 0 & 0 \cdots
 \\
 0 & 0 & 0 & 0 & 0 & 0 & 0 & 0 \cdots
 \\
 \frac{2}{3} \left\llbracket{-}\frac{{S_3^-}}{{S_5^+}}\right\rrbracket & 0 & 0 & 0 & 0 & 0 & 0 &  0 \cdots
 \\
 \frac{2}{3} \left\llbracket{-}\frac{{S_1^-}}{{S_6^+}}\right\rrbracket & 0 & 0 & 0 & 0 & 0 &  0 & 0 \cdots
 \\
 0 & 0 & 0 & 0 & 0 & 0 & 0 &  0 \cdots
 \\
 0 & 0 & \frac{1}{2} \left\llbracket\frac{{S_2^{++}}}{{S_3^-}}\right\rrbracket & 0 & 0 &  0 &  0 & 0 \cdots
 \\
 0 & \frac{1}{2} \left\llbracket\frac{{S_2^{--}}}{{S_3^+}}\right\rrbracket & 0 & \frac{1}{2} \left\llbracket{-}\frac{{S_2^{--}}}{{S_1^+}}\right\rrbracket & 0 & 0 & 0 &  0 \cdots
 \\
 0 & 0 & 0 & 0 & \frac{1}{2} \left\llbracket{-}\frac{{S_2^{-+}}}{{S_1^+}}\right\rrbracket & 0 & 0 &   0 \cdots
 \\
 0 & 0 & \frac{1}{2} \left\llbracket\frac{{S_2^{+-}}}{{S_3^+}}\right\rrbracket & 0 & 0 & 0 & 0 &  0 \cdots 
 \\
 0 & \frac{1}{2} \left\llbracket{-}\frac{{S_3^-}}{{S_2^{-+}}}\right\rrbracket & 0 & 0 & 0 & \frac{1}{2} \left\llbracket{-}\frac{{S_1^+}}{{S_2^{-+}}}\right\rrbracket & 0 &  0 \cdots
 \\
 0 & 0 & 0 & 0 & \frac{1}{2} \left\llbracket{-}\frac{{S_2^{++}}}{{S_1^-}}\right\rrbracket & 0 & 0 &  0 \cdots
 \\
 0 & 0 & 0 & \frac{1}{2} \left\llbracket{-}\frac{{S_2^{+-}}}{{S_1^-}}\right\rrbracket & 0 & 0 & \frac{1}{2} \left\llbracket\frac{{S_2^{+-}}}{{S_3^+}}\right\rrbracket   & 0 \cdots
 \\
 0 & 0 & 0 & 0 & 0 & \frac{1}{2} \left\llbracket{-}\frac{{S_1^-}}{{S_2^{++}}}\right\rrbracket & \frac{1}{2} \left\llbracket\frac{2 {S_3^-}}{{S_2^{++}}}\right\rrbracket   & 0 \cdots
 \\
\end{array}
}\right)\,.
\end{align}
To fit the connection  matrix on the page, we have introduced the shorthand notation $\llbracket \bullet \rrbracket := \dlog(\bullet)$ and defined 
\begin{align}\begin{aligned}
    S_1^\pm &= X_1 \pm Y_1
    \,,
    &
    S_2^{\pm\pm} &= X_2 \pm Y_1 \pm Y_2
    \,,
    &
    S_3^\pm &= X_3 \pm Y_2
    \,,
    \\
    S_4 &= X_1 + X_2 + X_3 
    \,,
    &
    S_5^\pm &= X_1 + X_2 \pm Y_2 
    \,,
    &
    S_6^\pm &= X_2 + X_3 \pm Y_1 
    \,.
\end{aligned}\end{align}
Of course, whenever all superscripts are a ``$+$'' sign, these reduce to the familiar linear forms of equation \eqref{eq:3siteS}.

\subsection{Physical basis for the tree-level 4-site chain \label{app:4SiteChain}}

Below we present a basis for $H_\phys^4$. 
To present the basis efficiently, we introduce the short hand $\dlog[i,j,\cdots] = \dlog(i) \wedge \dlog(j) \wedge \cdots$ with $i = B_i$ and $\mbf{i} = x_i$.

We use the following labeling for the tubings/hyperplanes:
\begin{center}
    \includegraphics[width=.35\textwidth]{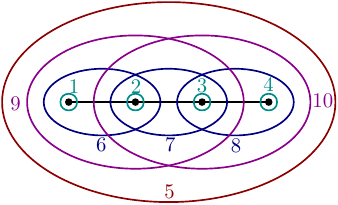}
    \,.
\end{center}
The flat-space wavefunction coefficient of the 4-site chain graph is (recall \eqref{eq:canForm}) 
\begin{align}\begin{aligned}\label{eq:4chainTri}
    \frac{\hat{\Omega}^{(0)}_{\text{4-chain}}}{8Y_1Y_2Y_3}
    = &\,
    \includegraphics[align=c,width=.18 \textwidth]{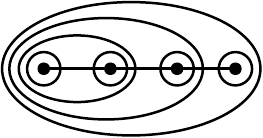}
    +\includegraphics[align=c,width=.18 \textwidth]{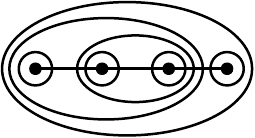}
    +\includegraphics[align=c,width=.18 \textwidth]{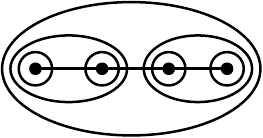}
    \\&\qquad
    +\includegraphics[align=c,width=.18 \textwidth]{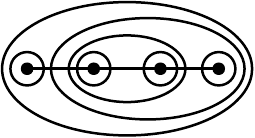}
    +\includegraphics[align=c,width=.18 \textwidth]{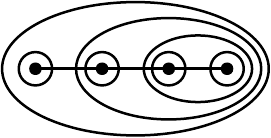}
    \,.
\end{aligned}\end{align}
For the 4-site chain graph, \eqref{eq:dualBasis} and \eqref{eq:frwBasis} produce an over complete set of 234 forms. 
However, the Euler characteristic or equivalently the rank of the intersection matrix is 213.
We choose a linearly independent set of 213 forms from the over complete set by removing the linearly dependent form from the three degenerate 3-boundaries and six degenerate 4-boundaries; the largest degenerate block being $12\times12$.
Constructing the analogous gauge transformation \eqref{eq:gauge}, we arrive at a differential equation with the physical and unphysical sectors separated. 

The physical basis contains a single 1-cut form, 
\begin{equation}\begin{aligned}
    \color{BrickRed}\vphi_1 &= \dlog{\textstyle[{\color{BrickRed}5},\frac{\mbf{1}}{\mbf{2}},\frac{\mbf{2}}{\mbf{3}},\frac{\mbf{3}}{\mbf{4}}]}
    \,,
\end{aligned}\end{equation}
and nine 2-cut forms
\begin{align}\begin{aligned}
    \color{BrickRed}\vphi_2 &= \dlog{\textstyle[{\color{BrickRed}5,1},\frac{\mbf{2}}{\mbf{3}},\frac{\mbf{3}}{\mbf{4}}]} ,
    &
    \color{BrickRed}\vphi_3 &=  \dlog{\textstyle [{\color{BrickRed}10,1},\frac{\mbf{2}}{\mbf{3}},\frac{\mbf{3}}{\mbf{4}}]},
    &
    \color{BrickRed}\vphi_4 &=  \dlog {\textstyle [{\color{BrickRed}5,4}, \frac{\mbf{1}}{\mbf{2}},\frac{\mbf{2}}{\mbf{3}}]},
    \\
    \color{BrickRed}\vphi_5 &= \dlog{\textstyle[{\color{BrickRed}9,4},\frac{\mbf{1}}{\mbf{2}},\frac{\mbf{2}}{\mbf{3}}]},
    &
    \color{BrickRed}\vphi_6 &= \dlog {\textstyle[{\color{BrickRed}5,9},\frac{\mbf{1}}{\mbf{2}},\frac{\mbf{2}}{\mbf{3}}]},
    &
    \color{BrickRed}\vphi_7 &= \dlog{\textstyle[{\color{BrickRed}5,10},\frac{\mbf{2}}{\mbf{3}},\frac{\mbf{3}}{\mbf{4}}]},
    \\
    \color{BrickRed}\vphi_8 &= \dlog{\textstyle[{\color{BrickRed}5,6},\frac{\mbf{1}}{\mbf{2}}, \frac{\mbf{3}}{\mbf{4}}]},
    &
    \color{BrickRed}\vphi_9 &= \dlog{\textstyle[{\color{BrickRed}5,8},\frac{\mbf{1}}{\mbf{2}}, \frac{\mbf{3}}{\mbf{4}}]},
    &
    \color{BrickRed}\vphi_{10} &= \dlog{\textstyle[{\color{BrickRed}6,8},\frac{\mbf{1}}{\mbf{2}}, \frac{\mbf{3}}{\mbf{4}}]}.
\end{aligned}\end{align}
Note that once restricted to the $56$-, $58$- and $68$-cuts, the form 
$\dlog\left(\frac{x_1}{x_2}, \frac{x_3}{x_4}\right)$
becomes the corresponding non-simplicial canonical form on that cut. 

The 3-cut forms corresponding to non-degenerate boundaries are
\begin{align}\begin{aligned}
    \color{BrickRed}\vphi_{11} &= \dlog{\textstyle[{\color{BrickRed}8,1,2},\frac{\mbf{3}}{\mbf{4}}]},
    &
    \color{BrickRed}\vphi_{12} &= \dlog{\textstyle[{\color{BrickRed}10,1,2},\frac{\mbf{3}}{\mbf{4}}]},
    &
    \color{BrickRed}\vphi_{13} &= \dlog{\textstyle[{\color{BrickRed}5,1,4},\frac{\mbf{2}}{\mbf{3}}]},
    \\
    \color{BrickRed}\vphi_{14} &= \dlog{\textstyle[{\color{BrickRed}7,1,4},\frac{\mbf{2}}{\mbf{3}}]},
    &
    \color{BrickRed}\vphi_{15} &= \dlog{\textstyle[{\color{BrickRed}9,1,4},\frac{\mbf{2}}{\mbf{3}}]},
    &
    \color{BrickRed}\vphi_{16} &= \dlog{\textstyle[{\color{BrickRed}10,1,4},\frac{\mbf{2}}{\mbf{3}}]},
    \\
    \color{BrickRed}\vphi_{17} &= \dlog{\textstyle[{\color{BrickRed}5,6,1},\frac{\mbf{3}}{\mbf{4}}]},
    &
    \color{BrickRed}\vphi_{18} &= \dlog{\textstyle[{\color{BrickRed}1,5,8}, \frac{\mbf{3}}{\mbf{4}}]},
    &
    \color{BrickRed}\vphi_{19} &= \dlog{\textstyle[{\color{BrickRed}5,9,1},\frac{\mbf{2}}{\mbf{3}}]},
    \\
    \color{BrickRed}\vphi_{20} &= \dlog{\textstyle[{\color{BrickRed}6,8,1},\frac{\mbf{3}}{\mbf{4}}]} ,
    &
    \color{BrickRed}\vphi_{21} &= \dlog{\textstyle[{\color{BrickRed}10,7,1},\frac{\mbf{2}}{\mbf{3}}]},
    &
    \color{BrickRed}\vphi_{22} &= \dlog{\textstyle[{\color{BrickRed}10,8,1},\frac{\mbf{3}}{\mbf{4}}]},
    \\
    \color{BrickRed}\vphi_{23} &= \dlog{\textstyle[{\color{BrickRed}8,6,2},\frac{\mbf{3}}{\mbf{4}}]},
    &
    \color{BrickRed}\vphi_{24} &= \dlog{\textstyle[{\color{BrickRed}6,3,4},\frac{\mbf{1}}{\mbf{2}}]},
    &
    \color{BrickRed}\vphi_{25} &= \dlog{\textstyle[{\color{BrickRed}9,3,4},\frac{\mbf{1}}{\mbf{2}}]},
    \\
    \color{BrickRed}\vphi_{26} &= \dlog{\textstyle[{\color{BrickRed}6,8,3},\frac{\mbf{1}}{\mbf{2}}]}, 
    &
    \color{BrickRed}\vphi_{27} &= \dlog{\textstyle[{\color{BrickRed}5,6,4},\frac{\mbf{1}}{\mbf{2}}}],
    &
    \color{BrickRed}\vphi_{28} &= \dlog{\textstyle[{\color{BrickRed}5,8,4},\frac{\mbf{1}}{\mbf{2}}},
    \\
    \color{BrickRed}\vphi_{29} &= \dlog{\textstyle[{\color{BrickRed}5,10,4},\frac{\mbf{2}}{\mbf{3}}},
    &
    \color{BrickRed}\vphi_{30} &= \dlog{\textstyle[{\color{BrickRed}6,8,4},\frac{\mbf{1}}{\mbf{2}}]},
    &
    \color{BrickRed}\vphi_{31} &= \dlog{\textstyle[{\color{BrickRed}9,6,4},\frac{\mbf{1}}{\mbf{2}}]},
    \\
    \color{BrickRed}\vphi_{32} &= \dlog{\textstyle[{\color{BrickRed}9,7,4},\frac{\mbf{2}}{\mbf{3}}]}, 
    &
    \color{BrickRed}\vphi_{33} &= \dlog{\textstyle[{\color{BrickRed}5,9,6},\frac{\mbf{1}}{\mbf{2}}]},
    &
    \color{BrickRed}\vphi_{34} &= \dlog{\textstyle[{\color{BrickRed}5,10,8},\frac{\mbf{3}}{\mbf{4}}]},
\end{aligned}\end{align}
while the 3-boundary forms corresponding to degenerate boundaries are
\begin{align}\begin{aligned}
    \color{Orange}\vphi_{35} &= \dlog{\textstyle[{\color{Orange}5,\frac{6}{10},2},\frac{\mbf{3}}{\mbf{4}}]},
    &
    \color{Orange}\vphi_{36} &= \dlog{\textstyle[{\color{Orange}5,\frac{8}{9},3},\frac{\mbf{1}}{\mbf{2}}]},
    &
    \color{Orange}\vphi_{37} &= \dlog{\textstyle[{\color{Orange}5,\frac{9}{10},7},\frac{\mbf{2}}{\mbf{3}}]} \,.
\end{aligned}\end{align}
The non-degenerate 4-boundary forms:
\begin{align}\begin{aligned}
    \color{BrickRed}\vphi_{38} &= \dlog{\textstyle[{\color{BrickRed}1,2,3,4}]},
    &
    \color{BrickRed}\vphi_{39} &= \dlog{\textstyle[{\color{BrickRed}8,1,2,3}]},
    &
    \color{BrickRed}\vphi_{40} &= \dlog{\textstyle[{\color{BrickRed}7,1,2,4}]},
    \\
    \color{BrickRed}\vphi_{41} &= \dlog{\textstyle[{\color{BrickRed}8,1,2,4}]},
    &
    \color{BrickRed}\vphi_{42} &= \dlog{\textstyle[{\color{BrickRed}10,1,2,4}]},
    &
    \color{BrickRed}\vphi_{43} &= \dlog{\textstyle[{\color{BrickRed}10,7,1,2}]},
    \\
    \color{BrickRed}\vphi_{44} &= \dlog{\textstyle[{\color{BrickRed}6,1,3,4}]},
    &
    \color{BrickRed}\vphi_{45} &= \dlog{\textstyle[{\color{BrickRed}7,1,3,4}]},
    &
    \color{BrickRed}\vphi_{46} &= \dlog{\textstyle[{\color{BrickRed}9,1,3,4}]},
    \\
    \color{BrickRed}\vphi_{47} &= \dlog{\textstyle[{\color{BrickRed}6,8,1,3}]},
    &
    \color{BrickRed}\vphi_{48} &= \dlog{\textstyle[{\color{BrickRed}5,6,1,4}]},
    &
    \color{BrickRed}\vphi_{49} &= \dlog{\textstyle[{\color{BrickRed}5,8,1,4}]},
    \\
    \color{BrickRed}\vphi_{50} &= \dlog{\textstyle[{\color{BrickRed}6,8,1,4}]},
    &
    \color{BrickRed}\vphi_{51} &= \dlog{\textstyle[{\color{BrickRed}9,6,1,4}]},
    &
    \color{BrickRed}\vphi_{52} &= \dlog{\textstyle[{\color{BrickRed}10,8,1,4}]},
    \\
    \color{BrickRed}\vphi_{53} &= \dlog{\textstyle[{\color{BrickRed}5,9,6,1}]},
    &
    \color{BrickRed}\vphi_{54} &= \dlog{\textstyle[{\color{BrickRed}6,2,3,4}]},
    &
    \color{BrickRed}\vphi_{55} &= \dlog{\textstyle[{\color{BrickRed}6,8,2,3}]},
    \\
    \color{BrickRed}\vphi_{56} &= \dlog{\textstyle[{\color{BrickRed}6,8,2,4}]},
    &
    \color{BrickRed}\vphi_{57} &= \dlog{\textstyle[{\color{BrickRed}9,7,3,4}]},
    &
    \color{BrickRed}\vphi_{58} &= \dlog{\textstyle[{\color{BrickRed}5,10,8,4}]}\,.
\end{aligned}\end{align}
The degenerate 4-boundary forms:
\begin{align}\begin{aligned}
\label{eq:degen4cuts4sitechain}
    {\color{Orange}\vphi_{59}} &= \dlog{\textstyle[{\color{Orange}5,\frac{8}{9},1,3}]},
    &    
     {\color{Orange} \vphi_{60}} &= \dlog{\textstyle[{\color{Orange}10,\frac{7}{8},1,3}]},
    \\
    {\color{Orange} \vphi_{61}} &= \dlog{\textstyle[{\color{Orange}5,\frac{6}{10},2,4}]},
    &
    {\color{Orange} \vphi_{62}} &= \dlog{\textstyle[{\color{Orange}9,\frac{6}{7},2,4}]},
    \\
    {\color{Orange} \vphi_{63}} &=
        \dlog{\textstyle[{\color{Orange}5,9,6,2}]} 
        {-} \dlog{\textstyle[{\color{Orange}5,9,7,2}]}
        {+} \dlog{\textstyle[{\color{Orange}5,10,7,2}]}
        ,
    \\
    {\color{Orange} \vphi_{64}} &=  
        \dlog{\textstyle[{\color{Orange}5,9,7,3}]}
        {-} \dlog{\textstyle[{\color{Orange}5,10,7,3}]}
        {+} \dlog{\textstyle[{\color{Orange}5,10,8,3}]}
    \,.
\end{aligned}\end{align}

The resulting differential equation for the whole 213-dimensional cohomology has the following structure:
\begin{align} \label{eq:A4chain}
    \mat{A}_{4\text{-chain}}
    = 
    \includegraphics[align=c, width=.55\textwidth]{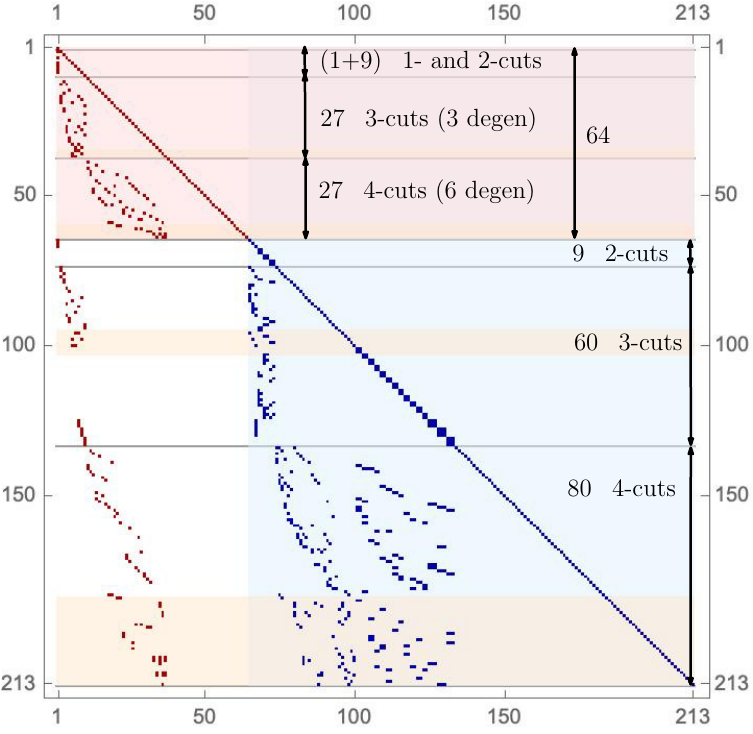}
    \,.
\end{align}
This procedure outlined in section \ref{sec:3SitePed} picks out a physical space with 64 elements exactly as the kinematic flow algorithm \cite{Arkani-Hamed:2023bsv, Arkani-Hamed:2023kig, Hang:2024xas, Baumann:2024mvm} and those obtained from a time integral perspective \cite{He:2024olr}.
Moreover, we have verified that the physical $64\times64$ submatrix of \eqref{eq:A4chain} is integrable and can be found in the \texttt{github} repository \github.

\subsection{Physical basis for the tree-level 4-site star \label{app:4SiteStar}}

For the 4-site star graph, we computed the intersection matrix of the over complete basis of 377 forms. 
The corresponding intersection matrix
\begin{align}
    \mat{C}_{\text{4-star}} = 
    \begin{pmatrix}
        \mat{C}_\text{1- and 2-cuts}
        & & 
        \\
        & \mat{C}_\text{3-cuts} & 
        \\
        & & \mat{C}_\text{4-cuts}
    \end{pmatrix}
    \,,
\end{align}
has rank 312 which is the dimension of the FRW cohomology. 
The intersection matrix corresponding to 1- and 2-cut forms is full rank and contains 10 elements that couple to the physical FRW form. 
There are three degenerate 3-cuts: manifesting as three $4{\times}4$ blocks in the intersection matrix:
\begin{gather}
    \mat{C}_{\text{1- and 2-cuts}} {=} \includegraphics[align=c,width=.3\textwidth]{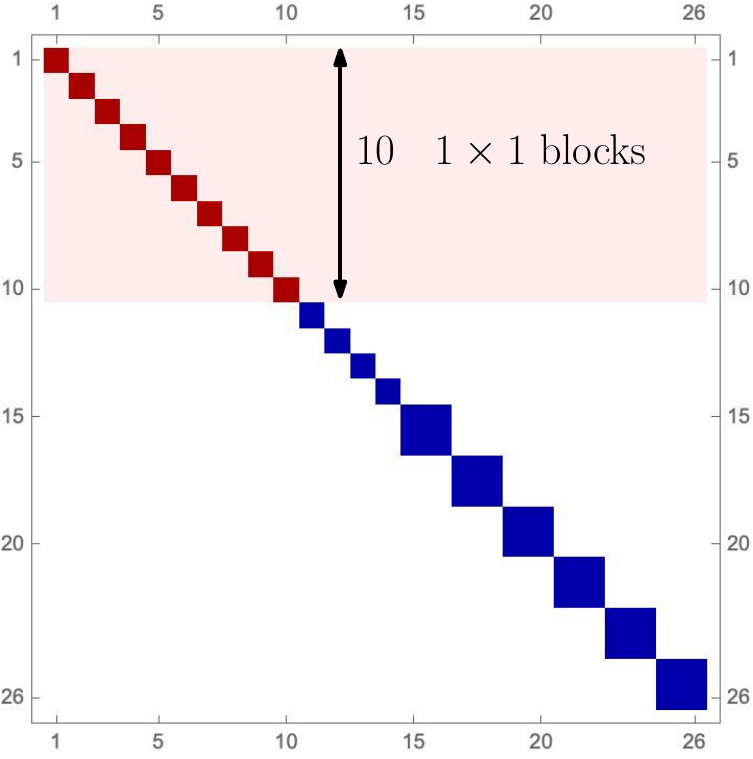}
    ,\,
    \mat{C}_{\text{3-cuts}} {=} \includegraphics[align=c,width=.3\textwidth]{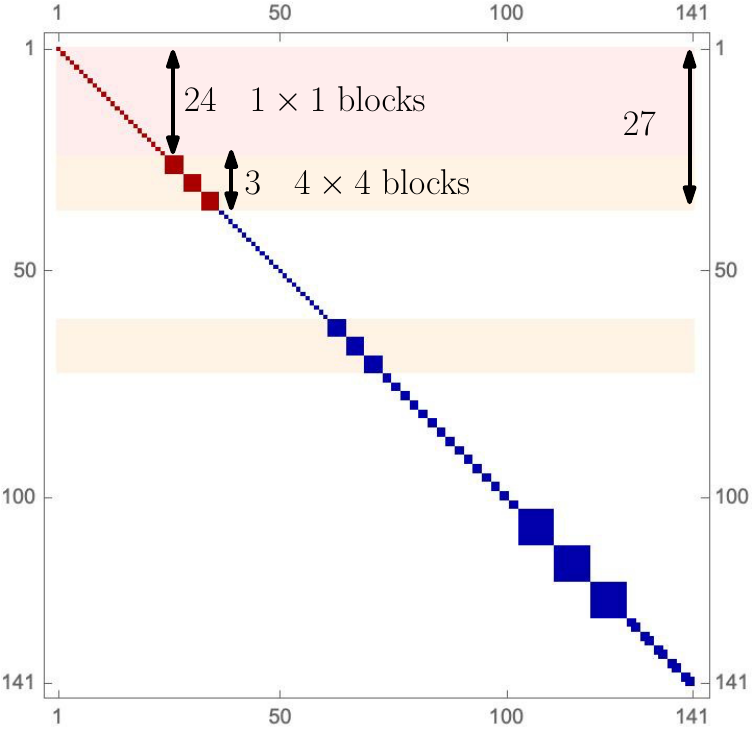}
    \,.
\end{gather}
There are seven degenerate 4-cuts: six $4{\times}4$ blocks and one $60{\times}60$ block that couple to $\Psi_{4\text{-star}}^{(0)}$
\begin{gather}
    \mat{C}_{\text{4-cuts}} = \includegraphics[align=c,width=.4\textwidth]{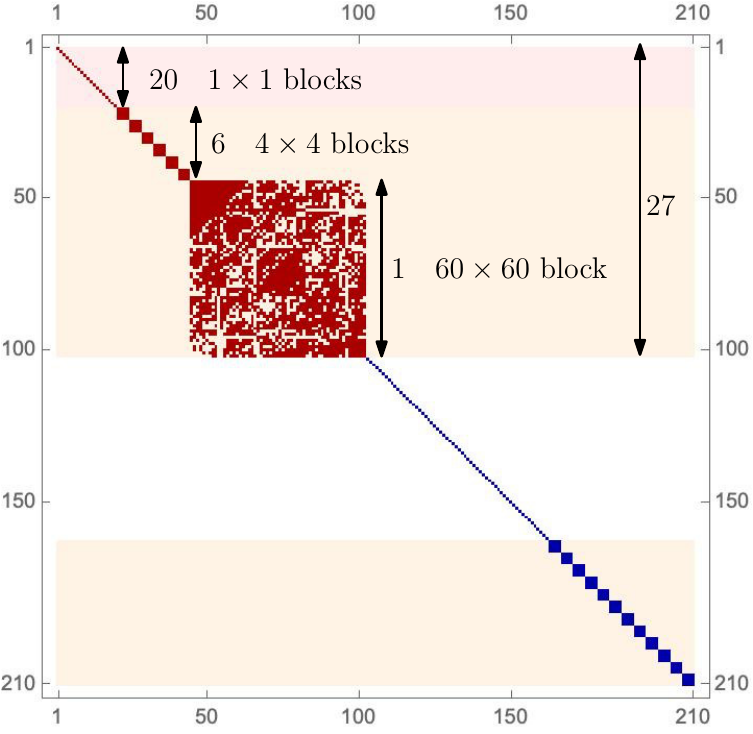}
    \,.
\end{gather}
Knowing that each degenerate block only contributes one element to the physical basis, we obtain 64 elements for the physical basis.

It is highly non-trivial that the degenerate blocks of different topologies conspire in just the right way to produce the $2^{n-1}$ counting for tree-level FRW correlators. 
It is a special feature of the hyperplane arrangement generated by the cosmological polytope.

\subsection{Physical basis for the 1-loop 3-gon \label{app:3Site1Loop}}

The hyperplane arrangement corresponding to the 1-loop 3-site diagram is generated by the 10 tubings below: 
\begin{center}
    \includegraphics[width=.5\textwidth]{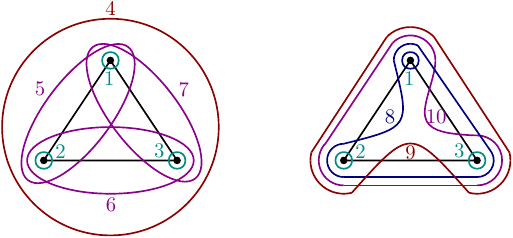}. 
\end{center}
The flat space wavefunction coefficient for the 1-loop 3-gon is (recall \eqref{eq:canForm}) 
\begin{align}\begin{aligned}\label{eq:3gonTri}
    \frac{\hat{\Omega}^{(1)}_{\text{3-gon}}}{8Y_1Y_2Y_3}
    &= \includegraphics[align=c,width=.18\textwidth]{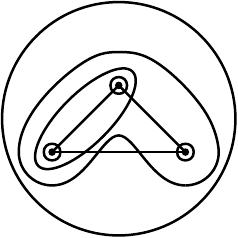}
    + 2\times\text{rotations}
    + \includegraphics[align=c,width=.18\textwidth]{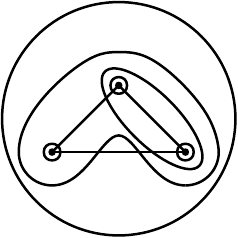}
    + 2\times\text{rotations}
    \,.
\end{aligned}\end{align}
The over complete set of forms \eqref{eq:dualBasis} and \eqref{eq:frwBasis} for the 1-loop 3-gon has 102 elements while the Euler characteristic/rank of the intersection matrix is 99. 
There are three degenerate $4\times4$ block (3-cuts). 

Unlike at tree-level, there are $1{+}n$ non-trivial 1-boundaries for any $n$-gon at 1-loop. 
In this example, $n=3$ and there are 4 non-trivial 1-boundary forms:
\begin{align}\begin{aligned}
    \color{BrickRed} \vphi_1 &= \dlog  {\textstyle [{\color{BrickRed}4},\frac{\mbf{1}}{\mbf{2}},\frac{\mbf{2}}{\mbf{3}} ]},
    &
    \color{BrickRed} \vphi_2 &= \text{dlog} {\textstyle [{\color{BrickRed}8},\frac{\mbf{1}}{\mbf{2}},\frac{\mbf{2}}{\mbf{3}} ]},
    &
    \color{BrickRed} \vphi_3 &= \text{dlog} {\textstyle [{\color{BrickRed}9},\frac{\mbf{1}}{\mbf{2}},\frac{\mbf{2}}{\mbf{3}} ]},
    &
    \color{BrickRed} \vphi_4 &= \text{dlog} {\textstyle [{\color{BrickRed}10},\frac{\mbf{1}}{\mbf{2}},\frac{\mbf{2}}{\mbf{3}} ]},
\end{aligned}\end{align}
There are 21 2-boundary forms:
\begin{align}\begin{aligned}
    \color{BrickRed} \vphi_5 &=\text{dlog} {\textstyle [{\color{BrickRed}4,1},\frac{\mbf{2}}{\mbf{3}} ]},
    &
    \color{BrickRed} \vphi_{6} &=\text{dlog} {\textstyle [{\color{BrickRed}6,1},\frac{\mbf{2}}{\mbf{3}} ]},
    &
    \color{BrickRed} \vphi_{7} &=\text{dlog} {\textstyle [{\color{BrickRed}8,1},\frac{\mbf{2}}{\mbf{3}} ]},
    &
    \color{BrickRed} \vphi_{8} &=\text{dlog} {\textstyle [{\color{BrickRed}10,1},\frac{\mbf{2}}{\mbf{3}} ]},
    \\
    \color{BrickRed} \vphi_{9} &=\text{dlog} {\textstyle [{\color{BrickRed}4,2},\frac{\mbf{1}}{\mbf{3}} ]},
    &
    \color{BrickRed} \vphi_{10} &=\text{dlog} {\textstyle [{\color{BrickRed}7,2},\frac{\mbf{1}}{\mbf{3}} ]},
    &
    \color{BrickRed} \vphi_{11} &=\text{dlog} {\textstyle [{\color{BrickRed}8,2},\frac{\mbf{1}}{\mbf{3}} ]},
    &
    \color{BrickRed} \vphi_{12} &=\text{dlog} {\textstyle [{\color{BrickRed}9,2},\frac{\mbf{1}}{\mbf{3}} ]},
    \\
    \color{BrickRed} \vphi_{13} &=\text{dlog} {\textstyle [{\color{BrickRed}4,3},\frac{\mbf{1}}{\mbf{2}} ]},
    &
    \color{BrickRed} \vphi_{14} &=\text{dlog} {\textstyle [{\color{BrickRed}5,3},\frac{\mbf{1}}{\mbf{2}} ]},
    &
    \color{BrickRed} \vphi_{15} &=\text{dlog} {\textstyle [{\color{BrickRed}9,3},\frac{\mbf{1}}{\mbf{2}} ]},
    &
    \color{BrickRed} \vphi_{16} &=\text{dlog} {\textstyle [{\color{BrickRed}10,3},\frac{\mbf{1}}{\mbf{2}} ]},
    \\
    \color{BrickRed} \vphi_{17} &=\text{dlog} {\textstyle [{\color{BrickRed}4,5},\frac{\mbf{1}}{\mbf{2}} ]},
    &
    \color{BrickRed} \vphi_{18} &=\text{dlog} {\textstyle [{\color{BrickRed}4,6},\frac{\mbf{2}}{\mbf{3}} ]},
    &
    \color{BrickRed} \vphi_{19} &=\text{dlog} {\textstyle [{\color{BrickRed}4,7},\frac{\mbf{1}}{\mbf{3}} ]},
    &
    \color{BrickRed} \vphi_{20} &=\text{dlog} {\textstyle [{\color{BrickRed}9,5},\frac{\mbf{1}}{\mbf{2}} ]},
    \\
    \color{BrickRed} \vphi_{21} &=\text{dlog} {\textstyle [{\color{BrickRed}10,5},\frac{\mbf{1}}{\mbf{2}} ]},
    &
    \color{BrickRed} \vphi_{22} &=\text{dlog} {\textstyle [{\color{BrickRed}8,6},\frac{\mbf{2}}{\mbf{3}} ]},
    &
    \color{BrickRed} \vphi_{23} &=\text{dlog} {\textstyle [{\color{BrickRed}10,6},\frac{\mbf{2}}{\mbf{3}} ]},
    &
    \color{BrickRed} \vphi_{24} &=\text{dlog} {\textstyle [{\color{BrickRed}8,7},\frac{\mbf{1}}{\mbf{3}} ]},
    \\
    \color{BrickRed} \vphi_{25} &=\text{dlog} {\textstyle [{\color{BrickRed}9,7},\frac{\mbf{1}}{\mbf{3}} ]}.
\end{aligned}\end{align}
There are 25 3-boundary forms:
\begin{align}\begin{aligned}
    \color{BrickRed} \vphi_{26} &=\text{dlog}{\textstyle [{\color{BrickRed}1,2,3}]},
    &
    \color{BrickRed} \vphi_{27} &=\text{dlog}{\textstyle [{\color{BrickRed}6,1,2}]},
    &
    \color{BrickRed} \vphi_{28} &=\text{dlog}{\textstyle [{\color{BrickRed}7,1,2}]},
    &
    \color{BrickRed} \vphi_{29} &=\text{dlog}{\textstyle [{\color{BrickRed}8,1,2}]},
    \\
    \color{BrickRed} \vphi_{30} &=\text{dlog}{\textstyle [{\color{BrickRed}5,1,3}]},
    &
    \color{BrickRed} \vphi_{31} &=\text{dlog}{\textstyle [{\color{BrickRed}6,1,3}]},
    &
    \color{BrickRed} \vphi_{32} &=\text{dlog}{\textstyle [{\color{BrickRed}10,1,3}]},
    &
    \color{BrickRed} \vphi_{33} &=\text{dlog}{\textstyle [{\color{BrickRed}5,1,4}]},
    \\
    \color{BrickRed} \vphi_{34} &=\text{dlog}{\textstyle [{\color{BrickRed}7,1,4}]},
    &
    \color{BrickRed} \vphi_{35} &=\text{dlog}{\textstyle [{\color{BrickRed}10,1,5}]},
    &
    \color{BrickRed} \vphi_{36} &=\text{dlog}{\textstyle [{\color{BrickRed}8,7,1}]},
    &
    \color{BrickRed} \vphi_{37} &=\text{dlog}{\textstyle [{\color{BrickRed}5,2,3}]},
    \\
    \color{BrickRed} \vphi_{38} &=\text{dlog}{\textstyle [{\color{BrickRed}7,2,3}]},
    &
    \color{BrickRed} \vphi_{39} &=\text{dlog}{\textstyle [{\color{BrickRed}9,2,3}]},
    &
    \color{BrickRed} \vphi_{40} &=\text{dlog}{\textstyle [{\color{BrickRed}4,5,2}]},
    &
    \color{BrickRed} \vphi_{41} &=\text{dlog}{\textstyle [{\color{BrickRed}4,6,2}]},
    \\
    \color{BrickRed} \vphi_{42} &=\text{dlog}{\textstyle [{\color{BrickRed}9,5,2}]},
    &
    \color{BrickRed} \vphi_{43} &=\text{dlog}{\textstyle [{\color{BrickRed}8,6,2}]},
    &
    \color{BrickRed} \vphi_{44} &=\text{dlog}{\textstyle [{\color{BrickRed}4,6,3}]},
    &
    \color{BrickRed} \vphi_{45} &=\text{dlog}{\textstyle [{\color{BrickRed}4,7,3}]},
    \\
    \color{BrickRed} \vphi_{46} &=\text{dlog}{\textstyle [{\color{BrickRed}10,6,3}]},
    &
    \color{BrickRed} \vphi_{47} &=\text{dlog}{\textstyle [{\color{BrickRed}9,7,3}]}
    \,.
    &
    &
\end{aligned}\end{align}
Note that there are three {\color{Orange}degenerate} blocks in this example
\begin{align}
    \color{Orange} \vphi_{48} &=\text{dlog} {\textstyle [{\color{Orange}8,\frac{7}{6},3}]},
    &
    \color{Orange} \vphi_{49} &=\text{dlog} {\textstyle [{\color{Orange}9,\frac{5}{7},1}]},
    &
    \color{Orange} \vphi_{50} &=\text{dlog} {\textstyle [{\color{Orange}10,\frac{6}{5},2}]}\,.
\end{align}

Computing the differential equation for the 99 linearly independent elements and partitioning into a physical and unphysical space we find 
\begin{align}\label{eq:A3gon}
    \mat{A}_{3\text{-gon}}^{1\text{-loop}}
    = 
    \includegraphics[align=c, width=.5\textwidth]{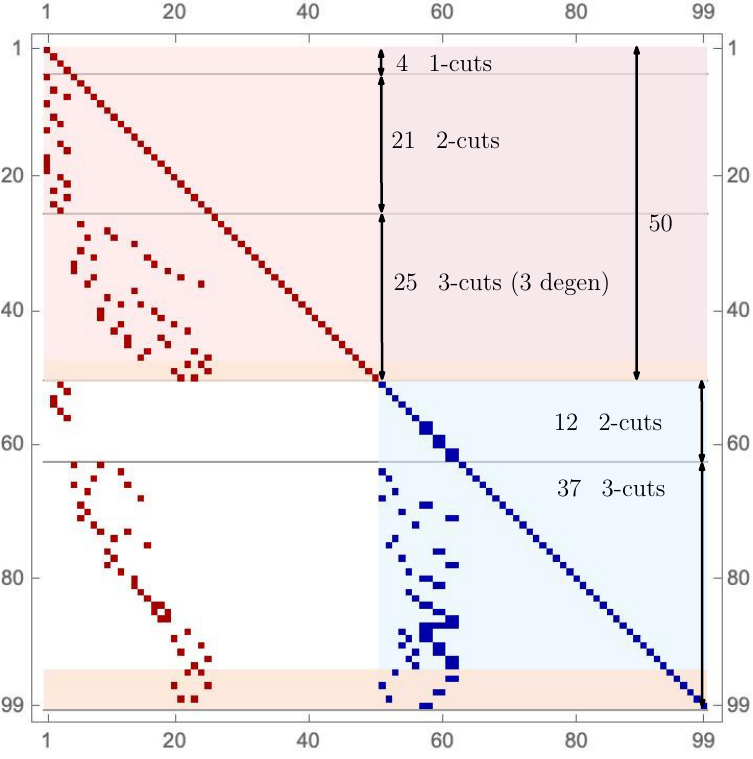}
    \,.
\end{align}
In particular, we find that the size of the physical basis is 50 elements, which agrees with \cite{He:2024olr}. 
Moreover, we have also checked that the $50\times50$ physical submatrix of \eqref{eq:A3gon} is integrable.
\texttt{Mathematica} files for this example can be found at \github.

\section{Diagrammatic partial fractions \label{app:partialFrac}}

Usually, the simplest way to compute the canonical form is by summing forms associated to the set of compatible complete tubings associated to a graph $\G$.\footnote{A complete non-crossing tubing is a tubing with $2n+\ell-1$ tubings that do not intersect.} 
Explicitly, if $\mathsf{T}$ is the set of all complete non-crossing tubings, the canonical form and function are 
\begin{align}
    \Omega^{(\ell)}_n &= 
    \hat{\Omega}^{(\ell)}_n
    \frac{\d^n\mbf{X} {\wedge} \d^{n{+}\ell{-}1}\mbf{Y}}{\text{GL}(1)}
    \,,
    &
    \hat{\Omega}^{(\ell)}_n &= 
    \left(\prod_{i=1}^{n+\ell-1} 2Y_i\right)
    \left(\sum_{\tau \in \mathsf{T}} 
    \prod_{t\in\tau} \frac{1}{S_t}\right)
    \,.
\end{align}
For FRW correlators, we are interested in the form \eqref{eq:psiphys} which is an $n$-form in $\mbf{x}$-space. 
Each complete non-crossing tubing contains $2n+\ell-1$ tubings or equivalently each term of \eqref{eq:canForm} contains $2n+\ell-1$ denominators. 

To isolate the specific linear combination of $\dlog$-forms of a degenerate boundary in the physical wavefunction, it is useful to think of the canonical function as a rational function of only the $\mbf{X}$ (or $\mbf{x}$ after the shift). 
From this perspective, each term in \eqref{eq:canForm} has more linear factors in the denominator than variables. 
Therefore, we partial fraction each term until there are only $n$ denominators. 
Thankfully, this can be done in a simple graphical way that leads to an analytic formula without any calculation!

Start by identifying a ``large'' tubing $S^\text{out}$ and a set of its subtubings that encircle all vertices in $S^\text{out}$ once
$S^\text{in}_1, \dots, S^\text{in}_m$
\begin{align} \label{eq:Sout}
    \includegraphics[align=c, width=.3\textwidth]{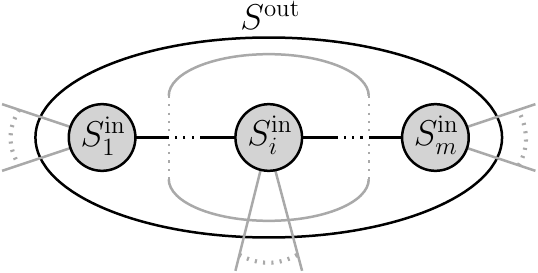}
    := \prod_{i : S_i  \text{ inside } S^\text{out}} 
    \frac{1}{S_i}
    \,. 
\end{align}
Here, each $S^\text{in}_i$ \emph{must be} a ``large'' tubing that contains further nested subtubings or a bare vertex. 
The gray external lines denote the possibility that $S^\text{out}$ is only a small piece of larger tubing. 
The gray internal tubing is a place holder for the possibility of tubings that encircle the $S^\text{in}_i$ as a whole. 
Then, 
\begin{align}\begin{aligned}
    \eqref{eq:Sout}
    &= \frac{1}{2Y^\text{out}} \left[
    \sum_{i=1}^m 
    \includegraphics[align=c, width=.3\textwidth]{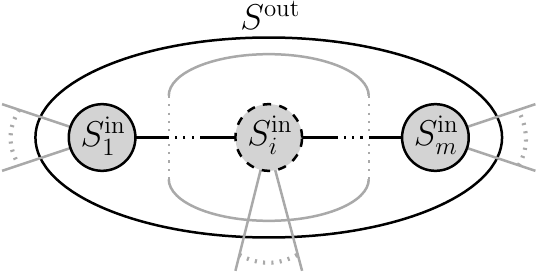}
    \right.
    \left.
    - \includegraphics[align=c, width=.3\textwidth]{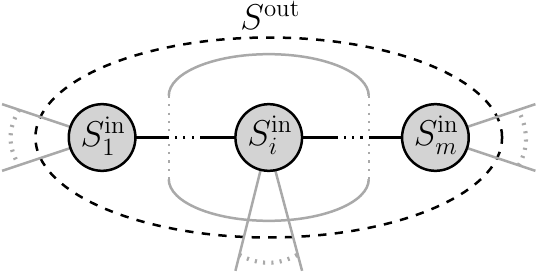}
    \right]
\end{aligned}\end{align}
where the dashed tubings above correspond to deleting that tubing, $Y^\text{out} = \sum_{e \in \mathcal{E}^\text{out}} Y_e$ and $\mathcal{E}^\text{out}$ is the set of edges that are \emph{wholly within} $S^\text{out}$ (i.e., not are crossed by $S^\text{out}$). 
It is easy to see that $\sum_{i=1}^m S^\text{in}_i - S^\text{out}$ is only a function of the $Y_i$ since each vertex is encircled the same number of times in the collection of the $S^\text{in}_i$ and $S^\text{out}$.
Then all edges that cross $S^\text{out}$ are also crossed by the $S^\text{in}_i$ and therefore cancel in the difference. 
Hence, we are left with twice\footnote{The factor of two comes from the fact that the collection of $S^\text{in}_i$ must cross each edge  wholly within $S^\text{out}$ twice.} the sum of all $Y_i$'s wholly within $S^\text{out}$.

The simplest example is the 2-site chain, which has one complete non-crossing tubing
\begin{align}
    \hat\Omega^{(0)}_2
    = \frac{2Y}{S_1 S_2 S_3}
    = \includegraphics[align=c,width=.12\textwidth]{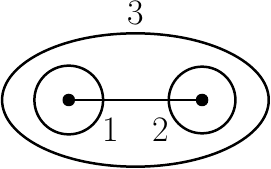}
    \,.
\end{align}
Choosing $S^\text{out}=S_3$ and $S^\text{in}_i = S_i$, we can multiply the canonical function by unity
\begin{align}
    \hat\Omega^{(0)}_2
    = \overset{=1}{\overbracket{
        \frac{S_1 + S_2 - S_3}{2Y}
    }} \hat\Omega^{(0)}_2
    = 
    \includegraphics[align=c,width=.1\textwidth]{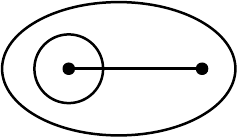}
    {+} \includegraphics[align=c,width=.1\textwidth]{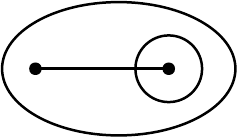}
    {-} \includegraphics[align=c,width=.1\textwidth]{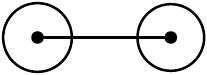}
    .
    \nn
\end{align}

Now that we have understood how to partial fraction the maximal non-crossing tubings it is fairly straightforward to show that the terms in the canonical function that correspond to maximal codimenson degenerate boundaries come with the same relative sign. 

While more complicated, it is also straightforward to verify that the $\dlog$-forms corresponding to non-maximal codimension degenerate boundaries always come in linear combinations without relative signs. 
For example, suppose we want to know the linear combination of the degenerate propagators
\begin{align}\begin{aligned} \label{eq:593and583}
   \frac{1}{S_5 S_9 S_3} =  \includegraphics[align=c,width=.2\textwidth]{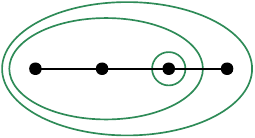}
   \,,
   \frac{1}{S_5 S_8 S_3} = \includegraphics[align=c,width=.2\textwidth]{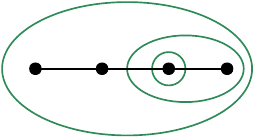}
   \,,
\end{aligned}\end{align}
that appear in the canonical form. 
From the set of compatible complete tubings \eqref{eq:4chainTri}, it is easy to see that the only $\dlog$ 4-forms containing both \eqref{eq:593and583} contain an $S_1$ or $S_2$. 
However, it turns out the the coefficients of $1/(S_5S_9S_3S_2)$ and $1/(S_5S_8S_3S_2)$ vanish. 
Therefore, we seek the coefficients of 
\begin{align}\begin{aligned} \label{eq:degenTubingMonomials}
    \frac{1}{S_5S_9S_3S_2} &= 
    \includegraphics[align=c,width=.2\textwidth]{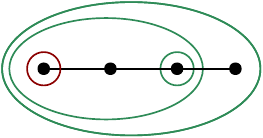}
    \,,
    &
    \frac{1}{S_5S_8S_3S_2} &= 
    \includegraphics[align=c,width=.2\textwidth]{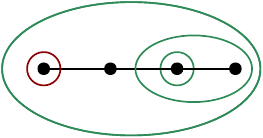}
    \,,
\end{aligned}\end{align}
Using crosses to denote the choice of $S^\text{out}$ and $S^\text{in}_i$ we partial fraction each term in the canonical function throwing away all terms the do not contain \eqref{eq:degenTubingMonomials}
\begin{align}
\label{eq:partialFracDecomp1}
    &\includegraphics[align=c,width=.18\textwidth]{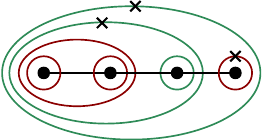}
    {\to} 
    \frac{1}{2Y_3}
    \includegraphics[align=c,width=.18\textwidth]{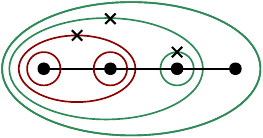}
    \nn\\&\qquad
    {\to} 
    \frac{1}{4Y_2Y_3}
    \includegraphics[align=c,width=.18\textwidth]{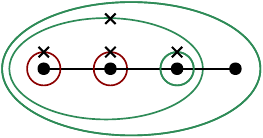}
    {\to} 
    \frac{1}{8(Y_1{+}Y_2)Y_2Y_3}
    \includegraphics[align=c,width=.18\textwidth]{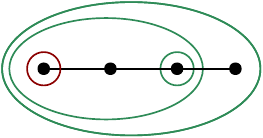}
    \,,
\end{align}
\begin{align}
    \label{eq:partialFracDecomp2}
    &\includegraphics[align=c,width=.18\textwidth]{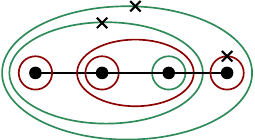}
    {\to} 
    \frac{1}{2Y_3}
    \includegraphics[align=c,width=.18\textwidth]{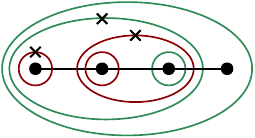}
    \nn\\&\qquad
    {\to} 
    \frac{1}{4Y_1Y_3}
    \includegraphics[align=c,width=.18\textwidth]{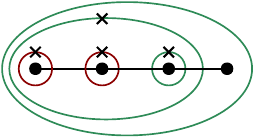}
    {\to} 
    \frac{1}{8(Y_1{+}Y_2)Y_1Y_3}
    \includegraphics[align=c,width=.18\textwidth]{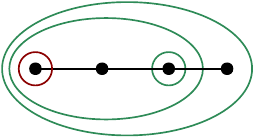}
    \,,
\end{align}
\begin{align}
    \label{eq:partialFracDecomp3}
    &\includegraphics[align=c,width=.18\textwidth]{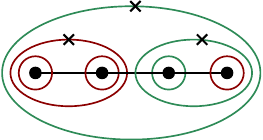}
    {\to} 
    \frac{1}{2Y_2}
    \includegraphics[align=c,width=.18\textwidth]{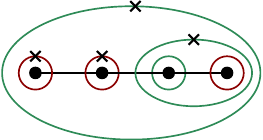}
    \nn\\&\qquad
    {\to} 
    \frac{1}{4(Y_1{+}Y_2)Y_2}
    \includegraphics[align=c,width=.18\textwidth]{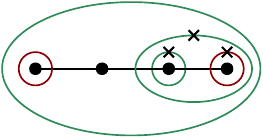}
    {\to} 
    \frac{1}{8(Y_1{+}Y_2)Y_2Y_3}
    \includegraphics[align=c,width=.18\textwidth]{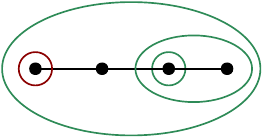}
    \,,
\end{align}
\begin{align}
    \label{eq:partialFracDecomp4}
    &\includegraphics[align=c,width=.18\textwidth]{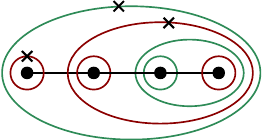}
    {\to} 
    \frac{-1}{2Y_1}
    \includegraphics[align=c,width=.18\textwidth]{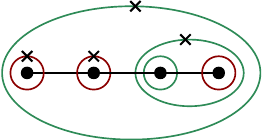}
    \nn\\&\qquad
    {\to} 
    \frac{-1}{4Y_1(Y_1{+}Y_2)}
    \includegraphics[align=c,width=.18\textwidth]{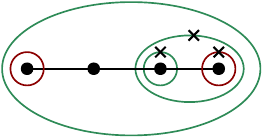}
    {\to} 
    \frac{-1}{8Y_1(Y_1{+}Y_2)Y_3}
    \includegraphics[align=c,width=.18\textwidth]{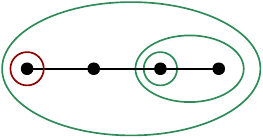}
    \,.
\end{align}
Adding equations \eqref{eq:partialFracDecomp1}-\eqref{eq:partialFracDecomp4} together, one finds 
\begin{align}
    \frac{1}{8 Y_1 Y_2 Y_3}
    \left[
    \includegraphics[align=c,width=.18\textwidth]{figs/PartialFractionExample/5931.pdf}
    {+}
    \includegraphics[align=c,width=.18\textwidth]{figs/PartialFractionExample/5831.pdf}
    \right]
    =
    \frac{1}{8 Y_1 Y_2 Y_3 S_1} \left(
        \frac{1}{S_5S_9S_3} 
        {+} 
        \frac{1}{S_5S_9S_3} 
    \right)
    .
\end{align}
To turn this into the $\dlog$-form of interest, we shift $\mbf{x}\to\mbf{X}$ and multiply by the volume form $\d^4\mbf{x}$
\begin{align}\begin{aligned}
    \frac{\d^4\mbf{x}}{B_1} \left(
        \frac{1}{B_5B_9B_3} 
        + \frac{1}{B_5B_8B_3} 
    \right)
    &
    =\frac{\d^4\mbf{x}}{
        \d B_5 \wedge
        \d B_9 \wedge
        \d B_3 \wedge
        \d B_1 
    }
    \dlog{\textstyle[5,9,3,1]}
    \\&\qquad
    + \frac{\d^4\mbf{x}}{
        \d B_5 \wedge
        \d B_8 \wedge
        \d B_3 \wedge
        \d B_1 
    }
    \dlog{\textstyle[5,8,3,1]}
    \,,
    \\&
    = \dlog{\textstyle[5,\frac{8}{9},3,1]}
    \,.
\end{aligned}\end{align}
The additive positivity of the partial fractioned canonical function is reflected in the fact that the pair of crossing hyperplanes in the degenerate boundary ($B_8$ and $B_9$) appear as a ratio in the above $\dlog$-form.


\bibliographystyle{JHEP}
\bibliography{refs}

\providecommand{\href}[2]{#2}\begingroup\raggedright\begin{thebibliography}{10}

\bibitem{Arkani-Hamed:2018kmz}
N.~Arkani-Hamed, D.~Baumann, H.~Lee and G.~L. Pimentel, \emph{{The Cosmological Bootstrap: Inflationary Correlators from Symmetries and Singularities}}, \href{https://doi.org/10.1007/JHEP04(2020)105}{\emph{JHEP} {\bfseries 04} (2020) 105}, [\href{https://arxiv.org/abs/1811.00024}{{\ttfamily 1811.00024}}].

\bibitem{Baumann:2022jpr}
D.~Baumann, D.~Green, A.~Joyce, E.~Pajer, G.~L. Pimentel, C.~Sleight et~al., \emph{{Snowmass White Paper: The Cosmological Bootstrap}},  in \emph{{Snowmass 2021}}, 3, 2022, \href{https://arxiv.org/abs/2203.08121}{{\ttfamily 2203.08121}}.

\bibitem{Baumann:2019oyu}
D.~Baumann, C.~Duaso~Pueyo, A.~Joyce, H.~Lee and G.~L. Pimentel, \emph{{The cosmological bootstrap: weight-shifting operators and scalar seeds}}, \href{https://doi.org/10.1007/JHEP12(2020)204}{\emph{JHEP} {\bfseries 12} (2020) 204}, [\href{https://arxiv.org/abs/1910.14051}{{\ttfamily 1910.14051}}].

\bibitem{Baumann:2020dch}
D.~Baumann, C.~Duaso~Pueyo, A.~Joyce, H.~Lee and G.~L. Pimentel, \emph{{The Cosmological Bootstrap: Spinning Correlators from Symmetries and Factorization}}, \href{https://doi.org/10.21468/SciPostPhys.11.3.071}{\emph{SciPost Phys.} {\bfseries 11} (2021) 071}, [\href{https://arxiv.org/abs/2005.04234}{{\ttfamily 2005.04234}}].

\bibitem{Pajer:2020wnj}
E.~Pajer, D.~Stefanyszyn and J.~Supe\l{}, \emph{{The Boostless Bootstrap: Amplitudes without Lorentz boosts}}, \href{https://doi.org/10.1007/JHEP12(2020)198}{\emph{JHEP} {\bfseries 12} (2020) 198}, [\href{https://arxiv.org/abs/2007.00027}{{\ttfamily 2007.00027}}].

\bibitem{Pajer:2020wxk}
E.~Pajer, \emph{{Building a Boostless Bootstrap for the Bispectrum}}, \href{https://doi.org/10.1088/1475-7516/2021/01/023}{\emph{JCAP} {\bfseries 01} (2021) 023}, [\href{https://arxiv.org/abs/2010.12818}{{\ttfamily 2010.12818}}].

\bibitem{ArkaniHamed2017}
N.~Arkani-Hamed, P.~Benincasa and A.~Postnikov, \emph{Cosmological polytopes and the wavefunction of the universe},  \href{https://arxiv.org/abs/1709.02813}{{\ttfamily 1709.02813}}.

\bibitem{Benincasa:2024leu}
P.~Benincasa and G.~Dian, \emph{{The Geometry of Cosmological Correlators}},  \href{https://arxiv.org/abs/2401.05207}{{\ttfamily 2401.05207}}.

\bibitem{Benincasa:2019vqr}
P.~Benincasa, \emph{{Cosmological Polytopes and the Wavefuncton of the Universe for Light States}},  \href{https://arxiv.org/abs/1909.02517}{{\ttfamily 1909.02517}}.

\bibitem{Hillman:2019wgh}
A.~Hillman, \emph{{Symbol Recursion for the dS Wave Function}},  \href{https://arxiv.org/abs/1912.09450}{{\ttfamily 1912.09450}}.

\bibitem{Salcedo:2022aal}
S.~A. Salcedo, M.~H.~G. Lee, S.~Melville and E.~Pajer, \emph{{The Analytic Wavefunction}}, \href{https://doi.org/10.1007/JHEP06(2023)020}{\emph{JHEP} {\bfseries 06} (2023) 020}, [\href{https://arxiv.org/abs/2212.08009}{{\ttfamily 2212.08009}}].

\bibitem{Lee:2023kno}
M.~H.~G. Lee, \emph{{From amplitudes to analytic wavefunctions}}, \href{https://doi.org/10.1007/JHEP03(2024)058}{\emph{JHEP} {\bfseries 03} (2024) 058}, [\href{https://arxiv.org/abs/2310.01525}{{\ttfamily 2310.01525}}].

\bibitem{De:2023xue}
S.~De and A.~Pokraka, \emph{{Cosmology meets cohomology}}, \href{https://doi.org/10.1007/JHEP03(2024)156}{\emph{JHEP} {\bfseries 03} (2024) 156}, [\href{https://arxiv.org/abs/2308.03753}{{\ttfamily 2308.03753}}].

\bibitem{He:2024olr}
S.~He, X.~Jiang, J.~Liu, Q.~Yang and Y.-Q. Zhang, \emph{{Differential equations and recursive solutions for cosmological amplitudes}},  \href{https://arxiv.org/abs/2407.17715}{{\ttfamily 2407.17715}}.

\bibitem{gasparotto2024}
F.~Gasparotto, P.~Mazloumi and X.~Xu, \emph{{Differential equations for tree--level cosmological correlators with massive states}},  \href{https://arxiv.org/abs/2411.05632}{{\ttfamily 2411.05632}}.

\bibitem{Arkani-Hamed:2023kig}
N.~Arkani-Hamed, D.~Baumann, A.~Hillman, A.~Joyce, H.~Lee and G.~L. Pimentel, \emph{{Differential Equations for Cosmological Correlators}},  \href{https://arxiv.org/abs/2312.05303}{{\ttfamily 2312.05303}}.

\bibitem{Arkani-Hamed:2023bsv}
N.~Arkani-Hamed, D.~Baumann, A.~Hillman, A.~Joyce, H.~Lee and G.~L. Pimentel, \emph{{Kinematic Flow and the Emergence of Time}},  \href{https://arxiv.org/abs/2312.05300}{{\ttfamily 2312.05300}}.

\bibitem{Baumann:2024mvm}
D.~Baumann, H.~Goodhew and H.~Lee, \emph{{Kinematic Flow for Cosmological Loop Integrands}},  \href{https://arxiv.org/abs/2410.17994}{{\ttfamily 2410.17994}}.

\bibitem{Hang:2024xas}
Y.~Hang, \emph{{A Note on Kinematic Flow and Differential Equations for Two-Site One-Loop Graph in FRW Spacetime}},  \href{https://arxiv.org/abs/2410.17192}{{\ttfamily 2410.17192}}.

\bibitem{Benincasa:2024ptf}
P.~Benincasa, G.~Brunello, M.~K. Mandal, P.~Mastrolia and F.~Vaz\~ao, \emph{{On one-loop corrections to the Bunch-Davies wavefunction of the universe}},  \href{https://arxiv.org/abs/2408.16386}{{\ttfamily 2408.16386}}.

\bibitem{Benincasa:2024lxe}
P.~Benincasa and F.~Vaz\~ao, \emph{{The Asymptotic Structure of Cosmological Integrals}},  \href{https://arxiv.org/abs/2402.06558}{{\ttfamily 2402.06558}}.

\bibitem{Chen:2024glu}
J.~Chen, B.~Feng and Y.-X. Tao, \emph{{Multivariate hypergeometric solutions of cosmological (dS) correlators by $\text{d} \log$-form differential equations}},  \href{https://arxiv.org/abs/2411.03088}{{\ttfamily 2411.03088}}.

\bibitem{Grimm:2024tbg}
T.~W. Grimm and A.~Hoefnagels, \emph{{Reductions of GKZ Systems and Applications to Cosmological Correlators}},  \href{https://arxiv.org/abs/2409.13815}{{\ttfamily 2409.13815}}.

\bibitem{Fevola:2024nzj}
C.~Fevola, G.~L. Pimentel, A.-L. Sattelberger and T.~Westerdijk, \emph{{Algebraic Approaches to Cosmological Integrals}},  \href{https://arxiv.org/abs/2410.14757}{{\ttfamily 2410.14757}}.

\bibitem{Fan:2024iek}
B.~Fan and Z.-Z. Xianyu, \emph{{Cosmological Amplitudes in Power-Law FRW Universe}},  \href{https://arxiv.org/abs/2403.07050}{{\ttfamily 2403.07050}}.

\bibitem{Xianyu:2023ytd}
Z.-Z. Xianyu and J.~Zang, \emph{{Inflation correlators with multiple massive exchanges}}, \href{https://doi.org/10.1007/JHEP03(2024)070}{\emph{JHEP} {\bfseries 03} (2024) 070}, [\href{https://arxiv.org/abs/2309.10849}{{\ttfamily 2309.10849}}].

\bibitem{Ema:2024hkj}
Y.~Ema and K.~Mukaida, \emph{{Cutting rule for in-in correlators and cosmological collider}},  \href{https://arxiv.org/abs/2409.07521}{{\ttfamily 2409.07521}}.

\bibitem{Ghosh:2024aqd}
D.~Ghosh, E.~Pajer and F.~Ullah, \emph{{Cosmological cutting rules for Bogoliubov initial states}},  \href{https://arxiv.org/abs/2407.06258}{{\ttfamily 2407.06258}}.

\bibitem{AguiSalcedo:2023nds}
S.~Agui~Salcedo and S.~Melville, \emph{{The cosmological tree theorem}}, \href{https://doi.org/10.1007/JHEP12(2023)076}{\emph{JHEP} {\bfseries 12} (2023) 076}, [\href{https://arxiv.org/abs/2308.00680}{{\ttfamily 2308.00680}}].

\bibitem{Tong:2021wai}
X.~Tong, Y.~Wang and Y.~Zhu, \emph{{Cutting rule for cosmological collider signals: a bulk evolution perspective}}, \href{https://doi.org/10.1007/JHEP03(2022)181}{\emph{JHEP} {\bfseries 03} (2022) 181}, [\href{https://arxiv.org/abs/2112.03448}{{\ttfamily 2112.03448}}].

\bibitem{Baumann:2021fxj}
D.~Baumann, W.-M. Chen, C.~Duaso~Pueyo, A.~Joyce, H.~Lee and G.~L. Pimentel, \emph{{Linking the singularities of cosmological correlators}}, \href{https://doi.org/10.1007/JHEP09(2022)010}{\emph{JHEP} {\bfseries 09} (2022) 010}, [\href{https://arxiv.org/abs/2106.05294}{{\ttfamily 2106.05294}}].

\bibitem{Goodhew:2021oqg}
H.~Goodhew, S.~Jazayeri, M.~H.~G. Lee and E.~Pajer, \emph{{Cutting cosmological correlators}}, \href{https://doi.org/10.1088/1475-7516/2021/08/003}{\emph{JCAP} {\bfseries 08} (2021) 003}, [\href{https://arxiv.org/abs/2104.06587}{{\ttfamily 2104.06587}}].

\bibitem{Melville:2021lst}
S.~Melville and E.~Pajer, \emph{{Cosmological Cutting Rules}}, \href{https://doi.org/10.1007/JHEP05(2021)249}{\emph{JHEP} {\bfseries 05} (2021) 249}, [\href{https://arxiv.org/abs/2103.09832}{{\ttfamily 2103.09832}}].

\bibitem{Goodhew:2020hob}
H.~Goodhew, S.~Jazayeri and E.~Pajer, \emph{{The Cosmological Optical Theorem}}, \href{https://doi.org/10.1088/1475-7516/2021/04/021}{\emph{JCAP} {\bfseries 04} (2021) 021}, [\href{https://arxiv.org/abs/2009.02898}{{\ttfamily 2009.02898}}].

\bibitem{Arkani-Hamed:2018bjr}
N.~Arkani-Hamed and P.~Benincasa, \emph{{On the Emergence of Lorentz Invariance and Unitarity from the Scattering Facet of Cosmological Polytopes}},  \href{https://arxiv.org/abs/1811.01125}{{\ttfamily 1811.01125}}.

\bibitem{Werth:2024mjg}
D.~Werth, \emph{{Spectral Representation of Cosmological Correlators}},  \href{https://arxiv.org/abs/2409.02072}{{\ttfamily 2409.02072}}.

\bibitem{Goodhew:2024eup}
H.~Goodhew, A.~Thavanesan and A.~C. Wall, \emph{{The Cosmological CPT Theorem}},  \href{https://arxiv.org/abs/2408.17406}{{\ttfamily 2408.17406}}.

\bibitem{aomoto2011theory}
K.~Aomoto, M.~Kita, T.~Kohno and K.~Iohara, \emph{Theory of Hypergeometric Functions}.
\newblock Springer Monographs in Mathematics. Springer Japan, 2011.

\bibitem{yoshida2013hypergeometric}
M.~Yoshida, \emph{Hypergeometric functions, my love: modular interpretations of configuration spaces}, vol.~32.
\newblock Springer Science \& Business Media, 2013.

\bibitem{Matsubara-Heo:2023ylc}
S.-J. Matsubara-Heo, S.~Mizera and S.~Telen, \emph{{Four Lectures on Euler Integrals}},  \href{https://arxiv.org/abs/2306.13578}{{\ttfamily 2306.13578}}.

\bibitem{Caron-Huot:2021iev}
S.~Caron-Huot and A.~Pokraka, \emph{{Duals of Feynman Integrals. Part II. Generalized unitarity}}, \href{https://doi.org/10.1007/JHEP04(2022)078}{\emph{JHEP} {\bfseries 04} (2022) 078}, [\href{https://arxiv.org/abs/2112.00055}{{\ttfamily 2112.00055}}].

\bibitem{Mastrolia:2018uzb}
P.~Mastrolia and S.~Mizera, \emph{{Feynman Integrals and Intersection Theory}}, \href{https://doi.org/10.1007/JHEP02(2019)139}{\emph{JHEP} {\bfseries 02} (2019) 139}, [\href{https://arxiv.org/abs/1810.03818}{{\ttfamily 1810.03818}}].

\bibitem{Caron-Huot:2021xqj}
S.~Caron-Huot and A.~Pokraka, \emph{{Duals of Feynman integrals. Part I. Differential equations}}, \href{https://doi.org/10.1007/JHEP12(2021)045}{\emph{JHEP} {\bfseries 12} (2021) 045}, [\href{https://arxiv.org/abs/2104.06898}{{\ttfamily 2104.06898}}].

\bibitem{Frellesvig:2019uqt}
H.~Frellesvig, F.~Gasparotto, M.~K. Mandal, P.~Mastrolia, L.~Mattiazzi and S.~Mizera, \emph{{Vector Space of Feynman Integrals and Multivariate Intersection Numbers}}, \href{https://doi.org/10.1103/PhysRevLett.123.201602}{\emph{Phys. Rev. Lett.} {\bfseries 123} (2019) 201602}, [\href{https://arxiv.org/abs/1907.02000}{{\ttfamily 1907.02000}}].

\bibitem{Mizera:2019vvs}
S.~Mizera and A.~Pokraka, \emph{{From Infinity to Four Dimensions: Higher Residue Pairings and Feynman Integrals}}, \href{https://doi.org/10.1007/JHEP02(2020)159}{\emph{JHEP} {\bfseries 02} (2020) 159}, [\href{https://arxiv.org/abs/1910.11852}{{\ttfamily 1910.11852}}].

\bibitem{Henn:2013pwa}
J.~M. Henn, \emph{{Multiloop integrals in dimensional regularization made simple}}, \href{https://doi.org/10.1103/PhysRevLett.110.251601}{\emph{Phys. Rev. Lett.} {\bfseries 110} (2013) 251601}, [\href{https://arxiv.org/abs/1304.1806}{{\ttfamily 1304.1806}}].

\bibitem{Lee:2023jby}
M.~H.~G. Lee, C.~McCulloch and E.~Pajer, \emph{{Leading loops in cosmological correlators}}, \href{https://doi.org/10.1007/JHEP11(2023)038}{\emph{JHEP} {\bfseries 11} (2023) 038}, [\href{https://arxiv.org/abs/2305.11228}{{\ttfamily 2305.11228}}].

\bibitem{Chowdhury:2023arc}
C.~Chowdhury, A.~Lipstein, J.~Mei, I.~Sachs and P.~Vanhove, \emph{{The Subtle Simplicity of Cosmological Correlators}},  \href{https://arxiv.org/abs/2312.13803}{{\ttfamily 2312.13803}}.

\bibitem{Arkani-Hamed:2017tmz}
N.~Arkani-Hamed, Y.~Bai and T.~Lam, \emph{{Positive Geometries and Canonical Forms}}, \href{https://doi.org/10.1007/JHEP11(2017)039}{\emph{JHEP} {\bfseries 11} (2017) 039}, [\href{https://arxiv.org/abs/1703.04541}{{\ttfamily 1703.04541}}].

\bibitem{coaction}
A.~McLeod, A.~Pokraka and L.~Ren, ``{A diagrammatic coaction for cosmological correlators (work in progress)}.''

\bibitem{Abreu:2019wzk}
S.~Abreu, R.~Britto, C.~Duhr, E.~Gardi and J.~Matthew, \emph{{From positive geometries to a coaction on hypergeometric functions}}, \href{https://doi.org/10.1007/JHEP02(2020)122}{\emph{JHEP} {\bfseries 02} (2020) 122}, [\href{https://arxiv.org/abs/1910.08358}{{\ttfamily 1910.08358}}].

\bibitem{hwa1966homology}
R.~C. Hwa and V.~L. Teplitz, \emph{Homology and feynman integrals}, {\emph{(No Title)} (1966) }.

\bibitem{matsumoto1998kforms}
K.~Matsumoto, \emph{Intersection numbers for logarithmic $k$-forms}, {\emph{Osaka J. Math.} {\bfseries 35} (1998) 873--893}.

\bibitem{Benincasa:2020aoj}
P.~Benincasa, A.~J. McLeod and C.~Vergu, \emph{{Steinmann Relations and the Wavefunction of the Universe}}, \href{https://doi.org/10.1103/PhysRevD.102.125004}{\emph{Phys. Rev. D} {\bfseries 102} (2020) 125004}, [\href{https://arxiv.org/abs/2009.03047}{{\ttfamily 2009.03047}}].

\bibitem{Benincasa:2021qcb}
P.~Benincasa and W.~J.~T. Bobadilla, \emph{{Physical representations for scattering amplitudes and the wavefunction of the universe}}, \href{https://doi.org/10.21468/SciPostPhys.12.6.192}{\emph{SciPost Phys.} {\bfseries 12} (2022) 192}, [\href{https://arxiv.org/abs/2112.09028}{{\ttfamily 2112.09028}}].

\bibitem{dupont2015orliksolomonmodelhypersurfacearrangements}
C.~Dupont, \emph{The orlik-solomon model for hypersurface arrangements},  2015.

\bibitem{orlik1980combinatorics}
P.~Orlik and L.~Solomon, \emph{Combinatorics and topology of complements of hyperplanes}, {\emph{Inventiones mathematicae} {\bfseries 56} (1980) 167--189}.

\bibitem{pham2011singularities}
F.~Pham, \emph{Singularities of integrals: Homology, hyperfunctions and microlocal analysis}.
\newblock Universitext. Springer London, 2011.

\bibitem{Bourjaily:2017wjl}
J.~L. Bourjaily, E.~Herrmann and J.~Trnka, \emph{{Prescriptive Unitarity}}, \href{https://doi.org/10.1007/JHEP06(2017)059}{\emph{JHEP} {\bfseries 06} (2017) 059}, [\href{https://arxiv.org/abs/1704.05460}{{\ttfamily 1704.05460}}].

\end{thebibliography}\endgroup


\end{document}